\documentclass[12pt,letterpaper]{article}
\usepackage[left=1in,top=1in,right=1in,nohead]{geometry}
\usepackage{enumerate}
\usepackage{graphicx}
\usepackage{amsmath}
\usepackage{amsthm}
\usepackage{mathrsfs}
\usepackage{amssymb}
\usepackage{multirow}
\usepackage{color}
\usepackage{bm}
\usepackage{subfig}
\usepackage[justification=justified]{caption}
\usepackage{comment}
\usepackage{jheppub}
\usepackage{slashed}
\usepackage[T1]{fontenc}
\usepackage{mathtools}
\usepackage{afterpage}
\usepackage{customcommands}
\usepackage[normalem]{ulem}
\usepackage{paralist}

\usepackage{cleveref}

\title{Does the Round Sphere Maximize the Free Energy of (2+1)-Dimensional QFTs?}

\author{Sebastian Fischetti$^1$,}
\author{Lucas Wallis$^2$,}
\author{and Toby Wiseman$^2$}

\affiliation{$^1$Department of Physics, McGill University, Montr\'eal, QC, H3A 2T8, Canada}
\affiliation{$^2$Theoretical Physics Group, Blackett Laboratory, Imperial College, London SW7 2AZ, UK}

\emailAdd{fischetti@physics.mcgill.ca}
\emailAdd{l.wallis17@imperial.ac.uk}
\emailAdd{t.wiseman@imperial.ac.uk}

\abstract{
We examine the renormalized free energy of the free Dirac fermion and the free scalar on a~(2+1)-dimensional geometry~$\mathbb{R} \times \Sigma$, with~$\Sigma$ having spherical topology and prescribed area.  Using heat kernel methods, we perturbatively compute this energy when~$\Sigma$ is a small deformation of the round sphere, finding that at any temperature the round sphere is a local maximum.  At low temperature the free energy difference is due to the Casimir effect.  We then numerically compute this free energy for a class of large axisymmetric deformations, providing evidence that the round sphere \textit{globally} maximizes it, and we show that the free energy difference relative to the round sphere is unbounded below as the geometry on~$\Sigma$ becomes singular.  Both our perturbative and numerical results in fact stem from the stronger finding that the difference between the heat kernels of the round sphere and a deformed sphere always appears to have definite sign.  We investigate the relevance of our results to physical systems like monolayer graphene consisting of a membrane supporting relativistic QFT degrees of freedom.
}

\begin{document}

\maketitle
\flushbottom


\section{Introduction}
\label{sec:intro}

The equilibrium configuration of a physical membrane is often determined by a competition between several physical effects.  For instance, the round shape of a soap bubble arises from a competition between the bubble's intrinsic surface tension, which energetically prefers to collapse it, and a pressure differential between the air inside and outside of the bubble.  Likewise, the bending modulus of simple lipid bilayers tends to flatten them, with thermodynamic effects potentially causing deformations.

Here we are specifically interested in membranes supporting relativistic quantum degrees of freedom living on them.  A fiducial example is that of a graphene monolayer, whose energetics in a Born-Oppenheimer-like approximation can be split into a sum of two contributions: one from the atomic background lattice and another from relativistic excitations that propagate on this background.  At scales well above the lattice spacing, the lattice can be treated as a continuous membrane, and the free energy will depend on its geometry.  The contribution to this free energy from the background, interpreted as a \textit{classical} contribution~$F_c$, is then captured by a Landau free energy constructed from its embedding into an ambient flat space~\cite{NelPel87,PacKar88,NelPir89} (see~\cite{MembraneReview} for a review).  The effective relativistic excitations are two free massless Dirac fermions (with effective speed of light~$c_\mathrm{eff}$ given by the Fermi velocity~$ \sim 10^6$~m/s), and their free energy -- interpreted as a \textit{quantum} contribution~$F_q$ that depends on the membrane's instrinsic geometry -- is computed via an appropriate path integral.  Since the classical contribution to the free energy is relatively well-understood, our goal is to understand the contribution from the Dirac fermions.

In fact, while graphene is our motivating physical system (and one to which we will often turn for physical interpretation), there 
exist other more exotic examples of relativistic quantum fields supported on membranes -- for instance, domain walls in cosmology~\cite{CosmoBook} or braneworld models of our universe~\cite{ArkaniHamed:1998rs,Randall:1999ee}.  Moreover, one could also imagine engineering other graphene-like two-dimensional crystalline materials that exhibit relativistic excitations living on them.  Consequently, in this paper our main objective is to study the free energy of more general classes of relativistic QFTs living on~$(2+1)$-dimensional geometries, with the massless Dirac fermion corresponding to graphene as a special case\footnote{We note that studying the free energy on Euclidean three-dimensional geometries is also interesting and has fascinating links to quantum cosmology~\cite{Anninos:2012ft,BobBue17}.}.

Interestingly, previous work has found that such relativistic quantum fields tend to energetically prefer \textit{deformed} geometries.  To briefly summarize, let us assume that this field theory lives on~$\mathbb{R} \times \Sigma$, where~$\Sigma$ is a two-dimensional spatial manifold with metric~$g$.  Per the discussion above,~$F_q$ will depend on the intrinsic geometry of~$\Sigma$, i.e.~on~$g$.  Because~$F_q$ is a free energy, it is extensive, and therefore in order to sensibly discuss its dependence on the \textit{shape} of~$\Sigma$ we should imagine keeping the volume of~$\Sigma$ (computed with respect to~$g$) fixed as we vary~$g$\footnote{In other words, of course one can always make~$F_q$ arbitrarily large or small by simply varying the volume of~$\Sigma$ arbitrarily, but this is a standard volume-dependence that can be eliminated by, say, a classical tension term in~$F_c$.  More details will be provided in Section~\ref{subsec:freeenergyreview} below.}.  We therefore consider the ``background-subtracted'' free energy~$\Delta F_q \equiv F_q[g] - F_q[\bar{g}]$, with~$\bar{g}$ taken to be some fiducial reference metric which endows~$\Sigma$ with the same volume as~$g$.  With this understanding, the relevant extant results are summarized in Table~\ref{tab:results}.  The key takeaway is that~$\Delta F_q$ is negative for many different theories, a result most well-established when the geometry on~$\Sigma$ is perturbatively close to the round sphere or the flat plane.  Note that~$\Delta F_q$ remains nonzero even at zero temperature, when it can be interpreted as a Casimir energy.

\begin{table}
\centering
\begin{tabular}{|c|c|c|c|c|}
\hline
Theory & $(\Sigma,g)$ & Temp. & Result & Ref. \\
\hline
Holographic CFT & $(S^2,g)$ or $(\mathbb{R}^2,g)$ & $T = 0$ & $\Delta F_q \leq 0$ & \cite{HicWis15,FisHic16,Cheamsawat:2020crr} \\
Holographic CFT & $(\mathbb{R}^2,\bar{g}+\eps h)$ & $T \geq 0$ & $\Delta F_q \leq 0$ & \cite{CheWal18} \\
Holographic CFT & $(\mathbb{R}^2,\mbox{long wavelength})$ & $T \geq 0$ & $\Delta F_q \leq 0$ & \cite{CheWal18} \\
Holographic CFT & $(T^2, g)$ & $T \geq 0$ & $\Delta E \leq 0$ & \cite{CheGib19} \\
Unitary CFT & $(S^2,\bar{g}+\eps h)$ or $(\mathbb{R}^2,\bar{g}+\eps h)$ & $T = 0$ & $\Delta F_q \leq 0$ & \cite{FisWis17} \\
Free scalar or fermion & $(\mathbb{R}^2,\bar{g}+\eps h)$ & $T \geq 0$ & $\Delta F_q \leq 0$ & \cite{FisWal18} \\
Free scalar or fermion & $(\mathbb{R}^2,\mbox{long wavelength})$ & $T \geq 0$ & $\Delta F_q \leq 0$ & \cite{CheWal18} \\
\hline
\end{tabular}
\caption{A summary of results for the free energy~$F_q$ (or just energy~$E$ in the fourth row) for various types QFTs at various temperatures. ``Holographic CFTs'' refer to CFTs dual to smooth geometries obeying Einstein's equations; the scalar refers to a scalar field of any mass and curvature coupling; and the fermion refers to a Dirac fermion of any mass.  $\bar{g}$ is always taken to be the maximally symmetric geometry on the manifold~$\Sigma$ (so~$\bar{g}$ is the round sphere metric when~$\Sigma = S^2$ and the flat metric when~$\Sigma = \mathbb{R}^2$ or~$T^2$), and when we write~$\bar{g} + \eps h$ it is understood that the result holds to leading nontrivial order in the perturbative expansion parameter~$\eps$.  ``Long wavelength'' refers to metrics whose curvature is small compared to the temperature and/or mass of the field.  All inequalities on~$\Delta F_q$ and~$\Delta E$ are saturated if and only if~$g = \bar{g}$.}
\label{tab:results}
\end{table}

These observations naturally lead to the following question: if the free energy~$F_q$ governs the equilibrium configuration of a membrane, does the fact that~$\Delta F_q$ is always negative lead to an instability of the round sphere or flat space?  If so, will a membrane settle down to some less-symmetric equilibrium configuration, or does this instability ultimately lead to a runaway process (which presumably breaks down once a UV scale is reached)?  Answering this question will of course depend on how~$\Delta F_q$ competes with other contributions to the free energy; returning to the case of graphene, in~\cite{FisWal18} we performed a parametric comparison of the competition between~$\Delta F_q$ and the classical bending free energy~$\Delta F_c$, finding that the typical curvature scale~$l_\mathrm{crit}$ at which the negative contribution of~$\Delta F_q$ becomes dominant over the (positive) contribution of~$\Delta F_c$ agrees with the ``rippling'' length scale~$l_\mathrm{rip}$ of graphene measured in experiments~\cite{GrapheneRippleExpt}.  However, the order of magnitude of this scale is only slightly above that of the lattice spacing, where the effective Dirac fermion description breaks down, so the validity of our estimates for~$\Delta F_q$ and~$\Delta F_c$ in this regime is suspect.

Moreover, the order-of-magnitude analysis of~\cite{FisWal18} was made subtle for two reasons.  First, since~$\Sigma$ is a plane, its volume is infinite; hence one must be careful in defining precisely what is meant by the condition that the volumes computed from~$g$ and~$\bar{g}$ match.  Second, the leading perturbation to~$\Delta F_c$ is linear in the deformation amplitude~$\eps$, while~$\Delta F_q$ is quadratic in~$\eps$; hence balancing these two contributions requires a careful accounting of various orders-of-limits.

Our first purpose here is therefore to repeat the perturbative analysis of~\cite{FisWal18} -- that is, the perturbative computation of~$\Delta F_q$ for the nonminimally coupled free scalar and the free Dirac fermion -- in the case where~$\Sigma$ is a topological sphere, rather than a plane.  This modification alleviates both of the issues just mentioned, since when~$\Sigma$ is a sphere it has finite volume and we also find that both the contributions~$\Delta F_c$ and~$\Delta F_q$ are quadratic in~$\eps$.  Again we find that the round sphere locally mazimizes~$F_q$.

Our second purpose is then to investigate the questions posed above: namely, is the round sphere a \textit{global} maximum of~$\Delta F_q$?  Does~$\Delta F_q$ eventually find some new equilibrium configuration after a sufficiently large deformation to the sphere, or can it decrease indefinitely?  To address these questions, we numerically compute~$\Delta F_q$ for large (axisymmetric) deformations of the round sphere, finding that it is always negative, and in fact that it can be made arbitrarily negative as the geometry becomes singular.  Our conclusion, therefore, is that the energetics of~$\Delta F_q$ favor geometries that are not smooth.  Surprisingly, we also find that the behavior of~$\Delta F_q$ for such large deformations is remarkably similar for the scalar and the fermion when normalized by its perturbative expression.

In fact, our main result is stronger: not only is~$\Delta F_q$ always negative for the deformations we study, but the heat kernel~$\Delta K(t)$ which computes~$\Delta F_q$ has definite sign for all~$t$.  This heat kernel will be introduced in more detail in Section~\ref{sec:setup}, but in short it is related to the eigenvalues~$\lambda_I$ of the operator~$L$ that defines the equations of motion of the free fields:
\be
\Delta K_L(t) = \sum_I \left(e^{-t\lambda_I} - e^{-t\bar{\lambda}_I}\right),
\ee
where~$\lambda_I$ and~$\bar{\lambda}_I$ are the eigenvalues of~$L$ on the deformed sphere and the round sphere, respectively.  The fact that~$\Delta K_L(t)$ apparently has fixed sign for all~$t$ is therefore a nontrivial statement about the behavior of the eigenvalues of~$L$.  The universality of this result leads us to conjecture that~$\Delta K_L(t)$ has fixed sign for \textit{any} free field theory and area-preserving deformation of the sphere. 

The order-of-magnitude analysis of the competition between~$\Delta F_c$ and~$\Delta F_q$ is provided immediately below, in Section~\ref{subsec:graphene}, for the sake of illustrating more clearly some of the concepts discussed so far.  We then establish our setup, conventions, and formalism in Section~\ref{sec:setup}, focusing specifically on the computation of~$\Delta F_q$ using heat kernels.  We then present the perturbative calculation of~$\Delta F_q$ for the free nonminimally coupled scalar and the free Dirac fermion in Section~\ref{sec:pert}, along with some checks showing that our results reproduce the CFT result of~\cite{FisWis17} and the flat-space result of~\cite{FisWal18} in appropriate limits.  We then present the numerical calculation of~$\Delta F_q$ for large deformations of the sphere in Section~\ref{sec:nonpert}, focusing for simplicity on axisymmetric perturbations.  Finding that~$F_q$ seems to decrease monotonically as the amplitude of the deformation is increased, in Section~\ref{sec:endstate} we analyze its behavior on extremely deformed geometries, showing that approaching a conical singularity allows~$\Delta F_q$ to become arbitrarily negative.  Section~\ref{sec:conc} concludes with a summary of our main conclusions and unexpected results.

\subsection{A Perturbative Example: Graphene}
\label{subsec:graphene}

For illustrative purposes, let us now study in some more detail the competition between~$\Delta F_c$ and~$\Delta F_q$ for two-dimensional crystalline materials such as monolayer graphene, focusing on the case where~$\Sigma$ is a small deformation of a round sphere (for large deformations, this competition will be discussed in Section~\ref{subsec:grapheneendstate}).  We note there is considerable technological interest in producing \textit{spherical} monolayer graphene (see for example~\cite{GrapheneSphere}).  As mentioned above, at scales much larger than the lattice spacing this crystal can be described as a smooth membrane, and the free energy will depend on the geometry of this membrane.  For simplicity, we will further assume that this effective description is diffeomorphism-invariant (although of course this is not expected to be the case for a crystalline material like graphene \cite{MembraneReview}).

For an order-of-magnitude estimate of perturbations to the round sphere, we also assume that the sphere minimizes the classical bending contribution~$\Delta F_c$ to the free energy.  Keeping only up to second derivatives, we may then write the Landau free energy as
\be
\label{eq:classicalfreeenergy}
\Delta F_c = \kappa \int d^2 x \, \sqrt{g} \left(K - \frac{2}{r_0}\right)^2,
\ee
where~$K$ is the mean curvature of~$\Sigma$,~$\kappa$ is a bending rigidity,~$r_0$ is the radius of the sphere that minimizes~$\Delta F_c$, and no term containing the scalar curvature of~$g$ appears because such a term is topological.  Working perturbatively around the sphere of radius~$r_0$, we write the ambient flat space in the usual spherical coordinates
\be
ds^2_{\mathbb{R}^3} = dr^2 + r^2 \left(d\theta^2 + \sin^2\theta \, d\phi^2 \right)
\ee
and take~$\{\theta,\phi\}$ as coordinates on~$\Sigma$ and embed~$\Sigma$ as~$r = r_0 (1 + \eps f(\theta,\phi))$, where~$\eps$ is a dimensionless expansion parameter.  To linear order in~$\eps$, the induced metric on~$\Sigma$ is in a gauge conformal to the round sphere,
\be
\label{eq:inducedmetric}
ds^2_\Sigma = r_0^2\left(1 + 2\eps f\right)\left(d\theta^2 + \sin^2\theta \, d\phi^2 \right) + \Ocal(\eps^2),
\ee
while the free energy~\eqref{eq:classicalfreeenergy} becomes\footnote{Breaking diffeomorphism invariance would allow for more general coefficients in front of the~$f^2$,~$f \overline{\grad}^2 f$, and~$(\overline{\grad}^2 f)^2$ terms.}
\be
\Delta F_c = \eps^2 \kappa \int d\theta \, d\phi \, \sin\theta \left(2f + \overline{\grad}^2 f \right)^2 + \Ocal(\eps^3),
\ee
where~$\overline{\grad}^2$ denotes the Laplacian on the round sphere of unit radius.  We then decompose~$f$ in spherical harmonics as
\be
\label{eq:fdecomposition}
f = \sum_{\ell,m} f_{\ell,m} Y_{\ell,m},
\ee
with the condition that the volume of~$\Sigma$ remain unchanged imposing that~$f_{0,0} = 0$.  We thus obtain
\be
\Delta F_c = \eps^2 \kappa \sum_{\ell,m} |f_{\ell,m}|^2 (\ell-1)^2(\ell+2)^2 + \Ocal(\eps^3).
\ee

The general contribution of quantum scalar or Dirac fermionic fields to~$\Delta F_q$ is obtained in Section~\ref{sec:pert} below.  To streamline the present analysis, let us take the relativistic quantum fields living on~$\Sigma$ to be a CFT; this is the case for graphene when it is slightly perturbed from a flat plane (additional gauge fields associated to the underlying lattice structure vanish when the the metric is in a conformally flat form)~\cite{Graphene,GrapheneDirac1,GrapheneDirac2,WagJua19}.  Here we assume the effective CFT description remains valid even for small perturbations of the round sphere.  At zero temperature, the contribution of these degrees of freedom to~$\Delta F_q$ is~\cite{FisWis17}
\be
\label{eq:Fquant}
\Delta F_q = -\eps^2 \frac{\pi^2 c_T \hbar c_\mathrm{eff}}{48 r_0} \sum_{\ell,m} |f_{\ell,m}|^2 \frac{(\ell^2-1)(\ell+2)}{\ell} \left( \frac{\Gamma\left(\frac{\ell+1}{2}\right)}{\Gamma\left(\frac{\ell}{2}\right)}\right)^2 + \Ocal(\eps^3),
\ee
where~$c_\mathrm{eff}$ is the effective speed of light for these relativistic degrees of freedom and~$c_T$ is the central charge (defined as the coefficient in the two-point function of the stress tensor); in our conventions, the central charges of a conformally coupled massless scalar field and of a massless Dirac fermion are~$c_T = (3/2)/(4\pi)^2$ and~$c_T = 3/(4\pi)^2$, respectively~\cite{CapCos89,BobBue17}.  Importantly, at large~$\ell$ the coefficients in the sum grow like~$\ell^3$, indicating that this contribution is non-local: that is, unlike~$\Delta F_c$ it does not arise from some local geometric functional.  Also note that while technically~\eqref{eq:Fquant} is only valid at zero temperature, the leading corrections to it go like~$e^{-l_T^2/(2l)^2}$, where~$l_T = \hbar c_\mathrm{eff}/(k_B T)$ is a thermal length scale and~$l$ is the typical length scale of the perturbation~$f$; hence~\eqref{eq:Fquant} holds for~$l_T \gtrsim 2l$.  (The corrections to the zero-temperature result will be discussed in Section~\ref{subsec:freeenergynonpert}.)

The combined contribution to the free energy from the classical and quantum contributions therefore goes like\footnote{We remark that the factor of~$r_0$ in~\eqref{eq:Ftotal} is illustrative of the aforementioned fact that in flat space, which can be obtained in the limit~$r_0 \to \infty$,~$\Delta F_c$ is of higher order in~$\eps$ than~$\Delta F_q$.}
\be
\label{eq:Ftotal}
\Delta F = \eps^2 \kappa \sum_{\ell,m} |f_{\ell,m}|^2 \left[A_\ell^{(c)} - \frac{\gamma}{r_0} \, A_\ell^{(q)} \right] + \Ocal(\eps^3),
\ee
where
\be
\label{eq:gamma}
\gamma \equiv \frac{\pi^2 c_T \hbar c_\mathrm{eff}}{48 \kappa}
\ee
is some characteristic length scale and
\be
A_\ell^{(c)} \equiv (\ell-1)^2(\ell+2)^2, \qquad A_\ell^{(q)} \equiv \frac{(\ell^2-1)(\ell+2)}{\ell} \left( \frac{\Gamma\left(\frac{\ell+1}{2}\right)}{\Gamma\left(\frac{\ell}{2}\right)}\right)^2.
\ee
We would like to investigate whether this combined expression can ever be \textit{negative} in its regime of validity.  This question can be investigated as follows: first note that at large~$\ell$,~$A_\ell^{(c)}$ goes like~$\ell^4$ while~$A_\ell^{(q)}$ only grows like~$\ell^3$, so the positive classical free energy will always dominate at sufficiently high angular momentum quantum number.  We must therefore investigate the behavior of the lowest modes:~$\ell = 0$ does not contribute since~$f_{0,0} = 0$, while the contribution of~$\ell = 1$ modes to both~$\Delta F_c$ and~$\Delta F_q$ vanishes due to the fact that such deformations correspond to infinitesimal diffeomorphisms.  However,~since $A_\ell^{(c)}$ vanishes quadratically around~$\ell = 1$ while~$A_\ell^{(q)}$ only vanishes \textit{linearly}, it is clear that for sufficiently small~$\ell - 1 > 0$,~$A_\ell^{(c)} - (\gamma/r_0) A_\ell^{(q)} < 0$.  Since~$\ell$ is an integer, making~$\Delta F$ negative therefore requires this to be true all the way to~$\ell = 2$, and hence
\be
\label{eq:gammaconstraint}
\gamma \geq \gamma_\mathrm{crit} \equiv \frac{A_{\ell=2}^{(c)}}{A_{\ell=2}^{(q)}} \, r_0 = \frac{32}{3\pi} \, r_0.
\ee
Since our analysis is only valid at scales well above the lattice spacing~$a$, we also require~$r_0 \gg a$, which implies~$\gamma \gg a$.

For the particular case of graphene, typically the bending rigidity is taken as~$\kappa \sim 1$~eV,~$a \sim 2.5$~\AA,~$c_\mathrm{eff} \sim 10^6$~m/s~\cite{GrapheneRippleMC}, and~$c_T = 2\times 3/(4\pi)^2$ (the factor of two coming from the two Dirac points in graphene's band structure), from which one finds~$\gamma/a \sim 0.1$ (note that the numerical prefactors matter: a purely parametric estimate would give~$\gamma/a \sim c_\mathrm{eff} \hbar /a \kappa \sim 10$).  Hence for graphene it does not seem likely that the quantum effect we have identified can ever compete with the classical bending energy to render the round sphere unstable, even if one were to keep more careful track of the precise form of the Landau free energy.  In the absense of fine-tuning, this result could have been expected: with no fine-tuning, the energy scale~$\kappa$ should be set by the lattice spacing and hence~$\kappa \sim \hbar c_\mathrm{eff}/a$, from which it would follow that~$\gamma/a$ is order unity\footnote{The large-$\ell$ scaling of~$a_q(\ell)$, which is necessary for this argument to go through, can be inferred by noting that perturbations with large~$\ell$ should be insensitive to the size of the sphere, and thus should behave as in flat space.  Interpreting~$k = \ell/r_0$ as a wave number for large~$\ell$, and knowing that~$\Delta F_q$ is quadratic in the perturbation~$f$, implies by dimensional analysis that~$\Delta F_q$ must go like~$k^3$.}.  We therefore interpret the condition~$\gamma \gg a$ as the required fine-tuning of the membrane parameters (e.g.~$\kappa$,~$c_\mathrm{eff}$) that makes it possible for~$\Delta F_q$ to dominate over~$\Delta F_c$.  Given the great current interest in monolayer graphene-like materials, conceivably such fine-tuned crystalline membranes could be engineered in a lab.


\section{Setup}
\label{sec:setup}

We consider thermal states of (2+1)-dimensional (unitary, relativistic) QFTs on the geometry~$\mathbb{R} \times \Sigma$, where~$\Sigma$ is a two-dimensional manifold with sphere topology.  The Euclidean continuation of this geometry is
\be
\label{eq:metric}
ds^2 = d\tau^2 + g_{ij}(x) dx^i dx^j,
\ee
with the period of Euclidean time~$\tau$ given by the inverse temperature~$\beta = 1/T$, and we have made explicit the fact that the spatial metric~$g_{ij}(x)$ on~$\Sigma$ is independent of~$\tau$.  The free energy is thus a functional of~$g_{ij}$ and of~$\beta$; to simplify notation, we will denote this free energy simply as~$F[\beta,g]$ (i.e.~without the subscript~$q$ as was used above).

\subsection{Free Energy}
\label{subsec:freeenergyreview}

The desired free energy~$F$ is determined by the Euclidean partition function~$Z$, which will depend on both~$\beta$ and the spatial geometry~$g_{ij}$:
\be
Z[\beta,g] = \int \Dcal \Phi \, e^{-S_E[\Phi;\beta,g]} = e^{-\beta F[\beta,g]},
\ee
where~$S_E$ is the Euclidean action and~$\Phi$ schematically stands for the QFT fields in the system.  Of course, as written~$Z$ (and thus~$F$) is UV-divergent, so we must regulate it.  Since we are only considering relativistic QFTs, any UV regulator (like, say, a lattice) cannot break diffeomorphism invariance in the IR, and hence for simplicity we may use a covariant UV regulator to ultimately compute UV-finite quantities.  To that end, note that for a UV cutoff~$\Lambda$, the most general covariant counterterms that can be added to the Euclidean action are
\be
S_\mathrm{ct} = \int d\tau \, \int d^2 x \sqrt{g} \, \left[c_1 \Lambda^3 + c_2 \mu \Lambda^2 + (c_3 \mu^2 + c_4 R) \Lambda\right],
\ee
where~$\mu$ schematically stands for any parameter in the QFT with dimensions of energy, if one exists (for instance, a mass),~$R$ is the Ricci scalar of~$g$, and the theory-dependent coefficients~$c_i$ are dimensionless and independent of~$\Lambda$ and of the geometry.  Hence the most general divergence structure of the free energy takes the form
\be
\label{eq:Fexpansion}
F[\beta,g] = \mathrm{Vol}[g] (c_1  \Lambda^3 + c_2 \mu \Lambda^2 + c_3 \mu^2 \Lambda) + 4\pi c_4 \chi_\Sigma \Lambda + F_\mathrm{fin}[\beta,g],
\ee
where~$\chi_\Sigma$ is the Euler characteristic of~$\Sigma$ and~$F_\mathrm{fin}[\beta,g]$ is finite as~$\Lambda \to \infty$.  Note in particular that the divergence structure depends on~$g$ only through the volume~$\mathrm{Vol}[g]$ of~$\Sigma$; in the context of two-dimensional crystalline lattices discussed in Section~\ref{sec:intro}, one can think of these terms as contributing to some (UV cutoff-dependent) tension in the classical membrane action.  In other words, we may interpret the volume preservation condition as merely a convenient way of grouping the leading-order divergences in~\eqref{eq:Fexpansion} with the couplings in the classical membrane action.

Physical information about the free energy is contained in the finite part~$F_\mathrm{fin}$, but this object is not uniquely defined by the expansion~\eqref{eq:Fexpansion} (since a general change in the UV cutoff can induce a change in~$F_\mathrm{fin}$).  However, the differenced free energy~$\Delta F \equiv F[\beta,g] - F[\beta,\bar{g}]$ discussed above (in which~$\bar{g}$ is a reference metric such that~$\mathrm{Vol}[\bar{g}] = \mathrm{Vol}[g]$) \textit{is} scheme-independent.  As shown in~\cite{FisWal18}, this differenced free energy can be defined via
\be
e^{-\beta \Delta F} = \ev{e^{-\Delta S_E}}_{\bar{g}},
\ee
where~$\Delta S_E$ is the difference of the Euclidean actions constructed from~$g$ and~$\bar{g}$ and the expectation value on the right-hand side is taken in the thermal vacuum state (of inverse temperature~$\beta$) associated to the geometry~$\bar{g}$.

\subsection{Heat Kernels}
\label{subsec:heatkernel}

Let us now restrict to the case where the QFT fields~$\Phi$ are free; in such a case, the Euclidean action is quadratic, and the partition function reduces to a functional determinant.  The free energy is then conveniently evaluated via heat kernel methods, which we now review.  The massive free scalar fields and Dirac fermions on which we focus have actions
\begin{subequations}
\label{eqs:actions}
\begin{align}
S_E[\phi] &= \frac{1}{2} \int d\tau \, \int d^2 x \sqrt{g} \, \phi (-\grad^2 + \xi R + M^2) \phi, \\
S_E[\bar{\psi},\psi] &= \int d\tau \, \int d^2 x \sqrt{g} \, \bar{\psi}(i \slashed{D} - iM) \psi,
\end{align}
\end{subequations}
where~$\xi$ is the curvature coupling of the scalar,~$M$ is a mass, and the spinor conventions are as in~\cite{FisWal18}.  Performing the path integral on the geometry~\eqref{eq:metric}, one obtains~\cite{FisWal18}
\be
\label{eq:L}
Z = (\det \Lcal)^\sigma \mbox{ with } \Lcal = -\partial_\tau^2 + L + M^2,
\ee
where~$\sigma = -1/2$ ($+1$) for the scalar (fermion) and~$L$ is a differential operator on~$\Sigma$.  For the non-minimally coupled scalar we have simply~$L = -\grad^2 + \xi R$, which acts on functions with spin weight zero.  The case of the Dirac fermion is slightly more complicated and we will give the full expression for~$L$ in~\eqref{eq:LDirac} below, but the key idea is that the square of the Dirac operator on the ultrastatic geometry~\eqref{eq:metric} is diagonal in the spinor indices and~$L$ is one of these two diagonal components, which acts on functions of spin weight~$1/2$.

We now define the heat kernel~$K_\Lcal(t) \equiv \Tr(e^{-t \Lcal})$, in terms of which the free energy is
\be
\beta F = - \ln Z = \sigma \int_0^\infty \frac{dt}{t} \, K_\Lcal(t).
\ee
This form of the free energy makes manifest its UV divergence structure, as UV divergences are associated with small~$t$ in the above integral.  More explicitly, by the heat kernel expansion~\cite{Vas03} the small-$t$ behavior of~$K_\Lcal(t)$ goes like
\be
\label{eq:Kexpansion}
K_\Lcal(t) = \frac{\beta}{\sqrt{4\pi} \, t^{3/2}} \sum_{n = 0}^\infty b_{2n} t^{n},
\ee
where the coefficients~$b_{2n}$ can be expressed as integrals of local geometric invariants on~$(\Sigma,g_{ij})$.  UV divergences are controlled by the leading and subleading coefficients~$b_0$ and~$b_2$, which depend only on the volume and topology of~$(\Sigma,g_{ij})$ (though they are otherwise theory-dependent):
\be
\label{eq:b0b2}
b_0 \propto \int d^2 x \, \sqrt{g} = \mathrm{Vol}[g], \qquad b_2 \propto \int d^2 x\, \sqrt{g} \, R = 4\pi \chi_\Sigma.
\ee
Thus the differenced free energy can be obtained directly from the difference~$\Delta K_\Lcal(t)$ between heat kernels corresponding to the spatial geometries~$(\Sigma,g_{ij})$ and~$(\Sigma,\bar{g}_{ij})$:
\be
\beta \Delta F = \sigma \int_0^\infty \frac{dt}{t} \, \Delta K_\Lcal(t).
\ee
It is clear from~\eqref{eq:Kexpansion} and~\eqref{eq:b0b2} that as long as~$(\Sigma,g_{ij})$ and~$(\Sigma,\bar{g}_{ij})$ have the same volume and topology,~$\Delta K_\Lcal$ is~$\Ocal(t^{1/2})$ at small~$t$, and thus that~$\Delta F$ is UV-finite, as expected from the arguments above.

Now we may specify to our case of interest: using the decomposition~\eqref{eq:L} for~$\Lcal$, we obtain
\be
\label{eq:DeltaFheatkernel}
\beta \Delta F = \sigma \int_0^\infty \frac{dt}{t} \, e^{-M^2 t} \Theta_\sigma(T^2 t) \Delta K_L(t),
\ee
where~$\Delta K_L(t) \equiv K_L(t) - K_{\overline{L}}(t)$ is the difference of heat kernels of the operators~$L$ on the two-dimensional geometries~$(\Sigma,g_{ij})$ and~$(\Sigma,\bar{g}_{ij})$ and we have defined
\be
\Theta_\sigma(\zeta) \equiv \sum_{n = -\infty}^\infty e^{-(2\pi)^2 (n - \sigma + 1/2)^2 \zeta},
\ee
which arises from a sum over Matsubara frequencies on the thermal circle.  Moreover, we will be concerned with the case where~$\Sigma$ is a (topological) sphere, in which case it is natural to take the reference metric~$\bar{g}_{ij}$ to be that of a round sphere.  Finally, we note that the heat kernel expansion~\eqref{eq:Kexpansion} for~$\Delta K_L(t)$ takes the form
\be
\label{eq:DeltaKexpansion}
\Delta K_L(t) = t \sum_{n = 0}^\infty \Delta b_{2n+4} t^n,
\ee
where the~$\Delta b_{2n}$ are the differences of the heat kernel coefficients between the geometries~$(\Sigma,g_{ij})$ and~$(\Sigma, \bar{g}_{ij})$.


\section{Perturbative Results}
\label{sec:pert}

The expression~\eqref{eq:DeltaFheatkernel} for the differenced free energy in terms of the heat kernel of~$L$ is convenient because it simply requires computing the variation in the spectrum of~$L$ as the spatial geometry~$g$ is varied:
\be
\label{eq:trace}
\Delta K_L(t) = \Tr(e^{-tL}) - \Tr(e^{-t\overline{L}}) = \sum_I \left(e^{-t \lambda_I} - e^{-t \bar{\lambda}_I}\right),
\ee
where~$I$ indexes the eigenvalues of~$L$ and~$\overline{L}$.  An explicit computation of this perturbed heat kernel was performed for deformations of flat space in~\cite{FisWal18}, with the key result that for both the fermion and the scalar, to leading nontrivial order~$\sigma \Delta K_L(t)$ is negative for all~$t$ (and hence~$\Delta F$ is negative for all perturbations).  In order to compare to our later results, we now repeat this calculation on the perturbed round sphere~\eqref{eq:conformalsphere}.  We remind the reader that the reasons for working on the sphere are twofold: first, since the sphere is compact we don't have to deal with IR divergences; second, we will find that for small perturbations of the round sphere, the free energy of quantum fields is of the same order as the contribution from the classical membrane free energy, and hence the two can consistently be compared.

In this Section, we will take the metric on~$\Sigma$ to be conformal to the round sphere, as in~\eqref{eq:inducedmetric}:
\be
\label{eq:conformalsphere}
ds^2_\Sigma = e^{2f} \left(d\theta^2 + \sin^2\theta \, d\phi^2\right),
\ee
where~$f$ is some scalar field on the sphere and where we are using units in which~$r_0 = 1$.  We expand~$f = \eps f^{(1)} + \eps^2 f^{(2)} + \Ocal(\eps^3)$; the reference metric corresponds to taking~$\eps = 0$.  The volume preservation condition thus requires that\footnote{The reason for giving~$f$ a nontrivial expansion in~$\eps$, rather than just defining~$\eps$ via~$f = \eps f^{(1)}$ exactly, is that the second-order volume preservation constraint fixes~$f^{(1)} = 0$ exactly unless a nonzero~$f^{(2)}$ is turned on as well.}
\be
\label{eq:volpreservation}
4\pi = \int d^2 x \sqrt{g} \quad \Rightarrow \quad \int d^2 x \sqrt{\bar{g}} \, f^{(1)} = 0 \text{ and } \int d^2 x \sqrt{\bar{g}} \, \left(\left(f^{(1)}\right)^2 + f^{(2)} \right) = 0.
\ee
Similarly, we write the resulting expansion of~$L$ and of its eigenvalues~$\lambda_I$ and eigenvectors~$h_I$ as
\bea
L &= \overline{L} + \eps L^{(1)} + \eps^2 L^{(2)} + \Ocal(\eps^3), \label{subeq:Ln} \\
h_I &= \bar{h}_I + \eps h^{(1)}_I + \eps^2 h^{(2)}_I + \Ocal(\eps^3), \\
\lambda_I &= \bar{\lambda}_I + \eps \lambda_I^{(1)} + \eps^2 \lambda_I^{(2)} + \Ocal(\eps^3),
\eea
where the eplicit expressions for~$L^{(1)}$ and~$L^{(2)}$ in terms of~$f^{(1)}$ and~$f^{(2)}$ are provided in Appendix~\ref{app:perttheory}.  Hence from~\eqref{eq:trace}, the perturbed heat kernel is
\begin{subequations}
\be
\Delta K_L(t) = \eps \, \Delta K^{(1)}(t) + \eps^2 \Delta K^{(2)}(t) + \Ocal(\eps^3),
\ee
where
\be
\Delta K^{(1)}(t) = - t \sum_I e^{-\bar{\lambda}_I t} \lambda_I^{(1)}, \qquad \Delta K^{(2)}(t) =  t \sum_I e^{-\bar{\lambda}_I t} \left(\frac{t}{2} \left(\lambda_I^{(1)}\right)^2 - \lambda_I^{(2)}\right).
\ee
\end{subequations}
Homogeneity of the round sphere implies that the leading variation of $\Delta K_L$ is quadratic, so $\Delta K^{(1)} = 0$ which we indeed find shortly \cite{CheWal18}.

Now, defining the matrix elements
\be
L^{(n)}_{IJ} \equiv \me{\bar{h}_I}{L^{(n)}}{\bar{h}_J} \equiv \int d^2 x \sqrt{\bar{g}} \, \bar{h}_I^* L^{(n)} \bar{h}_J,
\ee
a standard consistency condition in degenerate perturbation theory requires that~$L^{(1)}_{IJ}$ be diagonal on any degenerate subspaces of~$\overline{L}$ (that is, we must have~$L^{(1)}_{IJ} = 0$ for any~$I,J$ with~$I \neq J$ but~$\bar{\lambda}_I = \bar{\lambda}_J$)\footnote{Should~$L^{(1)}$ not be sufficient to break all degeneracy, then~$L^{(2)}_{IJ}$ must be diagonal on any remaining denerate subspaces, and so on to higher orders.  Here we will only need to worry about the diagonalization of~$L^{(1)}$.}.  Then standard perturbation theory yields the perturbations of the eigenvalues:
\be
\lambda^{(1)}_I = L^{(1)}_{II}, \qquad \lambda^{(2)}_I = \sum_{\substack{J \\ \bar{\lambda}_J \neq \bar{\lambda}_I}} \frac{L^{(1)}_{IJ} L^{(1)}_{JI}}{\bar{\lambda}_I - \bar{\lambda}_J} + L^{(2)}_{II}.
\ee
It is important to note that while consistency of the perturbation theory requires an appropriate choice of the unperturbed eigenfunctions~$\bar{h}_I$, in fact the final expression for the heat kernel is insensitive to this choice.  To see this, let us write the index~$I$ as the pair~$(\ell,m)$, with~$\ell$ labeling each degenerate subspace of degeneracy~$d_\ell$ and~$m$ indexing its elements\footnote{This choice of labels is of course in analogy with the indexing of the spherical harmonics~$Y_{\ell,m}$, which are eigenfunctions of the Laplacian on the round sphere with degenerate eigenvalues~$\lambda_\ell$, but at this point the discussion is still completely general.}.  Then we may relate the eigenfunctions~$\bar{h}_{\ell,m}$ to any other basis~$\tilde{h}_{\ell,m}$ by a unitary transformation on each degenerate subspace:
\be
\label{eq:hdef}
\bar{h}_{\ell,m} = \sum_{m'} c^\ell_{m'm} \tilde{h}_{\ell,m'},
\ee
where~$c^\ell_{m'm}$ are the components of a unitary matrix chosen to ensure that~$L^{(1)}_{\ell,m,\ell,m'} = 0$ for~$m \neq m'$.  We then have
\be
\label{eq:LLtilde}
\bm{L}^{(n)}_{\ell,\ell'} = (\bm{c}^\ell)^\dag \widetilde{\bm{L}}^{(n)}_{\ell,\ell'} \bm{c}^{\ell'}, \mbox{ where } \widetilde{L}^{(n)}_{\ell,m,\ell',m'} \equiv \me{\tilde{h}_{\ell m}}{L^{(n)}}{\tilde{h}_{\ell',m'}},
\ee
where bold characters denote matrices on the degenerate subspaces, so that e.g.~$\bm{c}^{\ell}$ is the~$d_\ell \times d_\ell$-dimensional matrix with elements~$c^\ell_{m'm}$,~$\bm{L}^{(1)}_{\ell,\ell'}$ is the~$d_\ell \times d_{\ell'}$-dimensional matrix with elements~$L^{(1)}_{\ell,m,\ell',m'}$, etc.  Hence
\be
\Delta K^{(1)} = -t \sum_{\ell} e^{-\bar{\lambda}_\ell t} \Tr\left(\bm{L}^{(1)}_{\ell,\ell}\right) = -t \sum_{\ell} e^{-\bar{\lambda}_\ell t} \Tr\left(\widetilde{\bm{L}}^{(1)}_{\ell,\ell}\right),
\ee
with the final expression following from the basis-independence of the trace.  Likewise, we have
\be
\Delta K^{(2)} = t \sum_{\ell}^\infty e^{-\bar{\lambda}_\ell t} \left[\frac{t}{2} \sum_{m} \left(L^{(1)}_{\ell,m,\ell,m}\right)^2 - \Tr\left(\bm{L}^{(2)}_{\ell,\ell} + \sum_{\ell', \ell' \neq \ell} \frac{\bm{L}^{(1)}_{\ell,\ell'} \bm{L}^{(1)}_{\ell',\ell}}{\bar{\lambda}_\ell - \bar{\lambda}_{\ell'}}\right)\right],
\ee
but since~$\bm{L}^{(1)}_{\ell,\ell}$ is required to be diagonal, the first sum in the square brackets can be written simply as~$\Tr((\bm{L}^{(1)}_{\ell,\ell})^2)$.  Then again using~\eqref{eq:LLtilde} and cyclicity of the trace, we find that
\be
\label{eq:K2perturb}
\Delta K^{(2)} = t \sum_{\ell} e^{-\bar{\lambda}_\ell t} \Tr\left[\frac{t}{2} \left(\widetilde{\bm{L}}^{(1)}_{\ell,\ell}\right)^2 - \widetilde{\bm{L}}^{(2)}_{\ell,\ell} - \sum_{\ell', \ell' \neq \ell} \frac{\widetilde{\bm{L}}^{(1)}_{\ell,\ell'} \widetilde{\bm{L}}^{(1)}_{\ell',\ell}}{\bar{\lambda}_\ell - \bar{\lambda}_{\ell'}}\right].
\ee
All dependence on~$\bm{c}^\ell$ has vanished due to the traces, and hence for the purposes of computing the heat kernel we may compute the matrix elements~$\widetilde{L}^{(n)}_{\ell,m,\ell',m'}$ in any desired basis~$\tilde{h}_{\ell,m}$.

\subsection{Scalar}
\label{subsec:scalarpert}

For the scalar, the operator~$L$ for general~$f$ is
\be
\label{eq:Lscalar}
L = e^{-2f}\left[-\overline{\grad}^2 + 2\xi\left(1 - \overline{\grad}^2 f\right)\right],
\ee
with~$\overline{\grad}_a$ the covariant derivative on the round sphere ($f = 0$).  The unperturbed operator~$\overline{L}$ is~$-\overline{\grad}^2 + 2\xi$ and has eigenvalues~$\bar{\lambda}_\ell = \ell(\ell+1) + 2\xi$, with~$\ell \in \{0, 1, 2, \ldots\}$ a non-negative integer.  For the computation of the matrix elements~$\widetilde{L}^{(n)}_{\ell,m,\ell',m'}$, we may take the eigenfunctions~$\tilde{h}_{\ell,m}$ to just be the usual spherical harmonics~$Y_{\ell,m}$.  Since the calculation is rather cumbersome and unilluminating, we relegate it to Appendix~\ref{app:perttheory}; in short, expanding~$f^{(1)}$ in spherical harmonics as
\be
\label{eq:fexpansion}
f^{(1)} = \sum_{\ell,m} f_{\ell,m} Y_{\ell,m},
\ee
for the non-minimally coupled scalar one ultimately obtains $\Delta K^{(1)} = 0$ and
\be
\label{eq:K2scalar}
\Delta K^{(2)}(t) = \sum_{\ell,m} a_\ell(t) |f_{\ell,m}|^2, \qquad a_\ell(t) \equiv t \sum_{\ell' = 0}^\infty e^{-\bar{\lambda}_{\ell'} t} \left(\alpha_{\ell,\ell'} + \beta_{\ell,\ell'} t\right),
\ee
with the general expressions for~$\alpha_{\ell,\ell'}$ and~$\beta_{\ell,\ell'}$ given in~\eqref{subeq:betascalar} and~\eqref{eq:alphascalarclosed} in the Appendix.  For the special case of odd~$\ell$, the expressions simplify substantially to\footnote{Technically this expression for~$\alpha_{\ell,\ell'}$, as well as that given in~\eqref{eq:alphascalarclosed} for general~$\ell$, was obtained by evaluating~\eqref{eq:alphascalarrewritten} (which expresses~$\alpha_{\ell,\ell'}$ as a finite sum) for various values of~$\ell$,~$\ell'$ and then inferring a closed-form formula by using built-in sequence finders in~\texttt{Mathematica}.  Although we have checked that the resulting formula is correct for all values of~$\ell$,~$\ell'$ from zero to~100, we are unable to provide a general derivation.}
\begin{subequations}
\label{eq:alphabetascalarodd}
\begin{align}
\alpha_{\ell,\ell'} &= \begin{dcases} \frac{(2\ell'+1)(\bar{\lambda}_{\ell'} - \xi \ell(\ell+1))^2}{\pi  \ell(\ell+1)} \, \frac{\left(\frac{2+\ell}{2}\right)_{\ell'} \left(\frac{\ell}{2}\right)_{-\ell'}}{\left(\frac{3+\ell}{2}\right)_{\ell'} \left(\frac{1+\ell}{2}\right)_{-\ell'}}, & \ell' < \frac{\ell}{2} \\
0 , & \ell' > \frac{\ell}{2} \end{dcases}, \\
\beta_{\ell,\ell'} &= 0,
\end{align}
\end{subequations}
where~$(x)_n \equiv \Gamma(x+n)/\Gamma(x)$ are Pochhammer symbols.  We will comment further on this expression in Section~\ref{subsec:negativeDeltaK} below.

\subsection{Dirac Fermion}
\label{subsec:Diracpert}

For the benefit of the reader, let us briefly summarize how to obtain the operator~$L$ for the fermion; more details can be found in~\cite{FisWal18}.  We first evaluate
\be
(i\slashed{D}+iM)(i\slashed{D}-iM) = -D^2 + \frac{1}{4} R + M^2,
\ee
where~$D_a = \, ^{(3)} \grad_a + \omega_{a\mu\nu} S^{\mu\nu}/2$ is the spinor covariant derivative, with~$^{(3)} \grad_a$ the usual Levi-Civita connection on the full (three-dimensional) Euclidean geometry,~$\omega_{a\mu\nu}$ the spin connection, and~$S^{\mu\nu}$ the generators of the Lorentz group.  Evaluating this object in the ultrastatic geometry~\eqref{eq:metric}, one finds that it is diagonal in its spinor indices:
\be
(i\slashed{D}+iM)(i\slashed{D}-iM) = \Lcal P_L + \Lcal^* P_R,
\ee
where~$\Lcal$ is the operator introduced in~\eqref{eq:L} and~$P_{L,R}$ are projectors onto left- and right-helicity Weyl spinors on the two-dimensional geometry~$\Sigma$; is it this decomposition that allows us to compute the fermion partition function from just the spectrum of the (non-spinorial) operator~$\Lcal$.  The explicit form of the operator~$L$ defining~$\Lcal$ can be given most easily by working in conformally flat coordinates on~$\Sigma$,
\be
\label{eq:confflatmetric}
ds^2_\Sigma = e^{2\tilde{f}}\left((dx^1)^2 + (dx^2)^2\right),
\ee
in which case
\be
L = -\grad^2 + \frac{1}{4} R - i \eps^{ab} \left(\partial_a \tilde{f}\right) \partial_b + \frac{1}{4} \left(\grad_a \tilde{f} \right)^2.
\ee
The expression adapted to the spherical coordinates of~\eqref{eq:conformalsphere} can be obtained easily by transforming from the conformally flat coordinates~$\{x^1, x^2\}$ to the spherical coordinates~$\{\theta, \phi\}$ via~$\sin\theta = \sech x^1$,~$\phi = x^2$; then since~$\tilde{f} = f + \ln\sin\theta$, in terms of the conformal factor~$f$ one ultimately obtains\footnote{A more covariant expression can be given by introducing the spin weight raising and lowering operators~$\edth$,~$\bar{\edth}$, in terms of which
\be
L = -e^{-2f} \left[\edth \bar{\edth} + \frac{1}{2}\left(\overline{\grad}^2 f + (\bar{\edth} f) \edth - (\edth f) \bar{\edth}\right) - \frac{1}{4} (\overline{\grad}_a f)^2 \right];
\ee
more details are presented in Appendix~\ref{app:perttheory}.}
\begin{multline}
\label{eq:LDirac}
L = -e^{-2f} \left[\overline{\grad}^2 -\frac{1}{2}\left(1 - \overline{\grad}^2 f\right) + i \bar{\eps}^{ab} (\overline{\grad}_a f) \overline{\grad}_b + i \cot\theta \csc\theta \, \partial_\phi \right. \\ \left. - \frac{1}{4} \left(\overline{\grad}_a f\right)^2 - \frac{1}{2} \cot\theta \, \partial_\theta f - \frac{1}{4} \cot^2\theta \right],
\end{multline}
where as before~$\overline{\grad}_a$ is the covariant derivative on the round sphere.  $L$ acts on functions with spin weight~$1/2$, and hence the unperturbed eigenfunctions~$\tilde{h}_{\ell,m}$ can be taken to be the spin-weighted spherical harmonics~$_{1/2} Y_{\ell,m}$ of spin weight~$1/2$, where~$\ell \in \{1/2, 3/2, 5/2, \ldots \}$ is a positive half odd integer and as usual~$m \in \{-\ell, -\ell+1, \ldots, \ell\}$.  The corresponding unperturbed eigenvalues are~$\bar{\lambda}_\ell = (\ell + 1/2)^2$.

Again we relegate the details of the computation of the heat kernel to Appendix~\ref{app:perttheory}; ultimately we obtain~$\Delta K^{(1)} = 0$ and
\be
\label{eq:K2dirac}
\Delta K^{(2)}(t) = \sum_{\ell,m} a_\ell(t) |f_{\ell,m}|^2, \qquad a_\ell(t) \equiv t \sum_{\ell' = 1/2}^\infty e^{-\bar{\lambda}_{\ell'} t} \left(\alpha_{\ell,\ell'} + \beta_{\ell,\ell'} t\right),
\ee
with the expressions for~$\alpha_{\ell,\ell'}$ and~$\beta_{\ell,\ell'}$ given in~\eqref{eqs:alphabetafermionraw}.  As for the scalar, taking~$\ell$ to be odd substantially simplifies them:
\begin{subequations}
\label{eq:alphabetafermionodd}
\begin{align}
\alpha_{\ell,\ell'} &= \begin{dcases} - \frac{(2\ell'+1)^3}{16\pi} \, \frac{\left(\frac{2+\ell}{2}\right)_{\ell'+1/2} \left(\frac{2+\ell}{2}\right)_{-(\ell'+1/2)} }{\left(\frac{1+\ell}{2}\right)_{\ell'+1/2} \left(\frac{1+\ell}{2}\right)_{-(\ell'+1/2)}}, & \ell' < \frac{\ell}{2} \\
0, & \ell' \geq \frac{\ell}{2}
\end{dcases}, \\
\beta_{\ell,\ell'} &= 0.
\end{align}
\end{subequations}

\subsection{Check: Conformal Field Theories}
\label{subsec:conformal}

As a simple check of our results, let us compare to the results of~\cite{FisWis17}, which computed the zero-temperature perturbative energy difference~$\Delta E^{(2)}$ for \textit{any} unitary conformal field theory.  There it was found that in any CFT, this leading-order energy difference is
\be
\label{eq:CFTpert}
\Delta E^{(2)}_\mathrm{CFT} = -\sum_{\ell,m} A^\mathrm{(CFT)}_\ell |f_{\ell,m}|^2, \quad A^\mathrm{(CFT)}_\ell = \frac{\pi^2 c_T}{48} \, \frac{(\ell^2 - 1)(\ell+2)}{\ell} \left( \frac{\Gamma\left(\frac{\ell+1}{2}\right)}{\Gamma\left(\frac{\ell}{2}\right)}\right)^2,
\ee
with~$c_T$ the central charge defined as the coefficient in the two-point function of the stress tensor.

We now show that our expressions~\eqref{eq:K2scalar} and~\eqref{eq:K2dirac} reproduce~\eqref{eq:CFTpert} with the correct central charges when the fields are conformal; we note that in this case,~$\bar{\lambda}_{\ell'} = (\ell' + 1/2)^2$ for both the scalar and the fermion (though the allowed values of~$\ell'$ of course still differ). To do so, first note that the free energy difference is given by inserting~\eqref{eq:K2scalar} and~\eqref{eq:K2dirac} into~\eqref{eq:DeltaFheatkernel}; for simplicity we will restrict to perturbations~$f_{\ell,m}$ with odd~$\ell$, so that we may use the more compact expressions~\eqref{eq:alphabetascalarodd} and~\eqref{eq:alphabetafermionodd}.  In the zero-temperature limit, the integral over~$t$ can be performed explicitly by noting that Poisson resummation gives (for both the scalar and the fermion)
\be
\label{eq:Thetazerotemperature}
\lim_{T \to 0} T \, \Theta_\sigma (T^2 t) = \frac{1}{\sqrt{4\pi t}}
\ee
for any~$t > 0$; hence the zero-temperature perturbative energy difference for odd $\ell$ is
\be
\Delta E^{(2)}|_{T = 0} = -\sum_{\ell,m} A_\ell |f_{\ell,m}|^2,
\ee
where
\be
A_\ell = -\frac{\sigma}{\sqrt{4\pi}} \int_0^\infty \frac{dt}{t^{3/2}} \, a_\ell(t) = -\frac{\sigma}{2} \sum_{\ell'} \frac{\alpha_{\ell,\ell'}}{(\bar{\lambda}_{\ell'})^{1/2}}, 
\ee
where we used the fact that the sum over~$\ell'$ is finite to integrate term-by-term (it is understood that the sum over~$\ell'$ runs over integers or half-integers depending on whether we are considering the scalar or the fermion, with~$\alpha_{\ell,\ell'}$ the corresponding expression; we also remind the reader that~$\sigma = -1/2$ for the scalar and~$\sigma = 1$ for the fermion).  We therefore have
\bea
A_\ell^\mathrm{(scal)} &= \frac{1}{2\pi} \sum_{\ell' = 0}^{(\ell-1)/2} \frac{((\ell'+1/2)^2-\ell(\ell+1)/8)^2}{\ell(\ell+1)} \, \frac{\left(\frac{2+\ell}{2}\right)_{\ell'} \left(\frac{\ell}{2}\right)_{-\ell'}}{\left(\frac{3+\ell}{2}\right)_{\ell'} \left(\frac{1+\ell}{2}\right)_{-\ell'}}, \\
A_\ell^\mathrm{(ferm)} &= \frac{1}{4\pi} \sum_{\ell' = 0}^{(\ell-1)/2} (\ell')^2 \, \frac{\left(\frac{2+\ell}{2}\right)_{\ell'} \left(\frac{2+\ell}{2}\right)_{-\ell'}}{\left(\frac{1+\ell}{2}\right)_{\ell'} \left(\frac{1+\ell}{2}\right)_{-\ell'}},
\eea
where in the expression for~$A_\ell^\mathrm{(ferm)}$ we shifted the index of summation by~$1/2$.  While we are unable to analytically show that these expressions reproduce the form~\eqref{eq:CFTpert} predicted by CFT perturbation theory, by computing these sums exactly we find that they do, with the correct central charges~$c_T = (3/2)/(4\pi)^2$ and~$c_T = 3/(4\pi)^2$ (we have checked up to~$\ell = 1001$).

Interestingly, if~$\ell$ is even then evaluating~$A_\ell$ by integrating term-by-term produces a divergent sum, presumably due to the fact that the (now infinite) sum over~$\ell'$ in~$a_\ell(t)$ doesn't commute with the integration over~$t$.  Nevertheless, the behavior of~$a_\ell(t)$ for even~$\ell$ makes clear that the integral is indeed finite when performed after the summation, and we have confirmed numerically that it reproduces~\eqref{eq:CFTpert} for a range of even~$\ell$.

\subsection{Check: The Flat Space Limit}
\label{subsec:flat}

As a final check of our results, let us consider the limit in which the radius of the sphere is taken to be very large, and only modes with large~$\ell$,~$m$ are excited.  In this limit, we expect the theory to be insensitive to the curvature of the sphere, and thus the heat kernel should reproduce its flat space behavior.  This behavior was computed for both the scalar field and the Dirac fermion in~\cite{FisWal18}, in which it was found that when the perturbed metric is in the conformally flat form~$ds^2 = e^{2f} \delta_{ij} dx^i dx^j$, the perturbation to the heat kernel is
\be
\label{eq:K2flat}
\Delta K^{(2)}(t) = t \int d^2 k \, k^4 \left|\hat{f}(\vec{k})\right|^2 I(k^2 t),
\ee
where~$\vec{k}$ is a wave vector defined by the Fourier decomposition of~$f^{(1)}$ as
\be
\label{eq:fFourier}
f^{(1)}(x) = \int d^2 k \, \hat{f}(\vec{k}) e^{i \vec{k} \cdot \vec{x}},
\ee
$k = |\vec{k}|$ is its magnitude, and the functions~$I(\zeta)$ are given for the scalar and fermion as
\be
\label{eq:Ifuncs}
I(\zeta) = \begin{cases} -\frac{\pi}{4 \zeta^2} \left[6 +  \zeta ( 1 - 8 \xi) \phantom{\Fcal\left(\frac{\sqrt{\zeta}}{ 2} \right)} \right. & \\ \left. \qquad \qquad - \left( 6 + 2 \zeta (1 - 4 \xi) + \frac{\zeta^2}{2}  (1 - 4 \xi)^2 \right) \Fcal\left( \frac{\sqrt{\zeta}}{ 2} \right) \right], & \mbox{scalar} \\
\frac{\pi}{4 \zeta^2} \left[\left(6 + \zeta\right) \Fcal\left(\frac{\sqrt{\zeta}}{ 2} \right) - 6 \right], & \mbox{fermion} \end{cases}
\ee
with~$\Fcal(\zeta) = \zeta^{-1} e^{-\zeta^2} \int_0^{\zeta} d\zeta' \, e^{(\zeta')^2}$.

We now introduce an appropriate flat-space scaling limit in which our expressions for~$\Delta K^{(2)}$ reproduce~\eqref{eq:K2flat}.  To do so, let us explicitly reintroduce the radius~$r_0$ of the sphere, so that the deformed sphere metric~\eqref{eq:conformalsphere} becomes
\be
ds^2 = r_0^2 e^{2f} \left(d\theta^2 + \sin^2\theta \, d\phi^2 \right).
\ee
The scaling limit is defined by ``zooming in'' on a point on the equator of the sphere by introducing new coordinates~$x = r_0 (\theta-\pi/2)$,~$y = r_0 \phi$ and then taking the limit~$r_0 \to \infty$ with~$x$,~$y$ held fixed.  The resulting metric is in the desired conformally flat form,
\be
ds^2 \to e^{2f} \left(dx^2 + dy^2\right),
\ee
with~$x$ and~$y$ having infinite range.  Restoring~$r_0$ to the expressions~\eqref{eq:K2scalar} and~\eqref{eq:K2dirac}, we obtain
\be
\label{eq:K2scalinglimit}
\Delta K^{(2)}(t) = \sum_{\ell,m} a_\ell(t) |f_{\ell,m}|^2, \qquad a_\ell(t) = \frac{t}{r_0^2} \sum_{\ell'} e^{-\bar{\lambda}_{\ell'} t/r_0^2} \left(\alpha_{\ell,\ell'} + \beta_{\ell,\ell'} \frac{t}{r_0^2}\right),
\ee
with~$\alpha_{\ell,\ell'}$ and~$\beta_{\ell,\ell'}$ unchanged.  Now let us again focus on the case where~$f^{(1)}$ only contains modes with odd~$\ell$, so~$\beta_{\ell,\ell'}$ vanishes and the sum over~$\ell'$ runs to~$\ell/2$.  In order to consider modes with large~$\ell$, we define~$k = \ell/r_0$ and~$k' = \ell'/r_0$ and keep~$k$ and~$k'$ fixed as we take~$r_0 \to \infty$.  As we show in Appendix~\ref{app:flatspace}, in this limit we find that 
\be
\label{eq:aflatspace}
a_{r_0 k}(t) \to \frac{r_0^2 \, t}{4\pi^2} \, k^4 I(k^2 t)
\ee
with~$I(\zeta)$ precisely the functions given in~\eqref{eq:Ifuncs}; assuming~$a_{r_0 k}(t)$ is continuous in~$k$ in this scaling limit, we may now remove the restriction to modes with odd~$r_0 k$.  We also find that as long as~$f^{(1)}(x,y)$ vanishes at large~$(x,y)$ (i.e.~$f^{(1)}(\theta,\phi)$ vanishes away from~$(\theta = \pi/2, \phi = 0)$), $f_{\ell,m} = f_{r_0 k,r_0 k_y}$ becomes
\be
\label{eq:fflatspace}
f_{r_0 k, r_0 k_y} \to \frac{2\pi}{r_0^2} \frac{\sqrt{k}}{(k^2 - k_y^2)^{1/4}} \left( \hat{f}(\sqrt{k^2 - k_y^2}, k_y) \pm \hat{f}(-\sqrt{k^2 - k_y^2}, k_y)\right),
\ee
where~$\hat{f}(k_x,k_y)$ is the Fourier transform of~$f^{(1)}(x,y)$,
 the upper (lower) signs correspond to even (odd)~$(k+k_y)r_0$, and we are neglecting an overall phase that will cancel out.  Inserting these expressions into~\eqref{eq:K2scalinglimit} and decomposing the sum over~$\ell = k r_0$ into sums over even and odd~$k r_0$, we finally obtain precisely the flat-space expression~\eqref{eq:K2flat} given in~\cite{FisWal18}:
\be
\label{eq:K2flatlimit}
\Delta K^{(2)} \to t \int d^2 k \, k^4 I(k^2 t) \left| \hat{f}(\vec{k}) \right|^2.
\ee

It is perhaps worth emphasizing that computing the perturbation to the free energy of a perturbation of flat space is rather subtle due to the requirement that the perturbed and unperturbed geometries have the same volume: since the volume of flat space is infinite, an IR divergence is introduced, and the volume preservation condition is interpreted as controlling this IR divergence to yield a finite differenced free energy.  In~\cite{FisWal18}, this problem was addressed by computing the heat kernel on a torus and then taking the limit in which the cycles of the torus go to infinity; this is analogous to the procedure performed here, where we computed the heat kernel on the sphere and then took a flat space scaling limit.  In these regularization schemes, the ``extra'' bits of the torus or the sphere that get sent to infinity in the flat space limit essentially deform in such a way as to ensure that the leading-order UV divergences in~\eqref{eq:Fexpansion} cancel out between the deformed and undeformed geometries.  It is, however, possible to ensure that the UV divergent terms in~\eqref{eq:Fexpansion} cancel out even without such a compactification: as shown in~\cite{CheWal18}, one can introduce a one-parameter family of large diffeomorphisms on flat space (that is, diffeomorphisms that don't vanish in the asymptotic region) in order to ensure that the differenced free energy is UV and IR finite.

Though the final result obtained in the flat-space limit of both the torus and the sphere is the same, we note the interesting difference that on the finite-size torus, the Dirac fermion has a negative mode for which $\Delta K_L$ does \textit{not} have fixed sign, and in fact even renders~$\Delta F$ \textit{positive}\footnote{Explicitly, consider the deformed torus~$ds^2 = e^{2f}[(dx^1)^2 + (dx^2)^2]$, where~$x^1$ and~$x^2$ both have periodicity~$\Delta x$, and we take~$f = \eps \cos(2\pi x^1/\Delta x) + \Ocal(\eps^2)$.  The perturbative heat kernel for the Dirac fermion on a deformed torus is computed in~\cite{FisWal18}, and in this case comes out to be
\be
\Delta K^{(2)} = -2t\left(\frac{2\pi}{\Delta x}\right)^2 \sum_{\mathclap{n_1,n_2 \in \mathbb{Z}}} e^{-(2\pi/\Delta x)^2(n_1^2+n_2^2) t}\left(\frac{(n_1^2 + n_2^2 - 1/4)^2 + n_2^2/4}{2n_1 - 1} + \frac{1}{16}\right).
\ee
This expression is positive for~$t > t_*$ for some~$t_*$, and with~$T = 0$,~$M = 0$ the differenced free energy~\eqref{eq:DeltaFheatkernel} comes out positive as well.}; as we now discuss, this is not the case for the sphere.

\subsection{Negativity of $\Delta K$}
\label{subsec:negativeDeltaK}

Our results imply that for any nontrivial deformation of the round sphere,~$\sigma \Delta K$ is strictly negative for all~$t$ to leading nontrivial order in~$\eps$, and hence so is~$a_\ell(t)$, and thus~$\Delta F$.  This is easiest to see when~$f^{(1)}$ contains only modes with odd~$\ell$: in this case, it is clear from the expressions~\eqref{eq:alphabetascalarodd} and~\eqref{eq:alphabetafermionodd} that~$\sigma \alpha_{\ell,\ell'}$ is negative when~$\ell$ is odd and greater than one, and hence so too is~$\sigma \Delta K^{(2)}(t)$ for all~$t$ (recall that~$\sigma = -1/2$ for the scalar and~$\sigma = 1$ for the fermion).  When~$\ell = 1$,~$a_\ell(t) = 0$ for both the scalar and the fermion, and hence~$\Delta K^{(2)}$ vanishes; this is due to the fact that~$\ell = 1$ deformations generate infinitesimal diffeomorphisms of the sphere and therefore do not change its intrinsic geometry to leading order in~$\eps$.  We will show this explicitly in Section~\ref{subsec:numerics}.

\begin{figure}[t]
\centering
\subfloat[][Minimally coupled scalar]{
\includegraphics[width=0.49\textwidth]{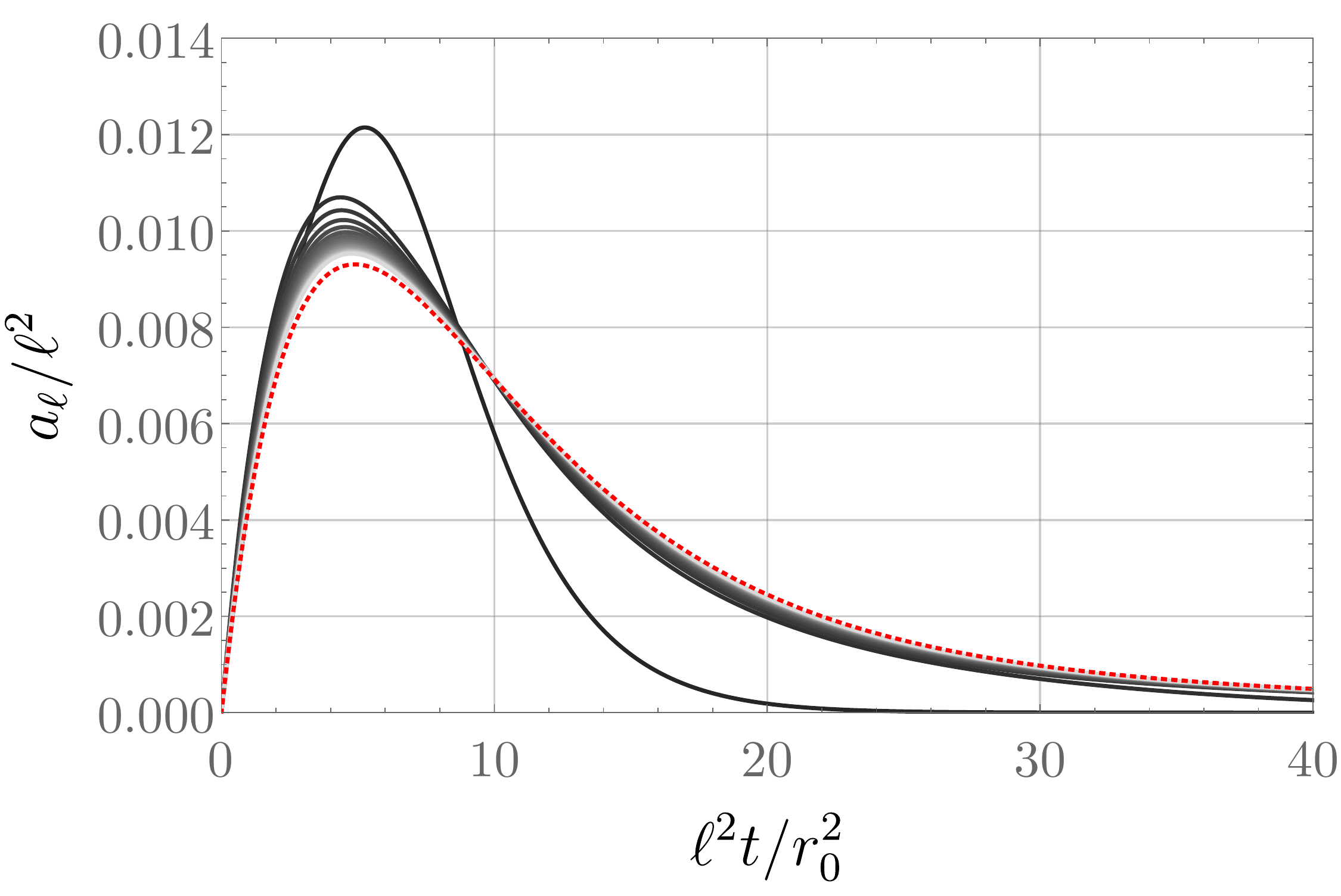}
\label{subfig:scalarflat}
}
\subfloat[][Fermion]{
\includegraphics[width=0.495\textwidth]{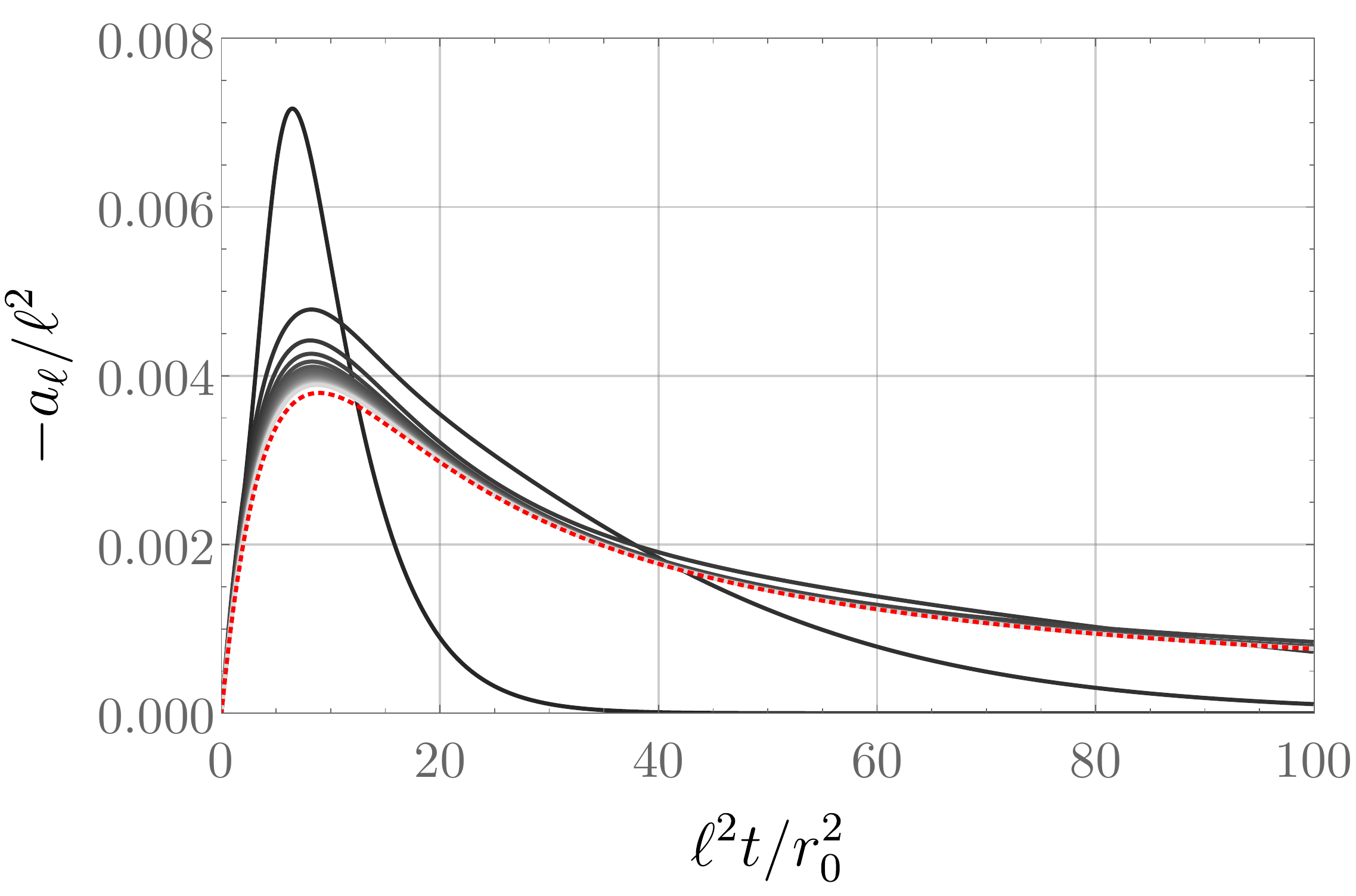}
\label{subfig:Diracflat}
}
\caption{The function~$a_\ell(t)$ for even~$\ell$ for the minimally coupled scalar (left) and the Dirac fermion (right); from dark to light gray, the curves correspond to~$\ell = 2$ to~$\ell = 40$.  The dashed red curve is the flat-space limit given by~\eqref{eq:aflatspace}.  The convergence of~$a_\ell(t)$ to the flat-space limit for the nonminimally coupled scalar is analogous.}
\label{fig:evenell}
\end{figure}

The case of even~$\ell$ is more subtle.  For the scalar, it follows from the full expression~\eqref{eq:alphascalarclosed} that~$\sigma \alpha_{\ell,\ell'}$ can be \textit{positive} for~$\ell' \geq \ell/2$; likewise, for the fermion it follows from~\eqref{subeq:Diracbetaraw} that~$\sigma \beta_{\ell,\ell'}$ is positive.  Hence in both cases the sign of~$a_\ell(t)$ is not immediately clear.  However, note that the large-$\ell$ behavior of~$a_\ell(t)$ can be obtained in the flat-space scaling limit discussed above, and is given in~\eqref{eq:aflatspace}; assuming~$a_{r_0 k}(t)$ is continuous in~$k$ as~$r_0 \to \infty$, we therefore conclude that for all large~$\ell$ (whether even or odd),~$\sigma a_\ell(t)$ is negative.  We therefore need only investigate the sign of~$\sigma a_\ell(t)$ for small even~$\ell$ (i.e.~before the transition to the flat-space behavior).  The result is shown in Figure~\ref{fig:evenell}, which verifies that~$\sigma a_\ell(t) < 0$ for all~$t$.

Thus~$\sigma \Delta K_L^{(2)}(t)$ is indeed negative for all~$t$.  This implies, of course, that small, nontrivial deformations of the round sphere all lower the free energy of the scalar and of the fermion (for any mass, temperature, and curvature coupling), but it is in fact a much stronger result: negativity of~$\Delta F$ \textit{does not} require that the heat kernel~$\Delta K_L(t)$ itself be everywhere negative.  We now investigate whether this stronger result continues to hold even for large area-preserving deformations of the sphere.

\section{Nonperturbative Results}
\label{sec:nonpert}

We have thus found that small perturbations of the round sphere always yield a negative~$\sigma \Delta K_L(t)$ (and hence also free energy~$\Delta F$) for both the scalar and fermion at any temperature, mass, or curvature coupling.  This observation naturally prompts a question: is the free energy maximized \textit{globally} by the round sphere, as it is for holographic CFTs at zero temperature \cite{HicWis15}?  Or do there exist sufficiently large deformations of the round sphere at which the free energy eventually increases above its value on the round sphere?  If the free energy is globally maximized, does the stronger result that the differenced heat kernel has fixed sign continue to hold?  Our purpose now is to examine these questions.  To do so, we will ultimately need to resort to numerics in order to evaluate the heat kernel~\eqref{eq:trace} for large deformations of the round sphere.  However, we will first examine the behavior of~$\Delta K_L(t)$ at large and small~$t$, which is tractable analytically even for large deformations.

\subsection{Heat Kernel Asymptotics}
\label{subsec:asymptotics}

Recall that the small-$t$ behavior of the differenced heat kernel is given by the heat kernel expansion~\eqref{eq:DeltaKexpansion}:
\be
\Delta K_L(t) = t \, \Delta b_4 + \Ocal(t^2).
\ee
The leading-order coefficient~$\Delta b_4$ is given by~\cite{Vas03}
\begin{subequations}
\label{eqs:Deltaa4}
\begin{align}
\Delta b_4^{\mathrm{(scal)}} &= \frac{1}{1440\pi}(5(6\xi-1)^2+1) \int d^2 x \left( \sqrt{g} \, R^2 - \sqrt{\bar{g}} \, \overline{R}^2 \right), \\
\Delta b_4^{\mathrm{(ferm)}} &= - \frac{1}{960\pi} \int d^2 x \left( \sqrt{g} \, R^2 - \sqrt{\bar{g}} \, \overline{R}^2 \right),
\end{align}
\end{subequations}
where~$\overline{R}$ is the Ricci scalar of the round sphere.  But it follows from volume preservation, the Gauss-Bonnet theorem, and the fact that~$\overline{R}$ is constant that
\be
\int d^2 x \left( \sqrt{g} \, R^2 - \sqrt{\bar{g}} \, \overline{R}^2 \right) = \int 
d^2 x \sqrt{g} (R - \overline{R})^2 \geq 0 
\ee
with equality if and only if~$R = \overline{R}$, i.e.~if~$g_{ij}$ is the metric of the round sphere.  Hence for both the scalar and fermion,~$\sigma \Delta K_L(t)$ is strictly \textit{negative} at sufficiently small~$t$ for any nontrivial deformation of the sphere, regardless of the size of the deformation.

To inspect the large-$t$ behavior, we instead recall that the differenced heat kernel can be expressed in terms of the eigenvalues~$\lambda_I$,~$\bar{\lambda}_I$ of the operators~$L$ and~$\overline{L}$:
\be
\Delta K_L(t) = \sum_I \left(e^{-t \lambda_I} - e^{-t \bar{\lambda}_I}\right).
\ee
The large-$t$ behavior of this expression -- particularly its sign -- is clearly dominated by the smallest eigenvalue of either~$L$ or~$\overline{L}$, so we must compare the low-lying spectra of these two operators.  For the scalar, this comparison can be performed by using a Rayleigh-Ritz formula for the lowest eigenvalue of~$L$:
\be
\lambda_\mathrm{min} = \inf_{\phi} J[\phi], \qquad J[\phi] \equiv \left[\int d^2 x \sqrt{g} \, \phi^2\right]^{-1}\int d^2 x \sqrt{g} \, \phi (-\grad^2+\xi R) \phi,
\ee
with the infimum taken over all (square-integrable) test functions~$\phi$.  Thus~$\lambda_\mathrm{min}$ can be bounded from above by taking~$\phi$ to be a constant function; then again using the Gauss-Bonnet theorem and volume preservation, we have
\be
\lambda_\mathrm{min} \leq J[\mathrm{const.}] =  \frac{4\pi \xi \chi_\Sigma}{\mathrm{Vol}[g]} =  \bar{\lambda}_\mathrm{min}
\ee
with equality if and only if a constant function is an eigenfunction of~$L$, which for~$\xi \neq 0$ is only the case if~$R$ is a constant and thus~$g_{ij}$ is the metric of the round sphere.  Hence for~$\xi \neq 0$ and a nontrivial perturbation of the sphere, the lowest eigenvalue of~$L$ is always strictly less than any eigenvalue of~$\overline{L}$, and~$\Delta K_L(t)$ is positive at sufficiently large~$t$.  On the other hand, when~$\xi = 0$ constant functions are always eigenfunctions of~$L = -\grad^2$, and hence the lowest eigenvalues of~$L$ and~$\overline{L}$ are identical.  The large-$t$ behavior of~$\Delta K_L(t)$ is then controlled by the next-lowest eigenvalue~$\lambda_\mathrm{next}$ of~$-\grad^2$, which is known to be bounded by~\cite{YanYau80}
\be
\lambda_\mathrm{next} \leq \frac{8\pi}{\mathrm{Vol}[g]} = \bar{\lambda}_\mathrm{next}.
\ee
Hence for the case~$\xi = 0$, we again find that~$\Delta K_L(t)$ is positive at sufficiently large~$t$.

We come to a similar conclusion for the fermion by invoking a theorem from~\cite{Baer1992}:  namely, given that~$\Sigma$ is a two-dimensional manifold of genus zero, all eigenvalues of the squared Dirac operator are bounded below by~${4\pi}/{\mathrm{Vol}[g]} = \bar{\lambda}_\mathrm{min}$, with equality only holding if~$g_{ij}$ is the metric of the round sphere.  Hence again we conclude that at sufficiently large~$t$,~$\sigma\Delta K_L(t)$ is negative.

We have therefore established that~$\sigma \Delta K_L(t)$ is always negative at sufficiently small or large~$t$, regardless the size of the perturbation to the sphere.  To analyze the intermediate-$t$ regime, and in particular to determine whether~$\Delta F$ decreases arbitrarily as the size of the perturbation grows, we turn to numerics.

\subsection{Numerical Results}
\label{subsec:numerics}

The advantage of using heat kernels to evaluate the differenced free energy is that computing the (differenced) heat kernel~\eqref{eq:trace} amounts to computing the spectrum of~$L$.  Moreover, only the smallest few eigenvalues of~$L$ are needed to obtain a good approximation for~$\Delta K_L(t)$ everywhere except near~$t = 0$ -- but small~$t$ is precisely the region in which the heat kernel expansion gives a good approximation.  The heat kernel expansion therefore provides both a check of the numerics as well as a tractable way of computing the differenced free energy (which requires the behavior of~$\Delta K_L(t)$ to be known for all~$t$).  In short, we compute~$\Delta K_L(t)$ numerically to a sufficient accuracy that at sufficiently small~$t$ it agrees with the leading linear behavior~$\Delta b_4 t$ of the heat kernel expansion, and we then sew these two behaviors together to perform the integration over all~$t$ that gives $\Delta F$.  We present more information on the numerical method used, as well as details of these checks, numerical errors, and computation of~$\Delta F$, in Appendix~\ref{app:numericalchecks}.  Here we instead describe the setup and the results.

First, note that on sufficiently deformed backgrounds the Ricci scalar will become negative somewhere, and hence the spectrum of~$L$ for the non-minimally coupled scalar may become negative.  If these eigenvalues are sufficiently larger in magnitude than~$M^2$ (as will always occur if~$M$ is fixed and the sphere is deformed more and more extremely), their presence introduces tachyonic instabilities, implying that the theory becomes ill-defined.  Consequently, we will restrict to numerical analysis of only the \textit{minimally coupled} scalar~$\xi = 0$, which as we showed in the previous section always has a non-negative spectrum.  No restriction is required on the fermion, since as mentioned above the spectrum of~$L$ for the fermion is always positive.

\afterpage{\clearpage}

\begin{figure}[t]
\captionsetup[subfigure]{justification=centering,labelformat=empty}
\centering
\subfloat[][$\ell = 1$ \\ $0 \leq \eps \leq 0.9\eps_\mathrm{max}$]{
\includegraphics[width=0.22\textwidth]{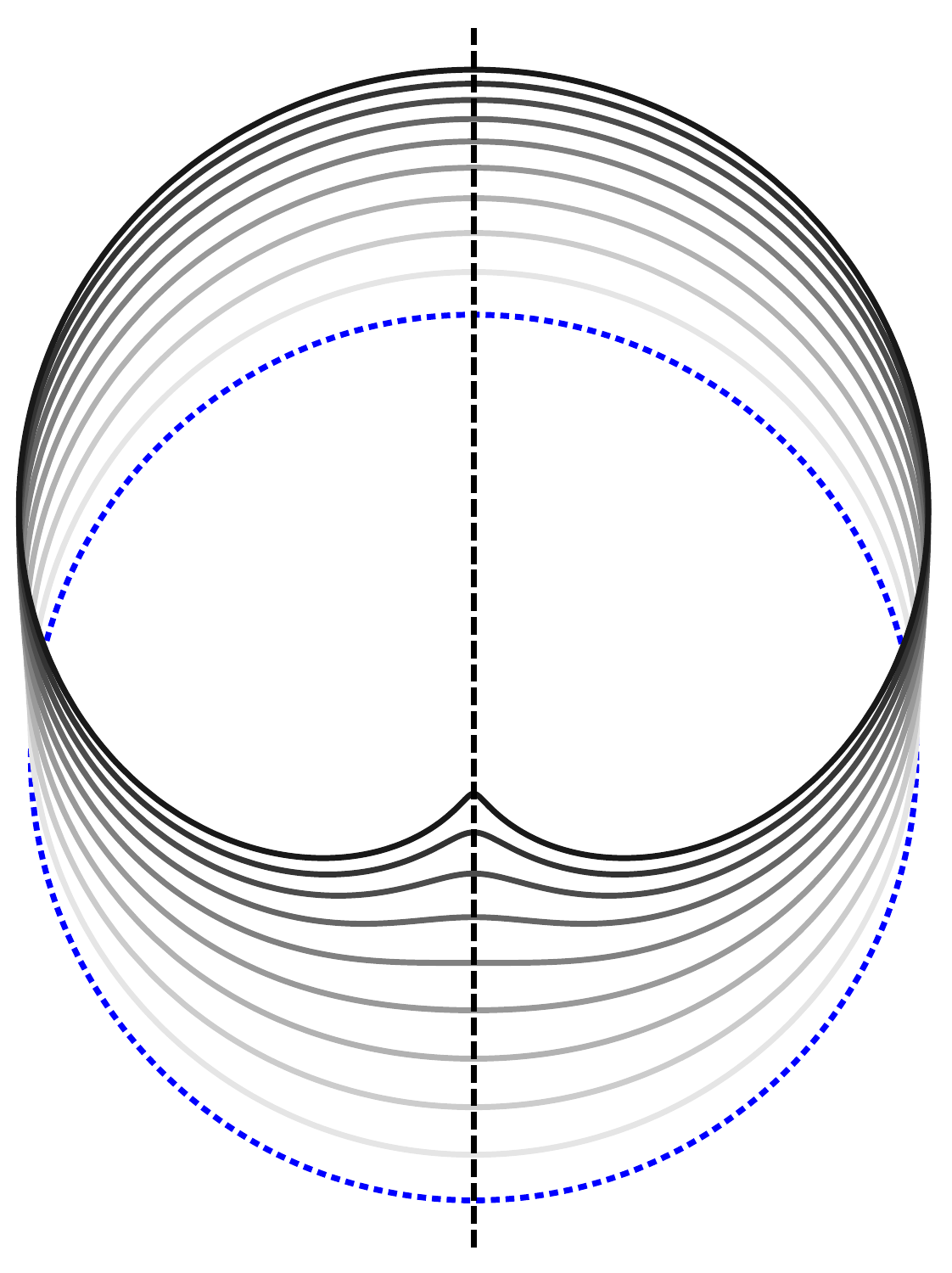}
}
\hspace{1cm}
\subfloat[][$\ell = 2$ \\ $0.9\eps_\mathrm{min} \leq \eps \leq 0$]{
\includegraphics[width=0.25\textwidth]{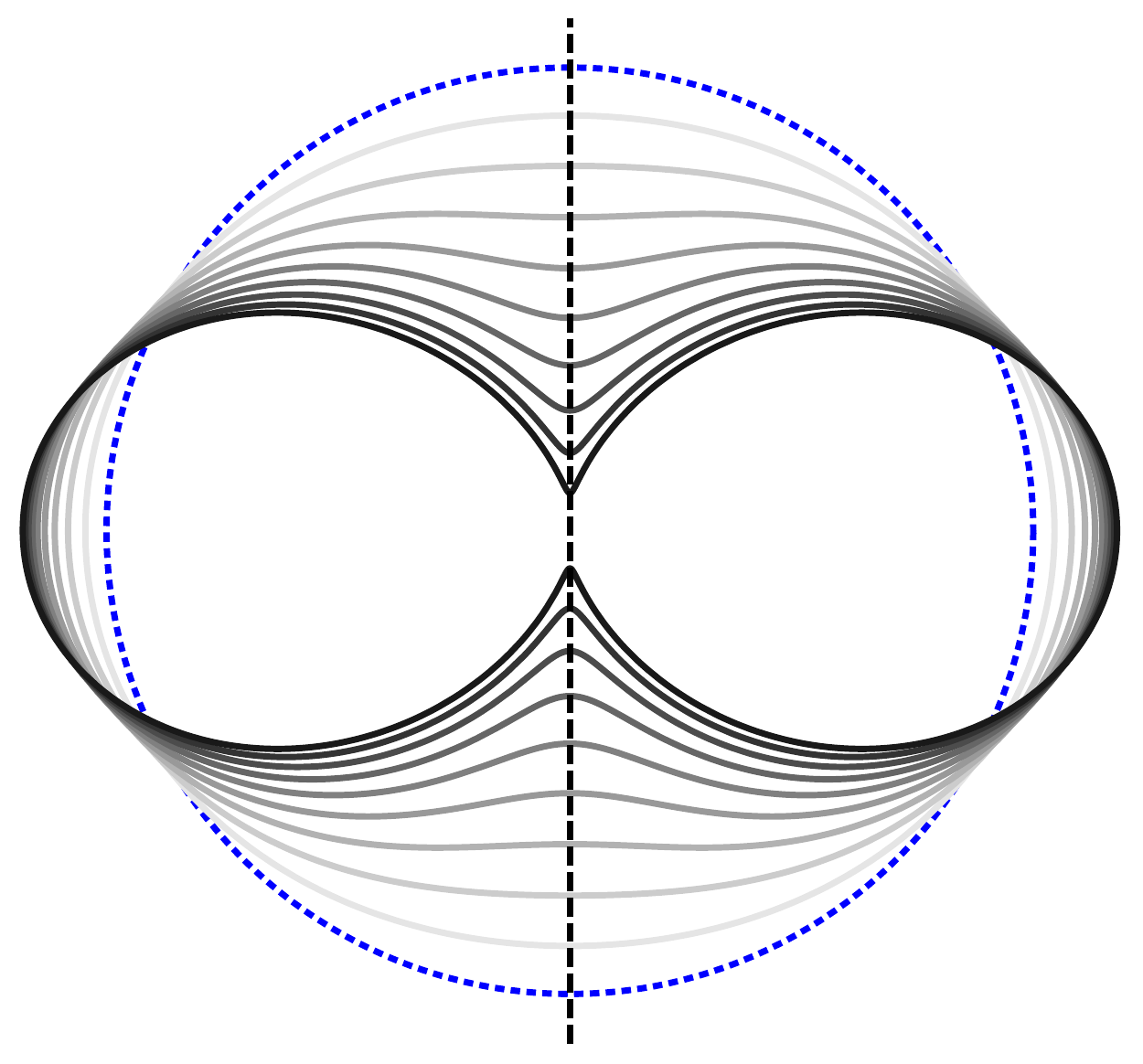}
}
\hspace{1cm}
\subfloat[][$\ell = 2$ \\ $0 \leq \eps \leq 0.9\eps_\mathrm{max}$]{
\includegraphics[width=0.17\textwidth]{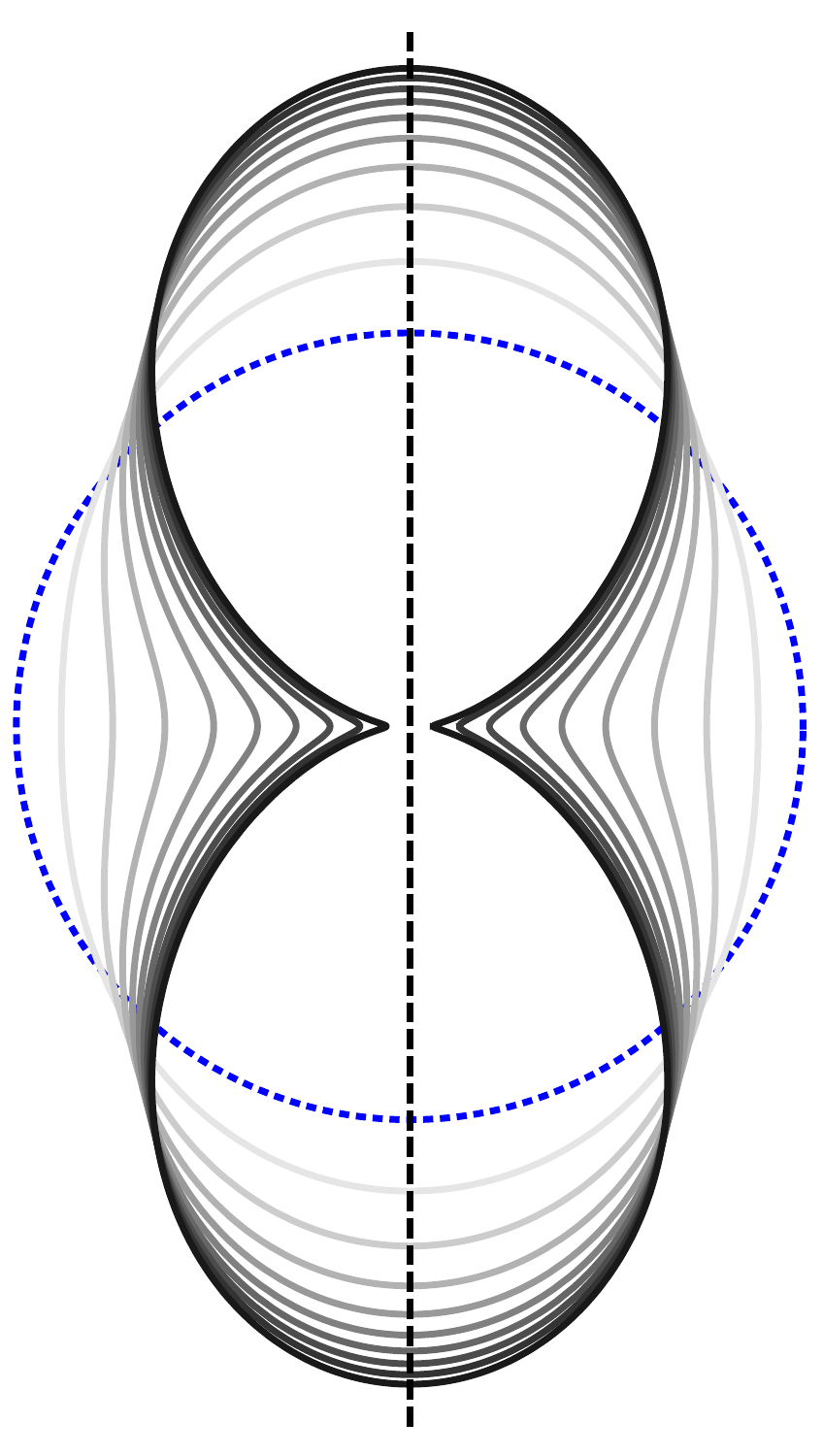}
}
\hspace{1.25cm}
\subfloat[][$\ell = 3$ \\ $0 \leq \eps \leq 0.9\eps_\mathrm{max}$]{
\includegraphics[width=0.22\textwidth]{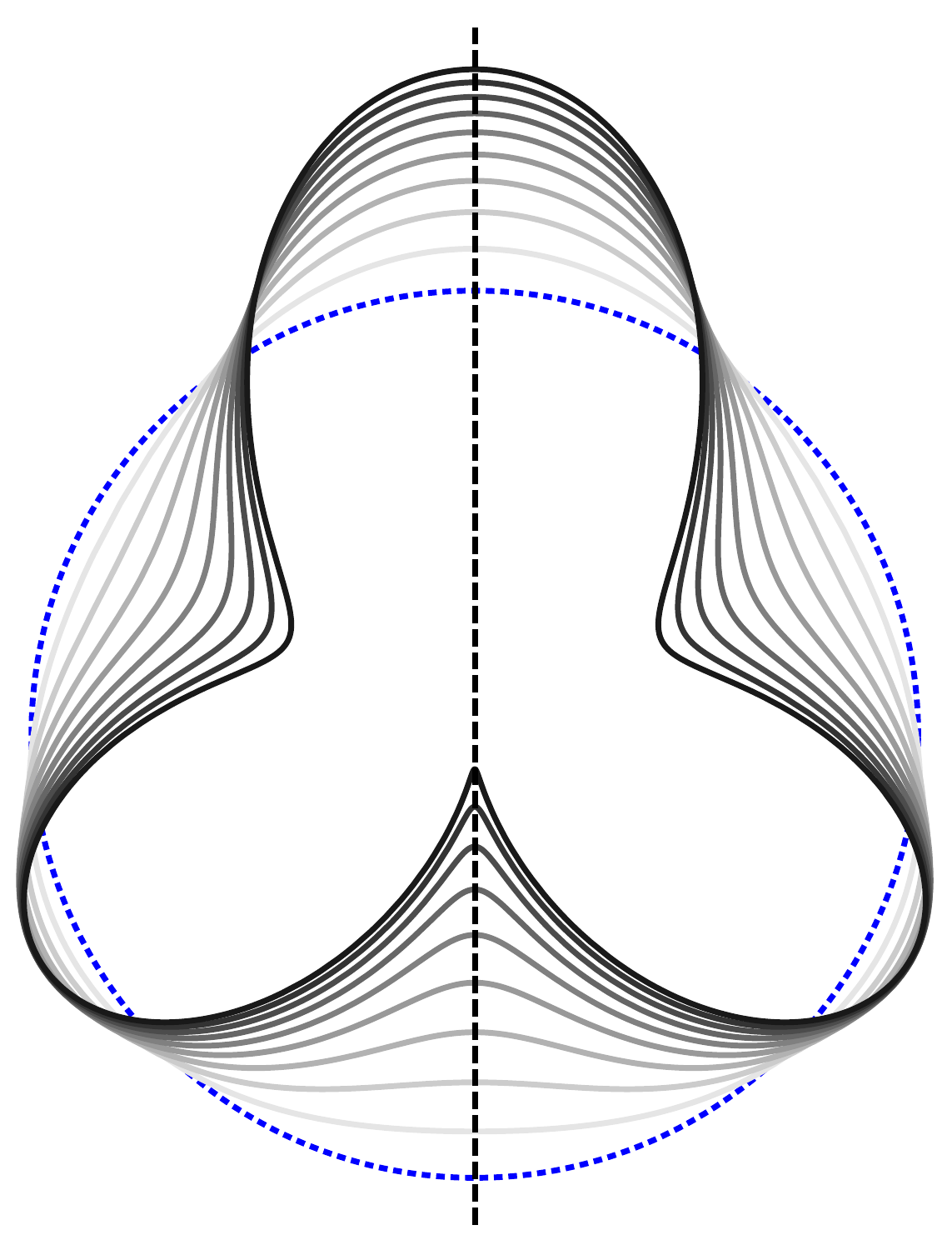}
}
\hspace{1.25cm}
\subfloat[][$\ell = 4$ \\ $0.9\eps_\mathrm{min} \leq \eps \leq 0$]{
\includegraphics[width=0.25\textwidth]{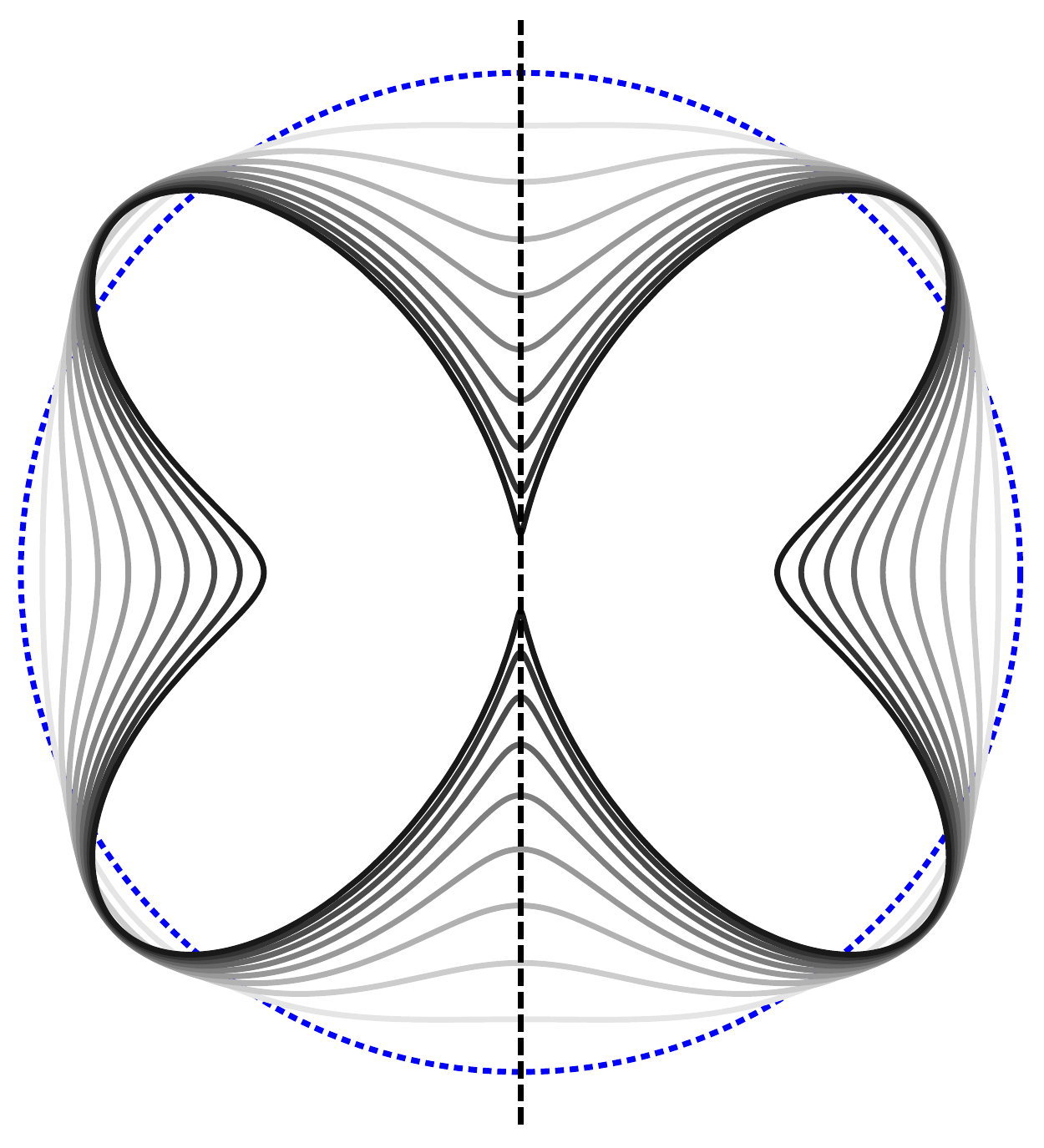}
}
\hspace{1.25cm}
\subfloat[][$\ell = 4$ \\ $0 \leq \eps \leq 0.9\eps_\mathrm{max}$]{
\includegraphics[width=0.17\textwidth]{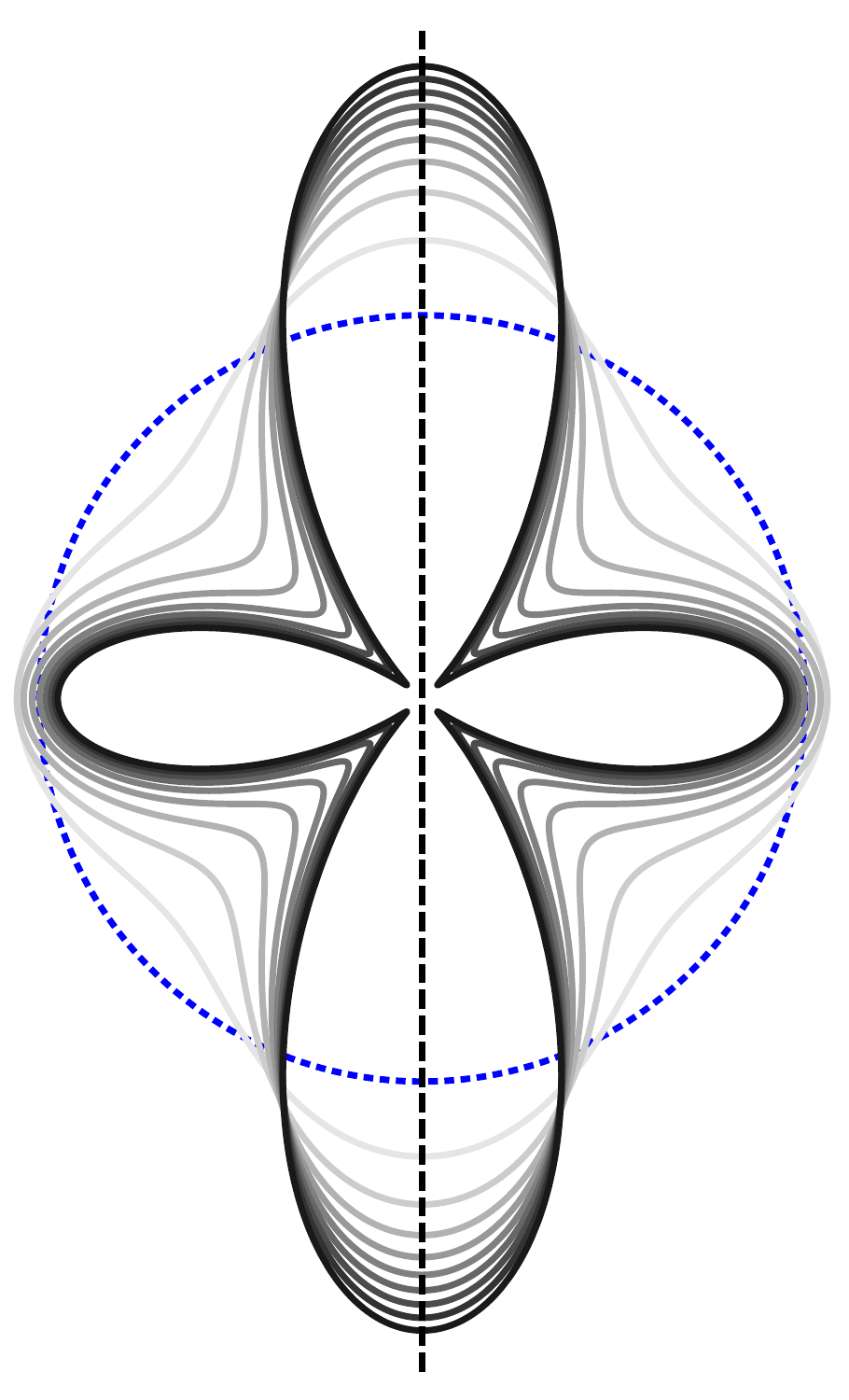}
}
\hspace{1.25cm}
\subfloat[][$\ell = 5$ \\ $0 \leq \eps \leq 0.9\eps_\mathrm{max}$]{
\includegraphics[width=0.22\textwidth]{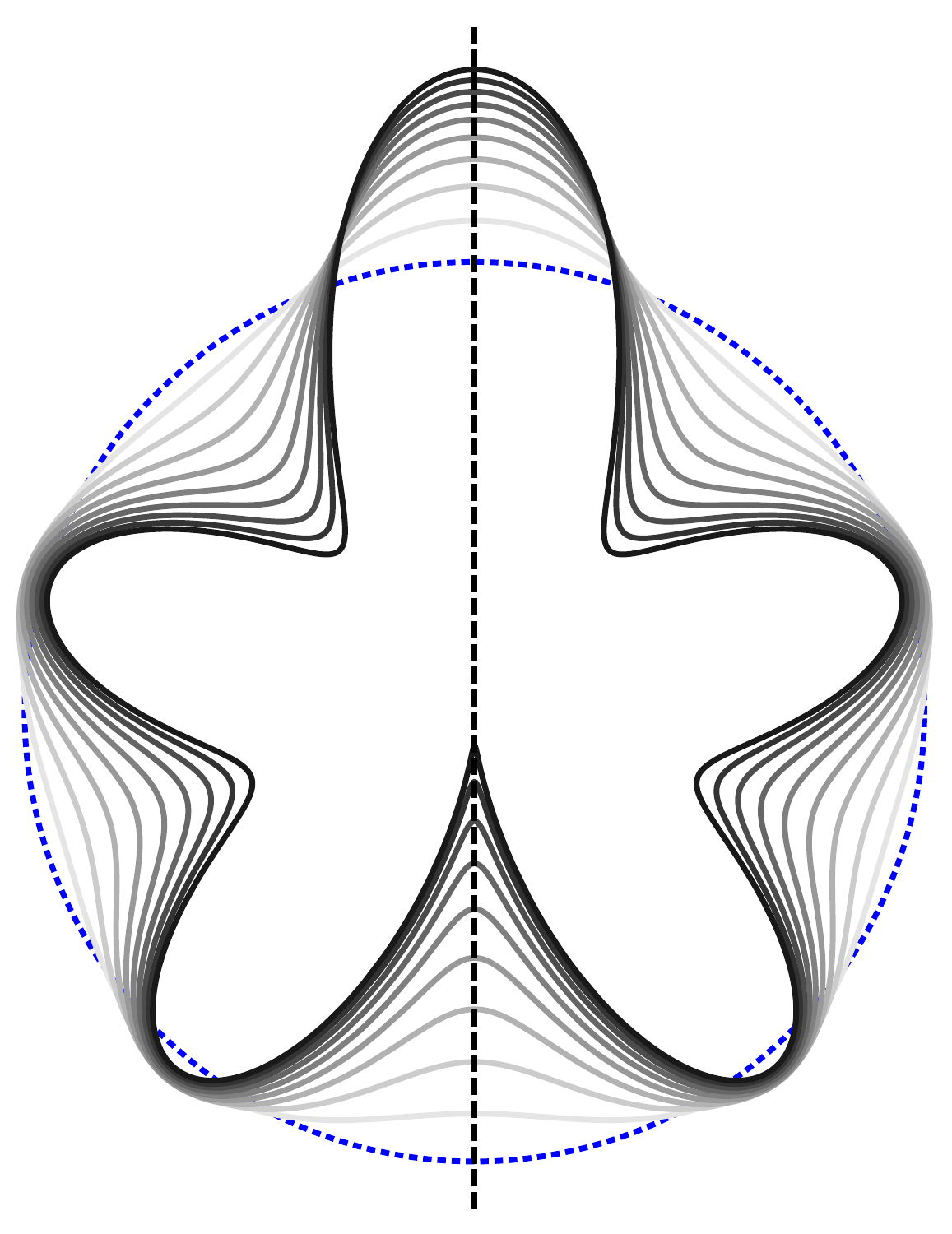}
}
\hspace{1.25cm}
\subfloat[][$\ell = 6$ \\ $0.9\eps_\mathrm{min} \leq \eps \leq 0$]{
\includegraphics[width=0.25\textwidth]{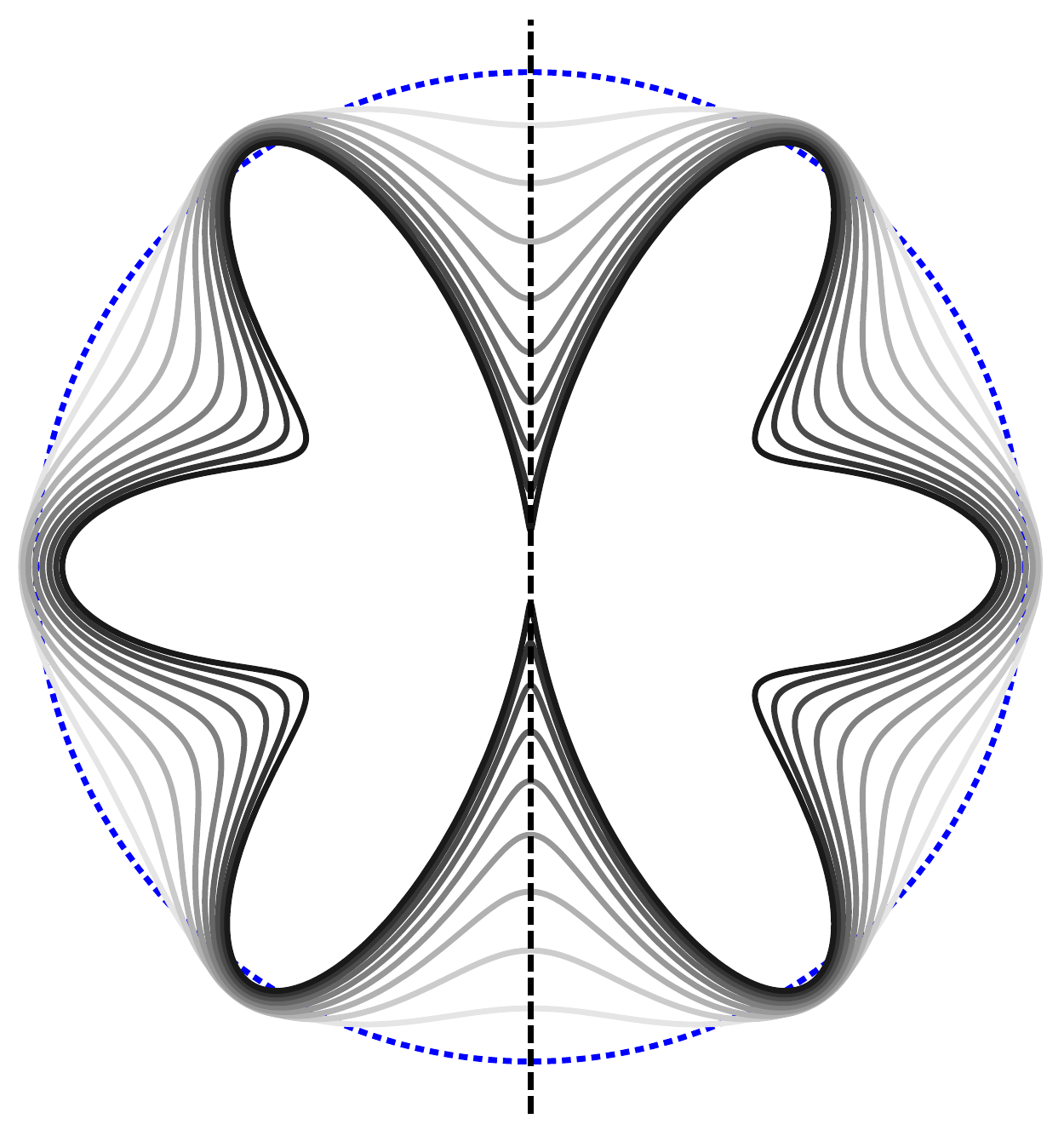}
}
\hspace{1.25cm}
\subfloat[][$\ell = 6$ \\ $0 \leq \eps \leq 0.9\eps_\mathrm{max}$]{
\includegraphics[width=0.17\textwidth]{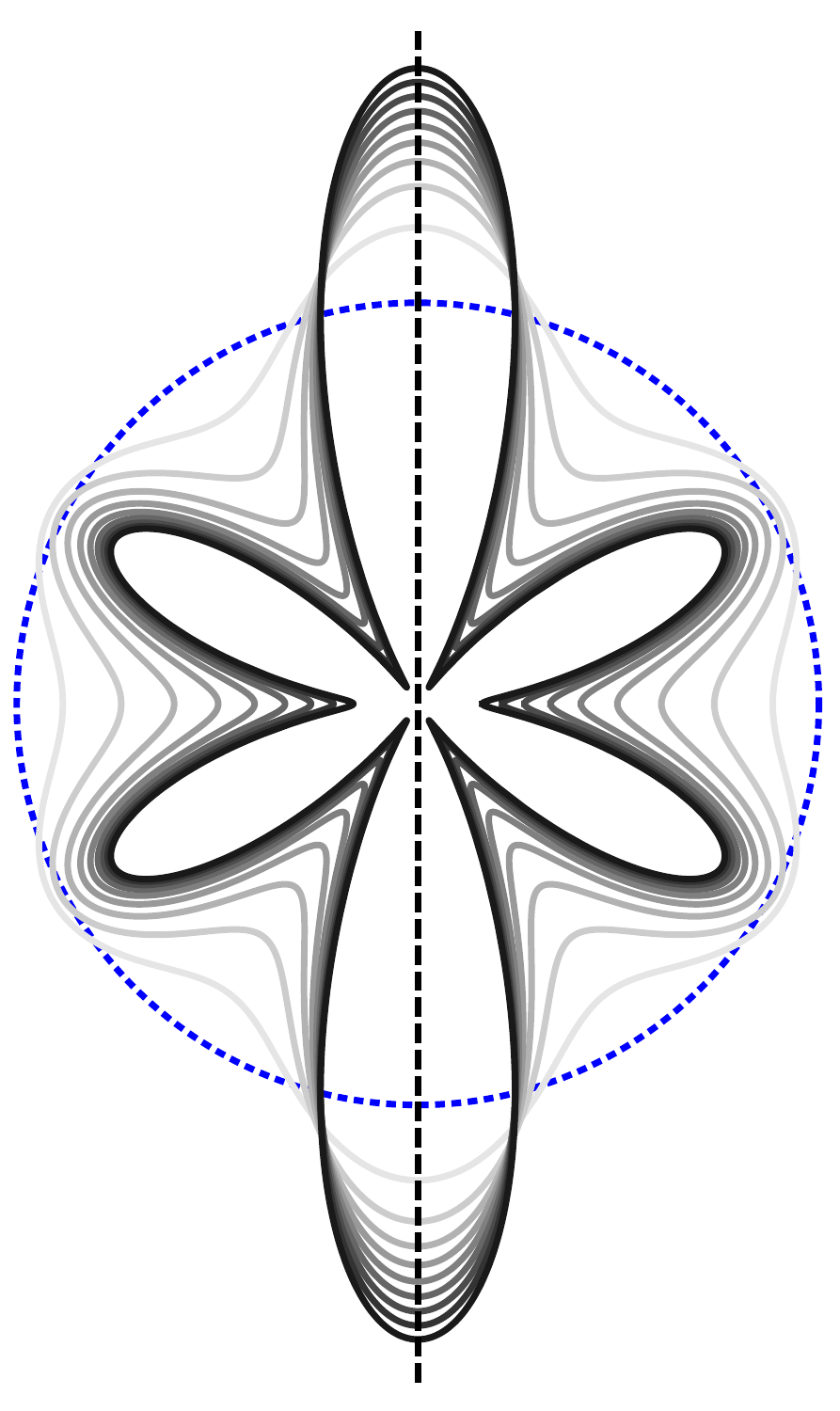}
}
\caption{Cross-sections of the geometries we consider; these should be rotated around the dotted axis to generate the corresponding deformed sphere.  The dotted blue circle is the unperturbed sphere; from light to dark gray, each curve corresponds to~$\eps$ ranging in steps of~$0.1 \eps_\mathrm{max}$ (or~$0.1\eps_\mathrm{min}$) from~$0.1\eps_\mathrm{max}$ ($0.1 \eps_\mathrm{min}$) to~$0.9 \eps_\mathrm{max}$ ($0.9 \eps_\mathrm{min}$).  For odd~$\ell$, negative~$\eps$ is related to positive~$\eps$ by a parity transformation which turns the cross-section ``upside-down'', leaving the geometry unchanged.}
\label{fig:embeddings}
\end{figure}

A numerical analysis can only by used to study a specific subset of deformations of the round sphere.  Here we will consider certain classes of axisymmetric deformations.  We begin by considering deformed spheres embedded in~$\mathbb{R}^3$ via~$r = R(\theta)$, corresponding to the induced metric\footnote{Such an embedding restricts~$\Sigma$ to be star-shaped in the technical sense that any ray fired from~$r = 0$ intersects~$\Sigma$ precisely once.} 
\be
\label{eq:NonPerturbativeDeformations}
ds^2 = R(\theta)^2 \left[\left(1 + \frac{R'(\theta)^2}{R(\theta)^2}\right)d\theta^2 + \sin^2\theta \, d\phi^2\right].
\ee
Specifically, we will take
\be
\label{eq:Rellembedding}
R_{\ell,\eps}(\theta) = c_{\ell,\eps} \left(1 + \eps \, Y_{\ell,0}(\theta,0) \right),
\ee
where~$c_{\ell,\eps}$ is a (positive) constant that ensures the volume of the sphere remains unchanged as~$\eps$ is varied.  It is straightforward to see that to linear order in~$\eps$, the metric~\eqref{eq:NonPerturbativeDeformations} obtained from these embedding functions is in the form~\eqref{eq:conformalsphere} conformal to the round sphere, and hence the behavior of~$\Delta K_L(t)$ to leading nontrivial order in~$\eps$ should be the same as that obtained in Section~\ref{sec:pert} (with~$f^{(1)} = Y_{\ell,0}$).  However, higher-order effects in~$\eps$ break the conformal form of the metric.  We will consider deformations~\eqref{eq:Rellembedding} with~$\ell=1,2,\ldots, 6$, while the range of~$\eps \in (\eps_\mathrm{min}, \eps_\mathrm{max})$ is fixed by the condition that~$R_{\ell,\eps} > 0$ everywhere; we show cross-sections of the embeddings of these surfaces into~$\mathbb{R}^3$ in Figure~\ref{fig:embeddings}.  Note that for odd~$\ell$ it suffices to consider only~$\eps > 0$, since positive and negative~$\eps$ are related by a parity transformation: the transformation~$(\theta,\phi) \to (\pi - \theta, \phi + \pi)$ sends~$Y_{\ell,0} \to (-1)^\ell Y_{\ell,0} = - Y_{\ell,0}$, and thus since~$Y_{\ell,0}$ always appears with a factor of~$\eps$, we find that~$R_{\eps,\ell} \to R_{-\eps, \ell}$ for odd~$\ell$.  It is also worth noting that the~$\ell = 1$ embedding doesn't appear to change the shape of the sphere much at all until~$\eps$ is relatively large; as mentioned above, this is because the~$\ell = 1$ deformation is an infinitesimal diffeomorphism, and thus the deformation of the intrinsic geometry is trivial to linear order in~$\eps$.  This can be seen explicitly by noting that since~$Y_{1,0}(\theta) = p \cos\theta$ (with~$p = \sqrt{3/4\pi}$), the induced metric~\eqref{eq:NonPerturbativeDeformations} with~$R = R_{1,\eps}$ becomes
\be
ds^2 = c_{1,\eps}^2 (1 + p \eps \cos\theta)^2 \left[(1+ p^2 \eps^2 \sin^2\theta) d\theta^2 + \sin^2\theta \, d\phi^2 \right] + \Ocal(\eps^3);
\ee
converting to a new coordinate~$\vartheta$ defined by~$\theta = \vartheta - p \eps \sin\vartheta + p^2 \eps^2 \sin\vartheta \cos\vartheta + \Ocal(\eps^3)$, we get
\be
ds^2 = \left(1 - \frac{\eps^2}{\sqrt{20\pi}} Y_{2,0}(\vartheta)\right) \left[d\vartheta^2 + \sin^2\vartheta \, d\phi^2 \right] + \Ocal(\eps^3),
\ee
so the induced metric to linear order in~$\eps$ is diffeomorphic to the round sphere, as claimed.  The nontrivial perturbation comes in at order~$\eps^2$ and takes the form of those considered in Section~\ref{sec:pert} with~$f^{(1)} = -Y_{2,0}/\sqrt{20\pi}$; the differenced heat kernel should thus be~$\Ocal(\eps^4)$.

\begin{figure}[t]
\captionsetup[subfigure]{justification=centering,labelformat=empty}
\centering
\includegraphics[width=0.33\textwidth]{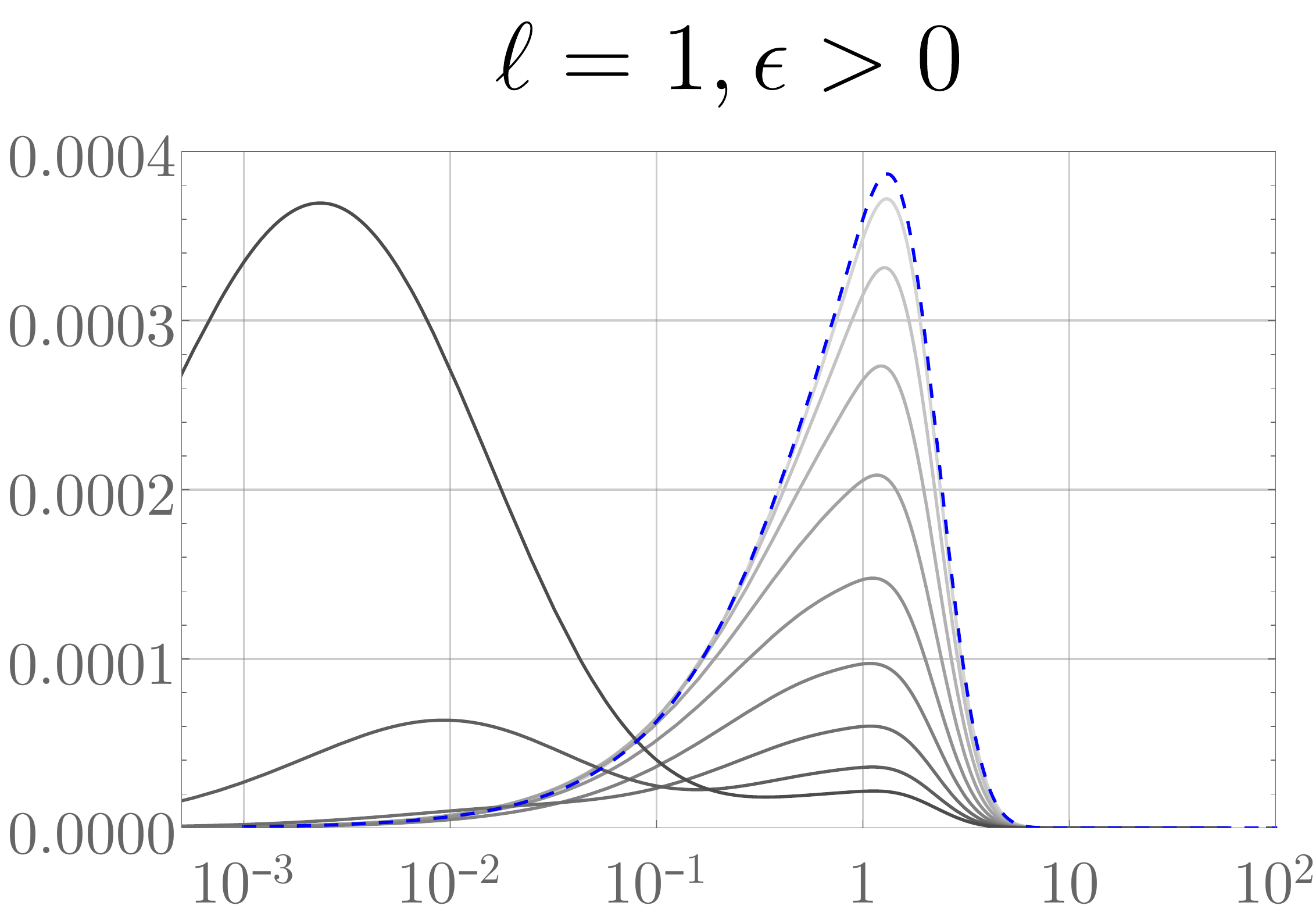}
\includegraphics[width=0.33\textwidth]{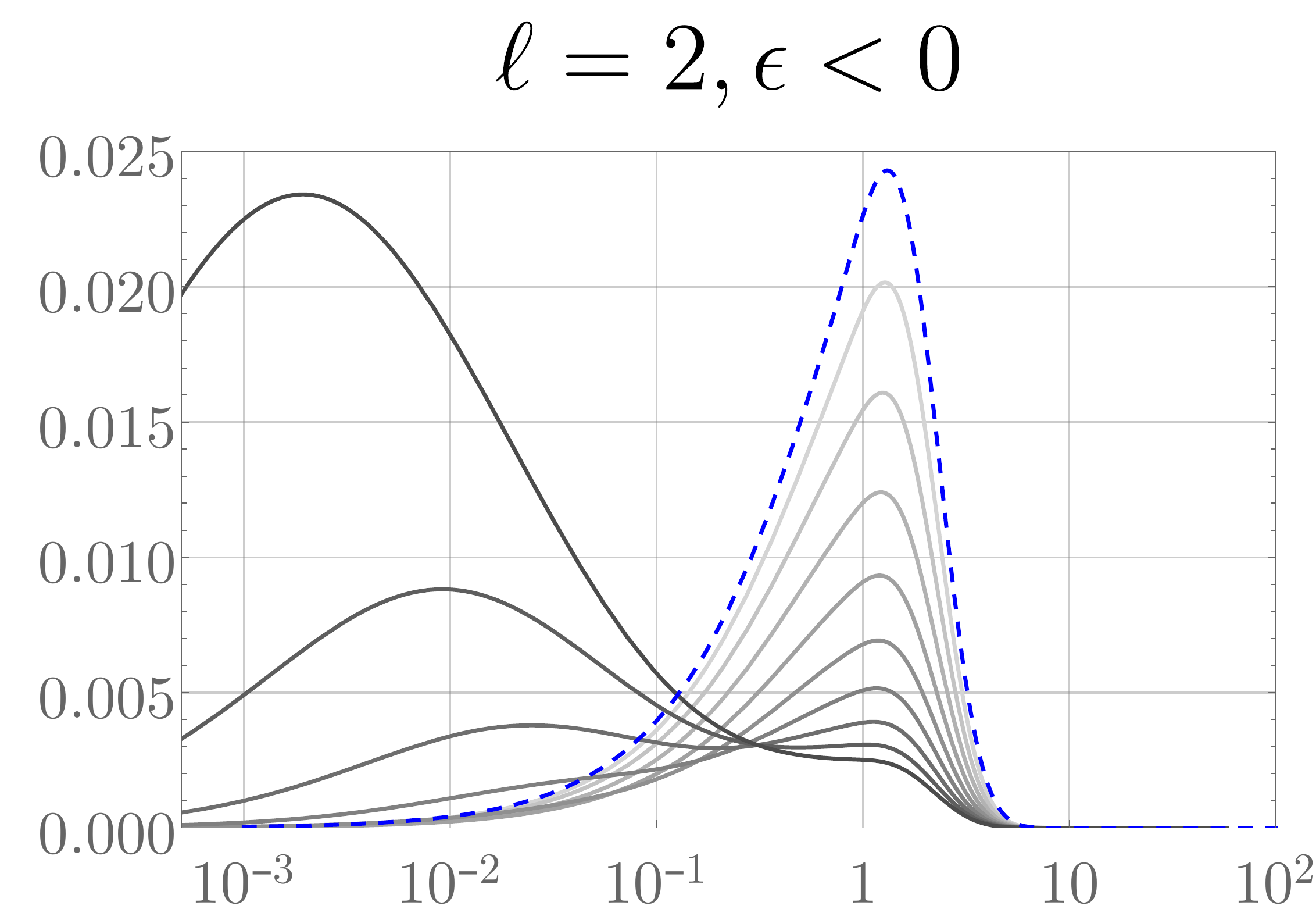}%
\includegraphics[width=0.33\textwidth]{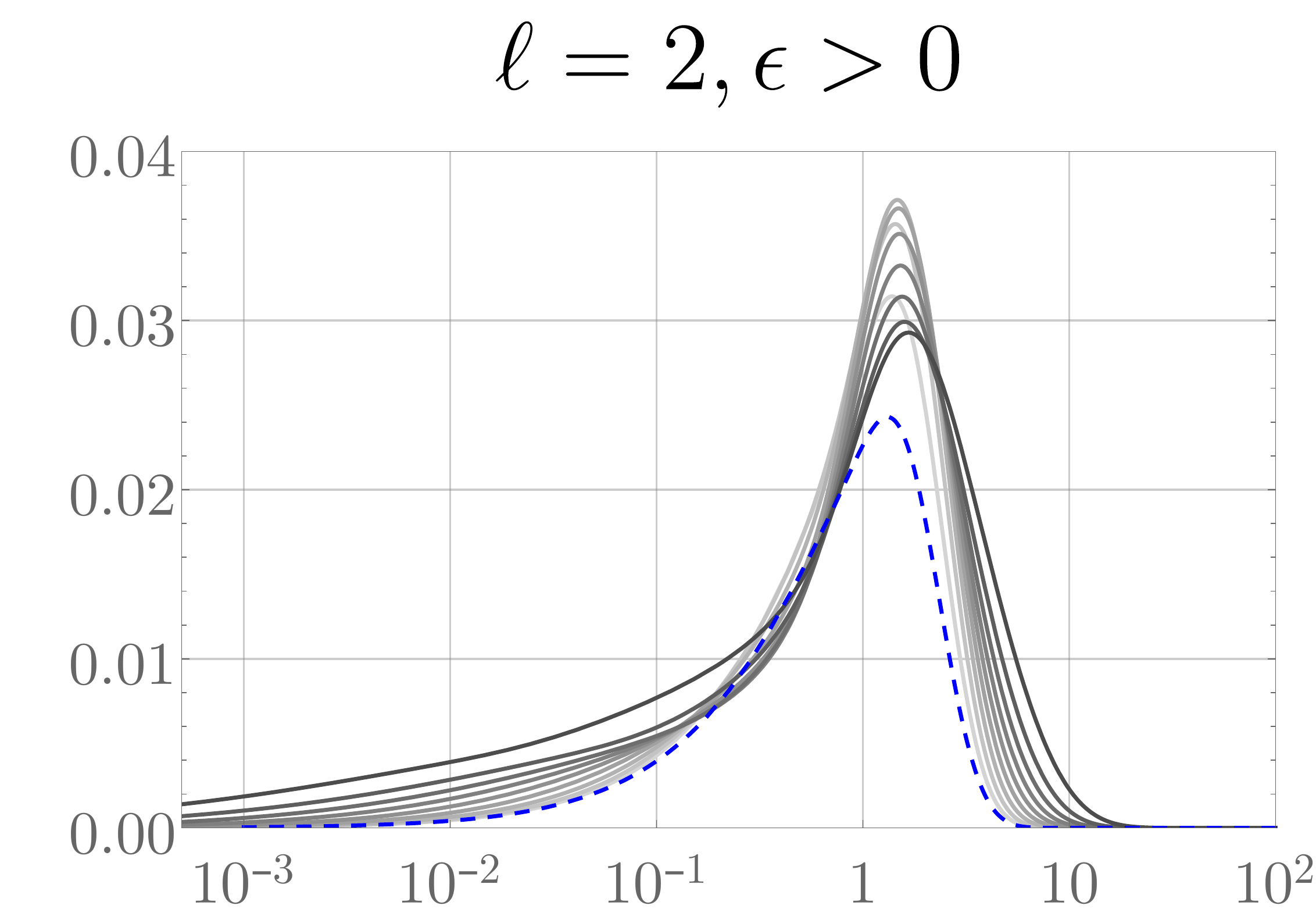}

\vspace{0.2cm}

\includegraphics[width=0.33\textwidth]{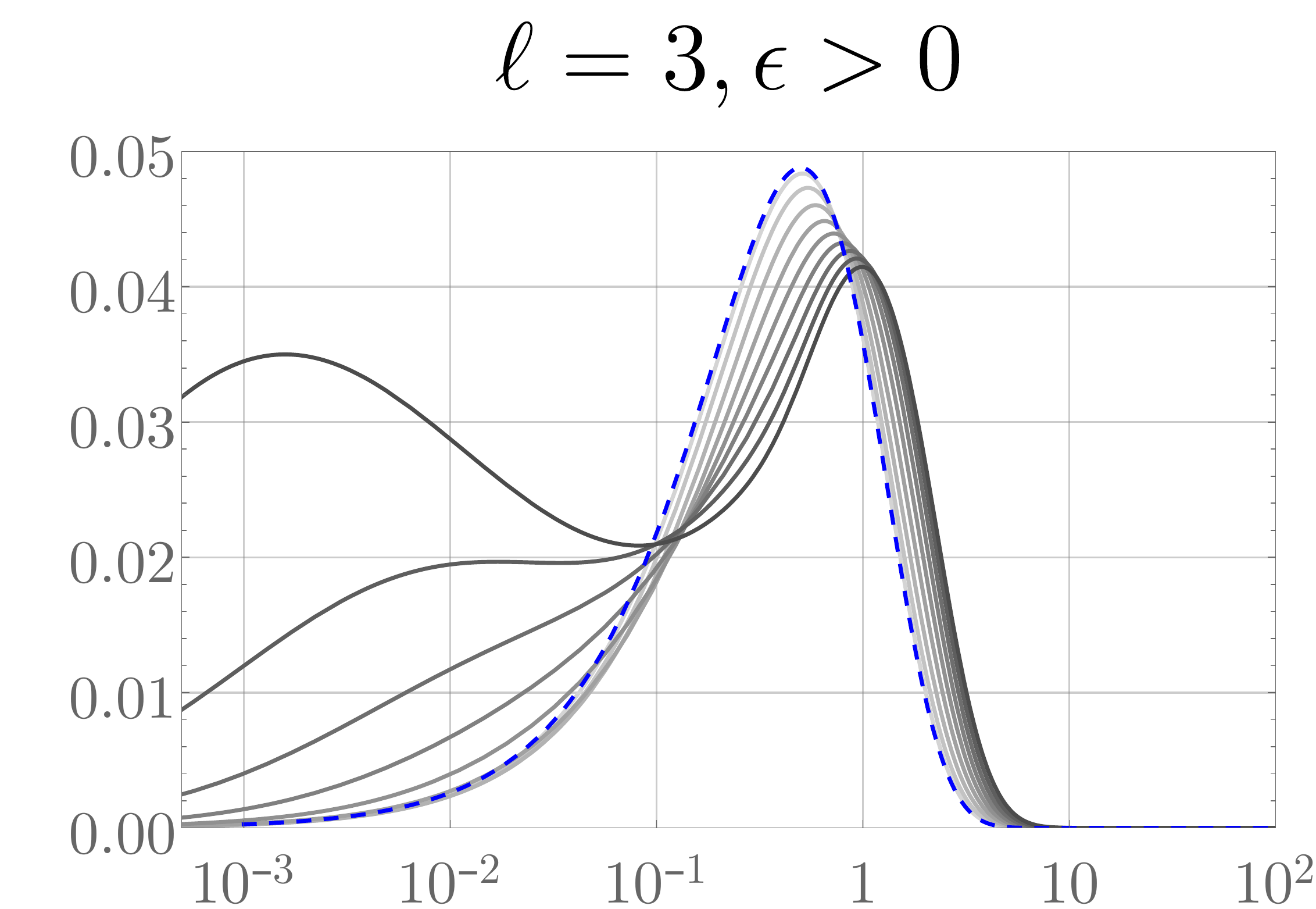}
\includegraphics[width=0.33\textwidth]{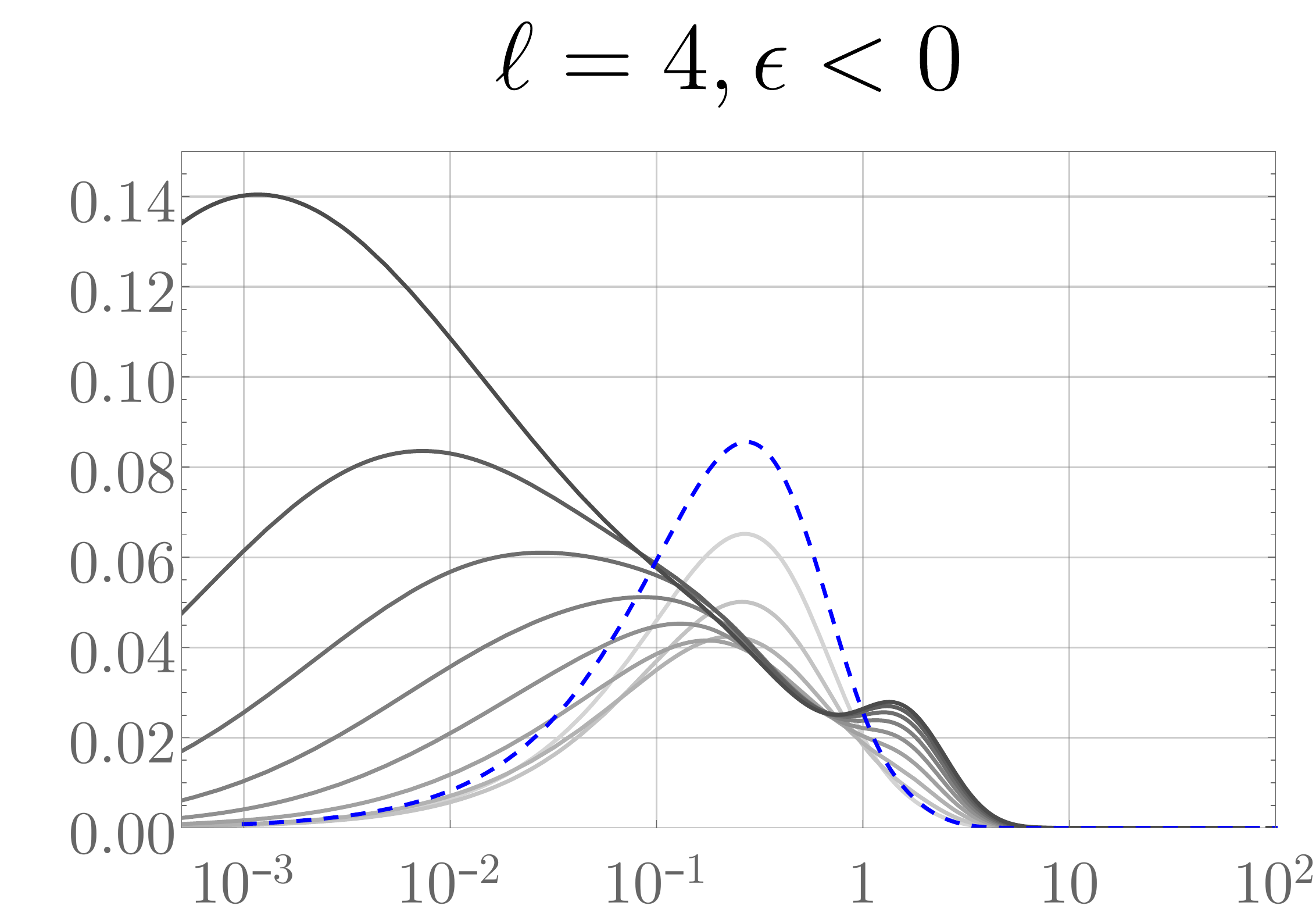}%
\includegraphics[width=0.33\textwidth]{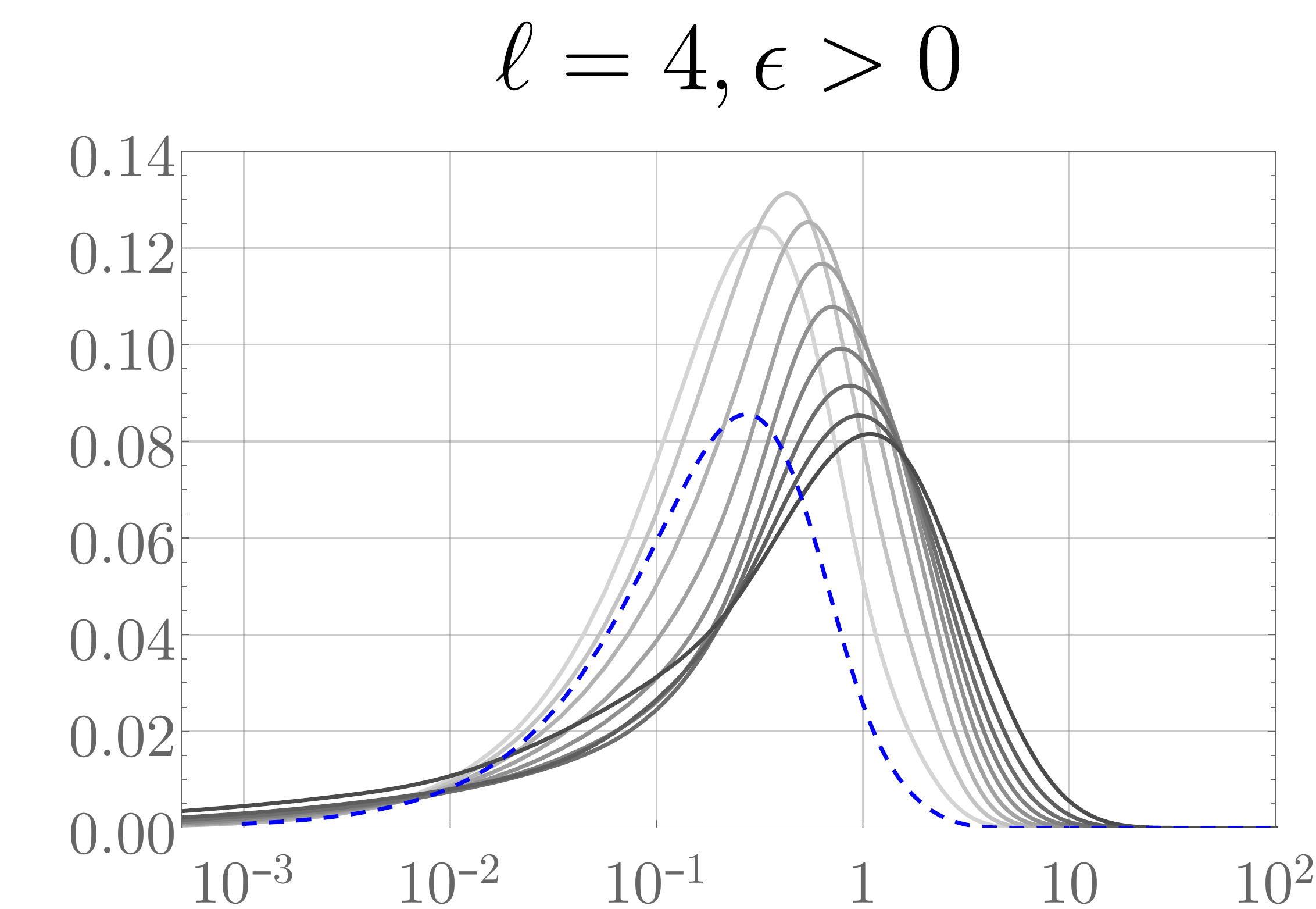}

\vspace{0.2cm}

\includegraphics[width=0.33\textwidth]{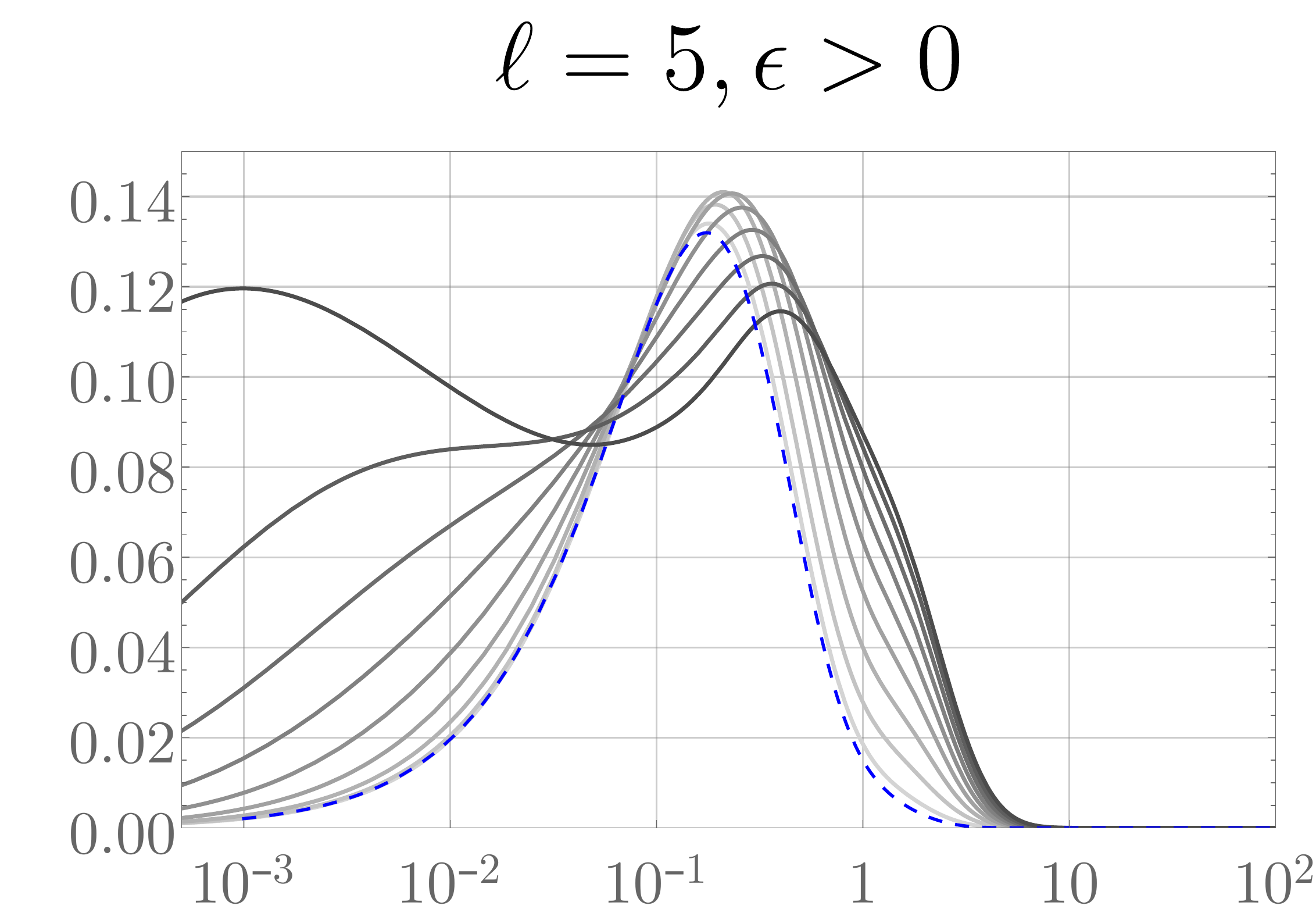}
\includegraphics[width=0.33\textwidth]{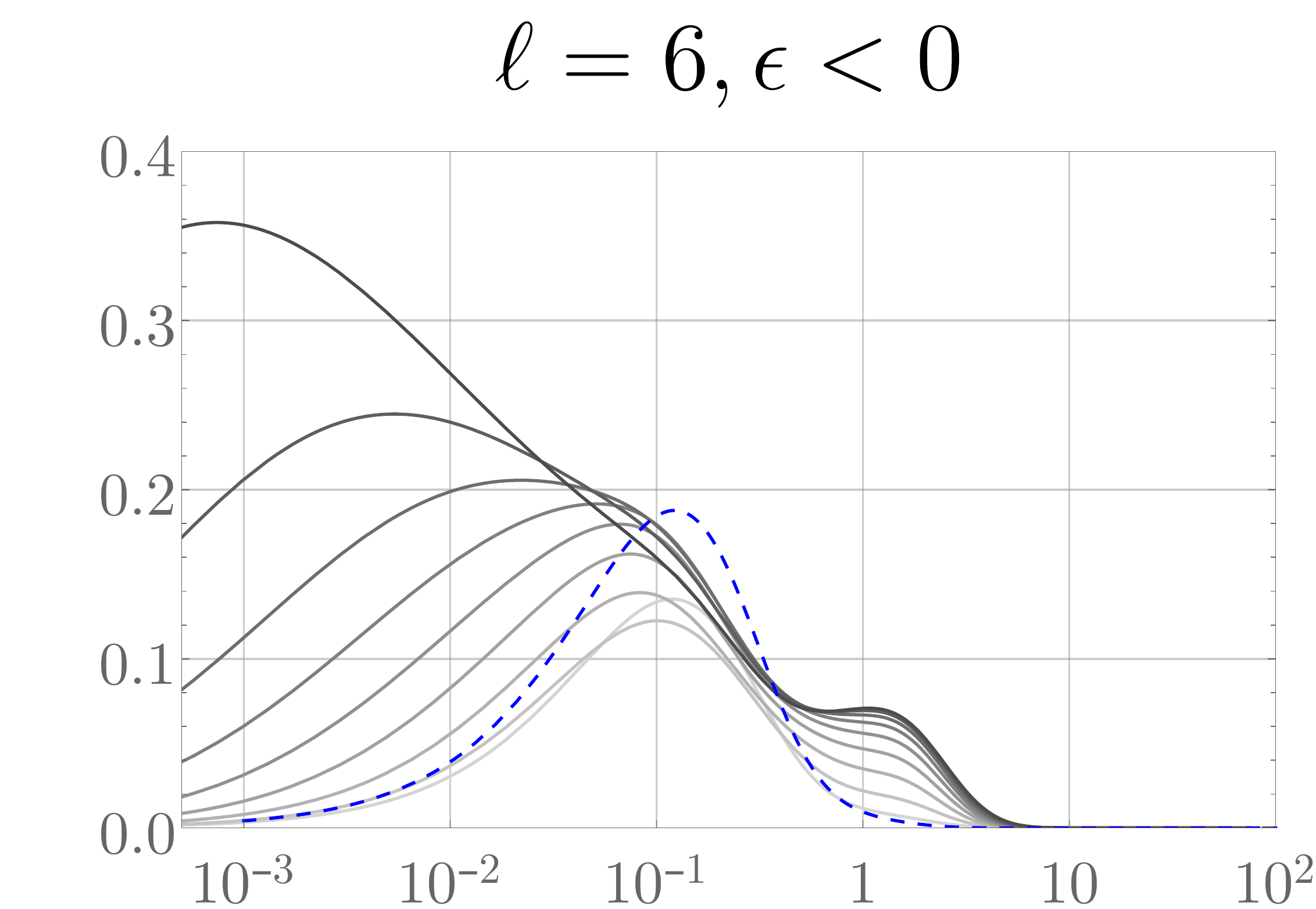}%
\includegraphics[width=0.33\textwidth]{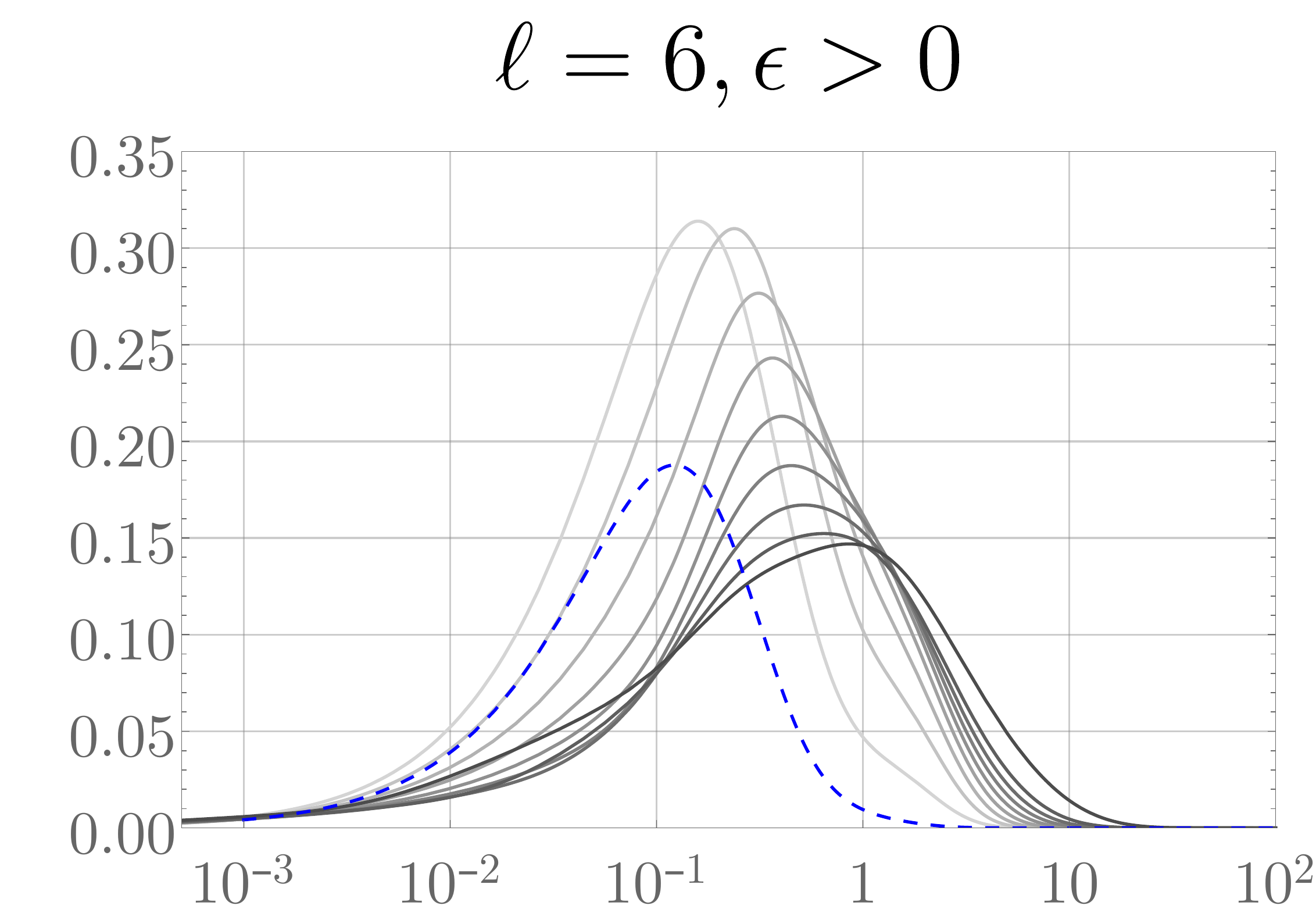}

\caption{The differenced heat kernel~$\Delta K_L(t)$ as a function of~$t$ for the minimally-coupled scalar on the deformed spheres given by~\eqref{eq:Rellembedding}.  Each plot shows the rescaled heat kernel~$-\sigma \Delta K_L/\eps^2$ (except for~$\ell = 1$, which shows~$-\sigma \Delta K_L/\eps^4$), with the dashed blue line corresponding to the perturbative result~\eqref{eq:K2scalar} and the gray lines to the numerical results for the deformations shown in Figure~\ref{fig:embeddings} (hence light to dark gray corresponds to increasing~$|\eps|$, with~$\eps \in [0.9\eps_\mathrm{min}, 0.9\eps_\mathrm{max}]$).}
\label{fig:scalarDeltaK}
\end{figure}

\begin{figure}[t]
\captionsetup[subfigure]{justification=centering,labelformat=empty}
\centering
\includegraphics[width=0.33\textwidth]{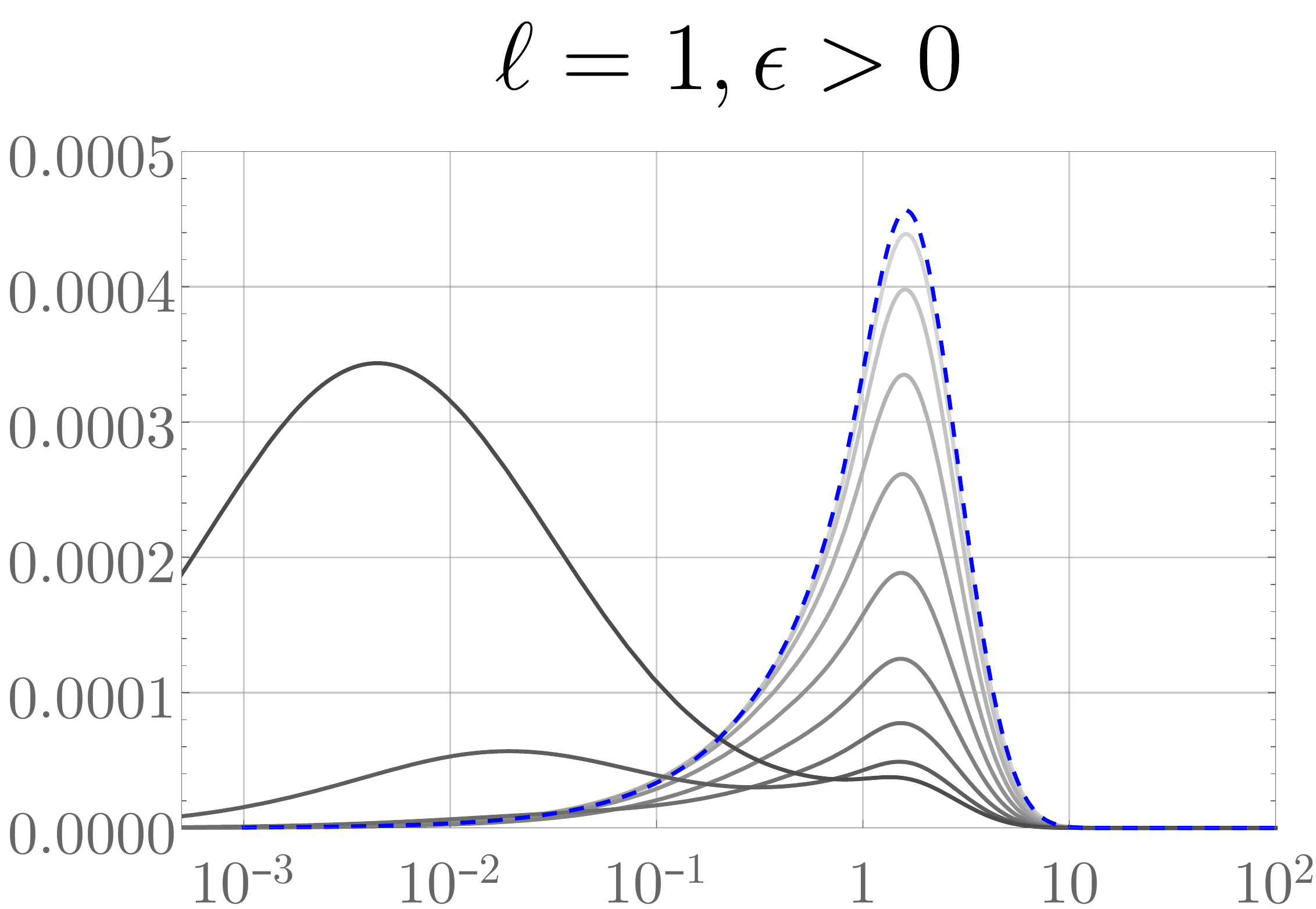}
\includegraphics[width=0.33\textwidth]{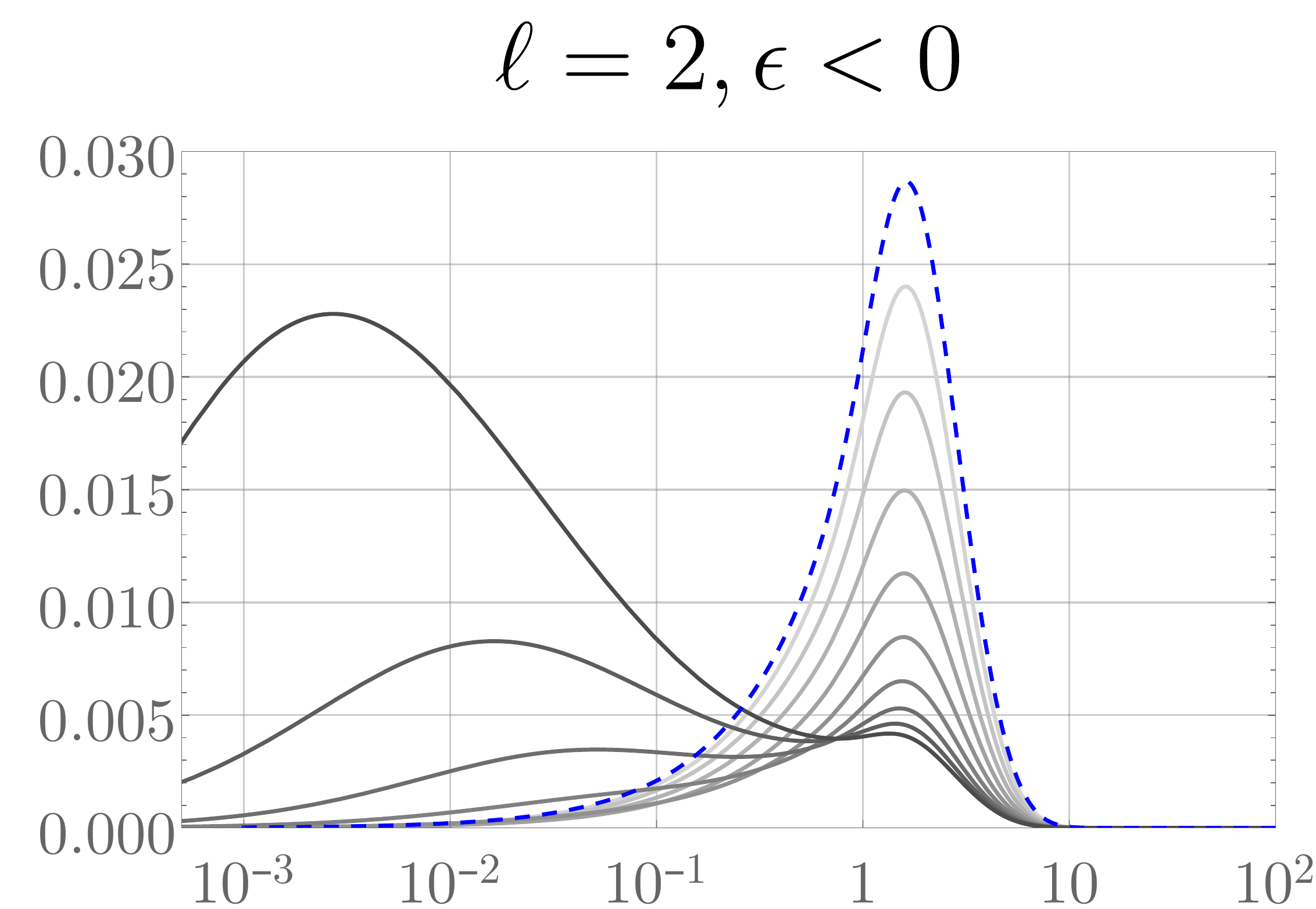}%
\includegraphics[width=0.33\textwidth]{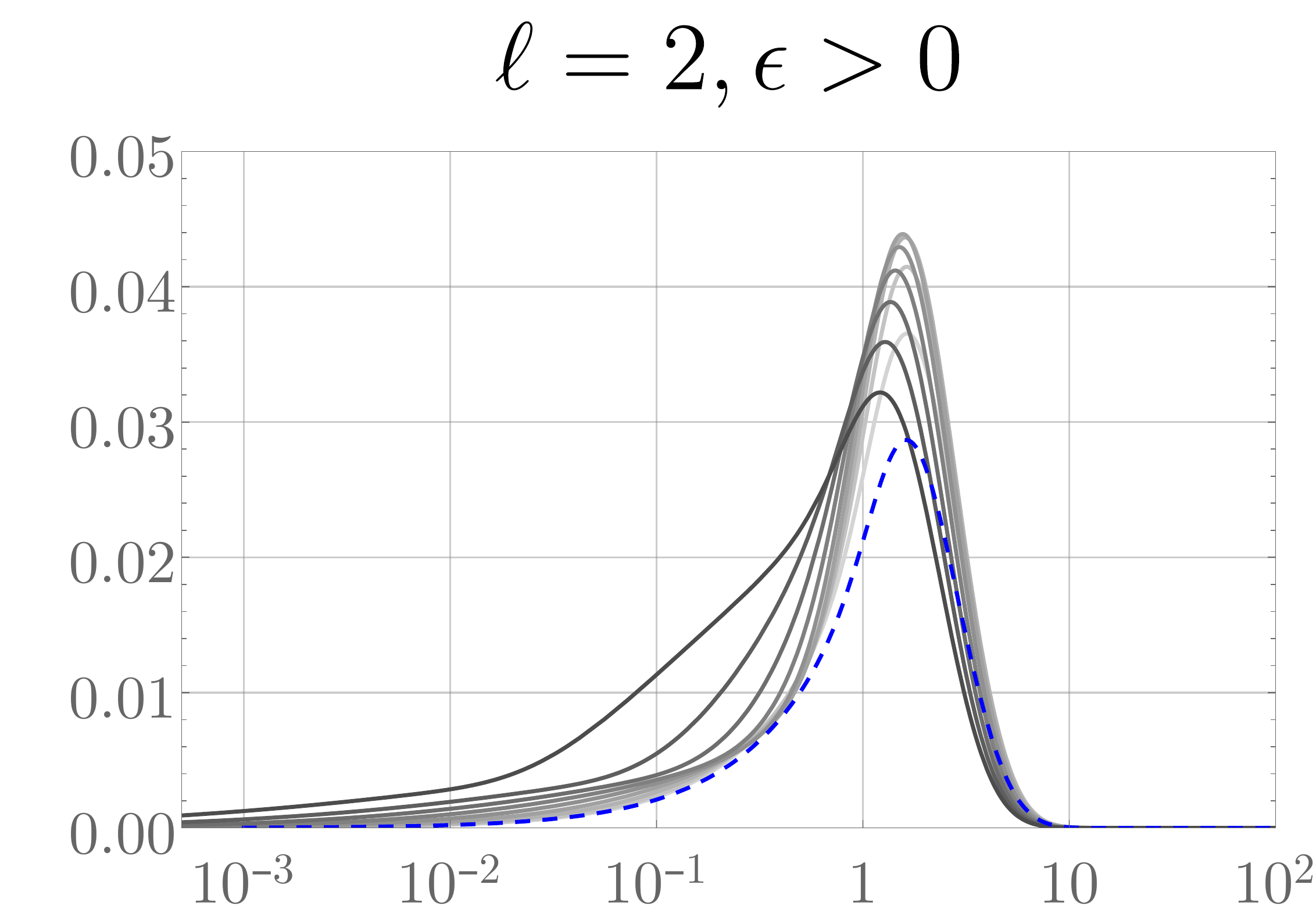}

\vspace{0.2cm}

\includegraphics[width=0.33\textwidth]{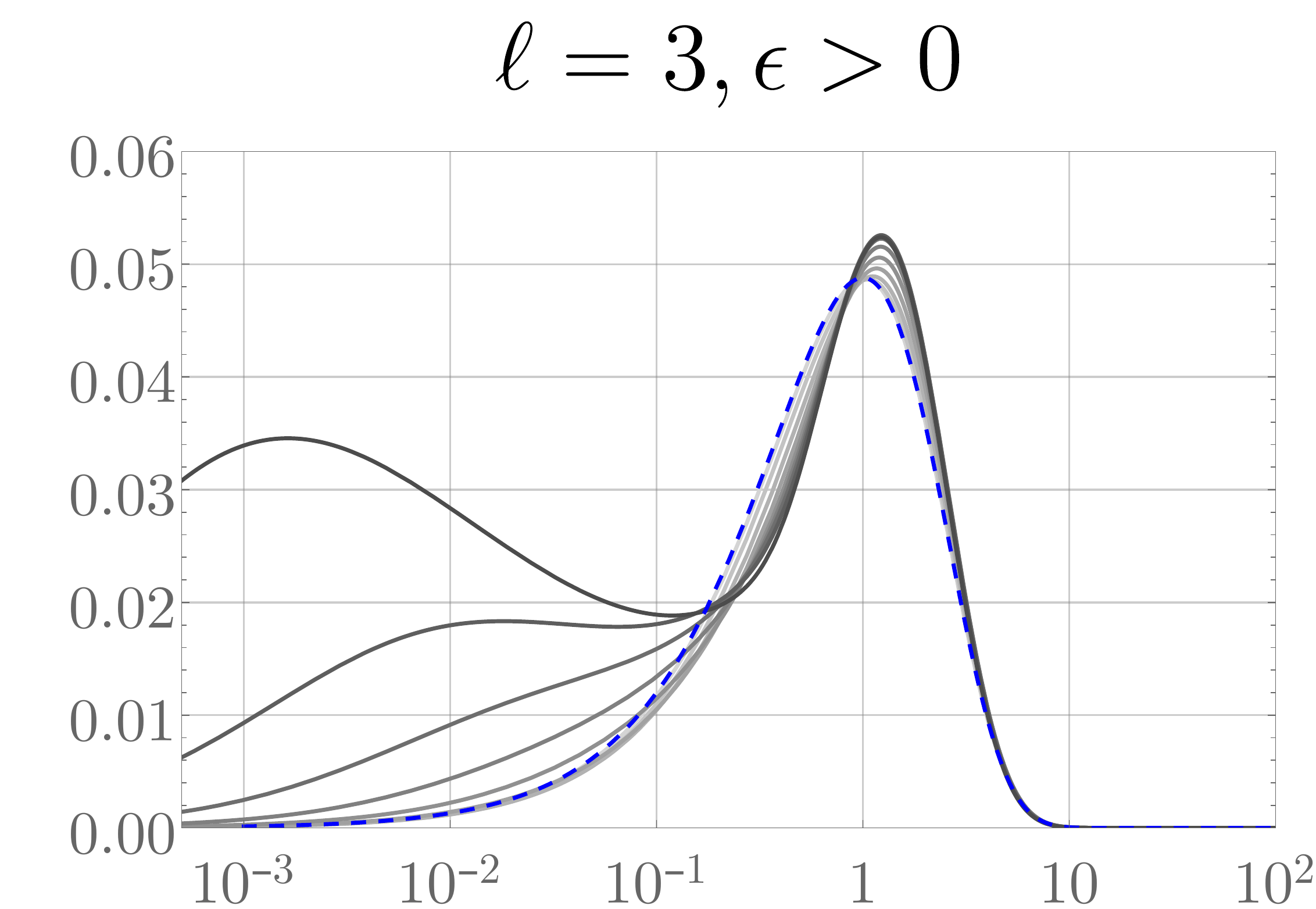}
\includegraphics[width=0.33\textwidth]{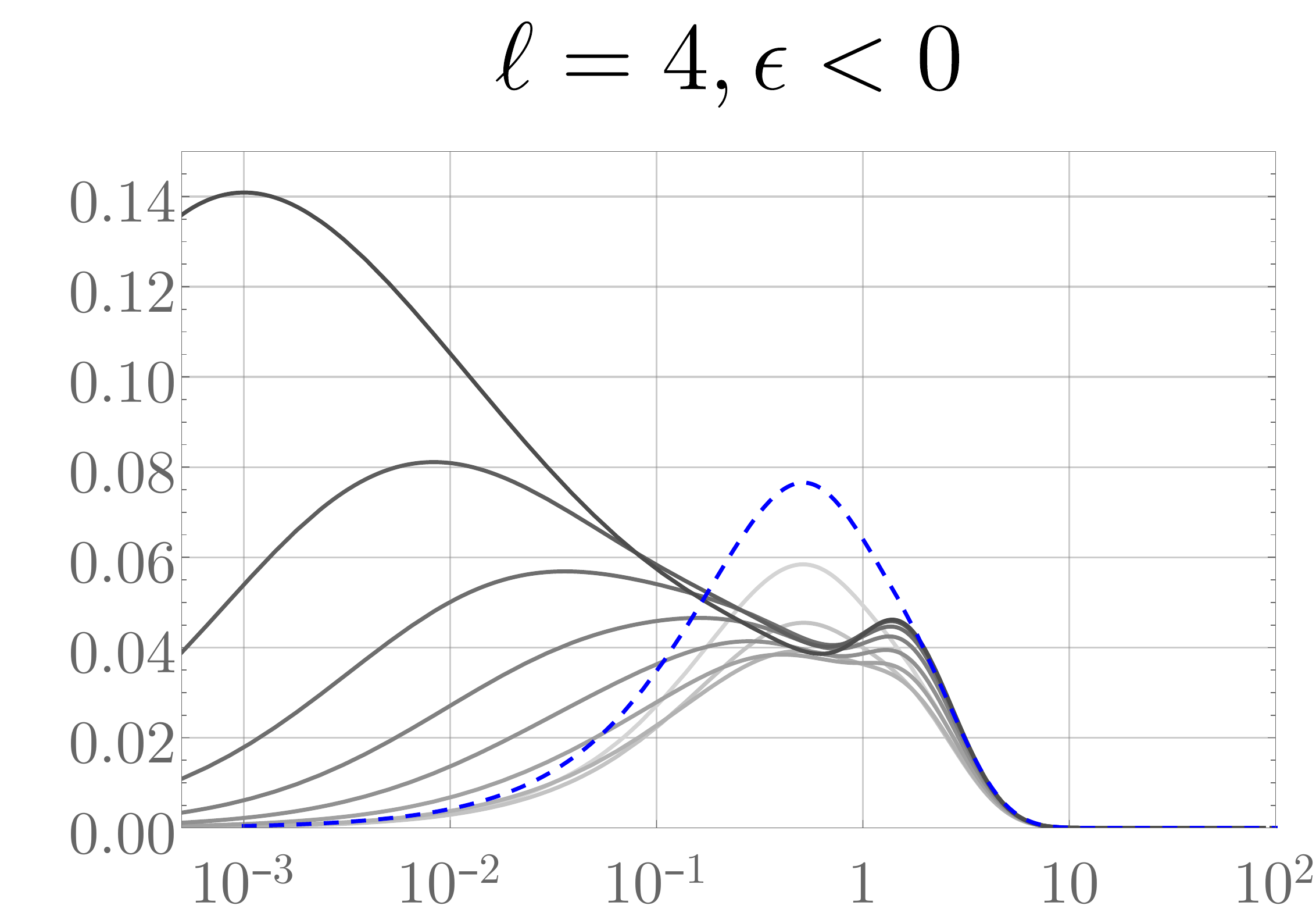}%
\includegraphics[width=0.33\textwidth]{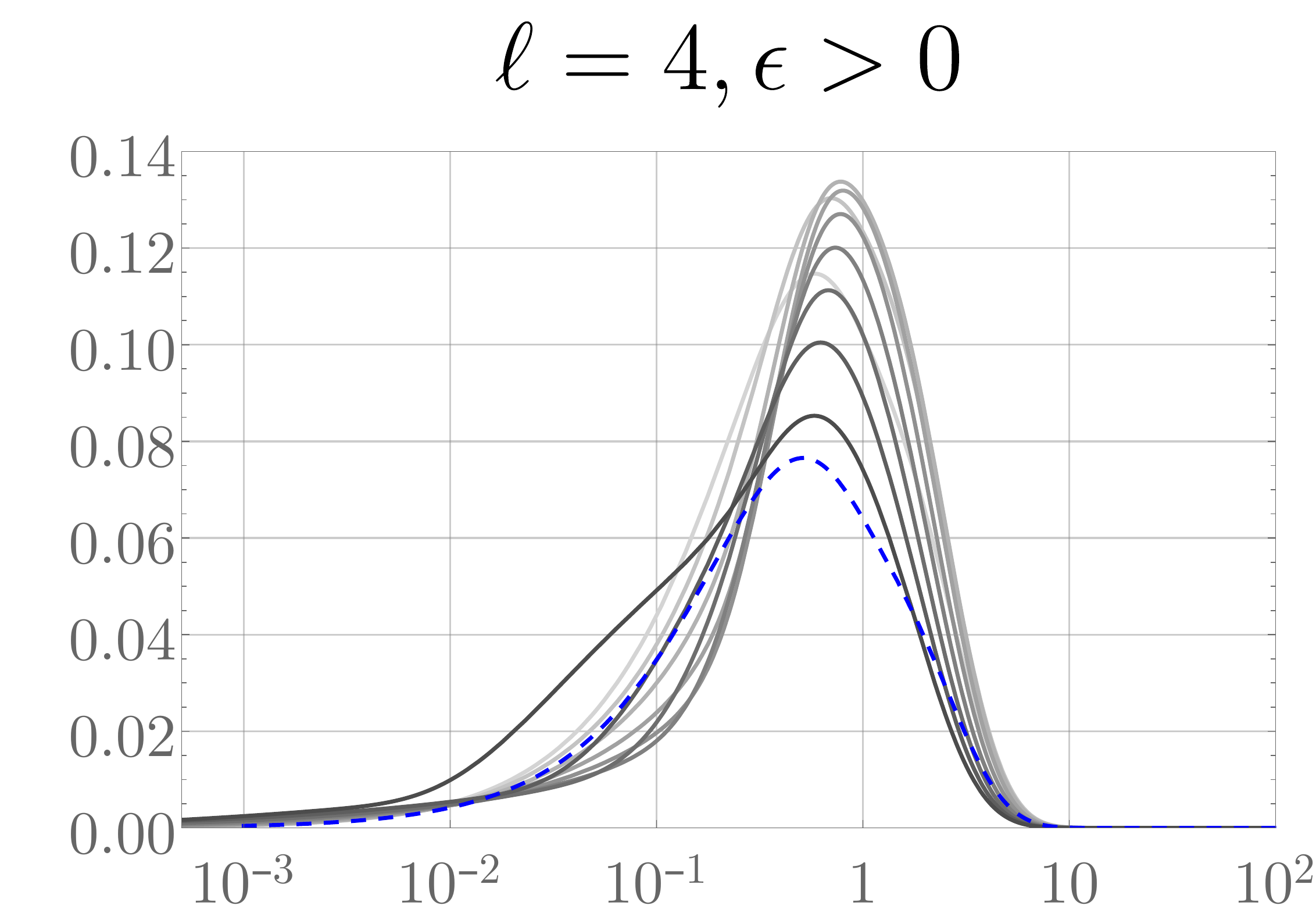}

\vspace{0.2cm}

\includegraphics[width=0.33\textwidth]{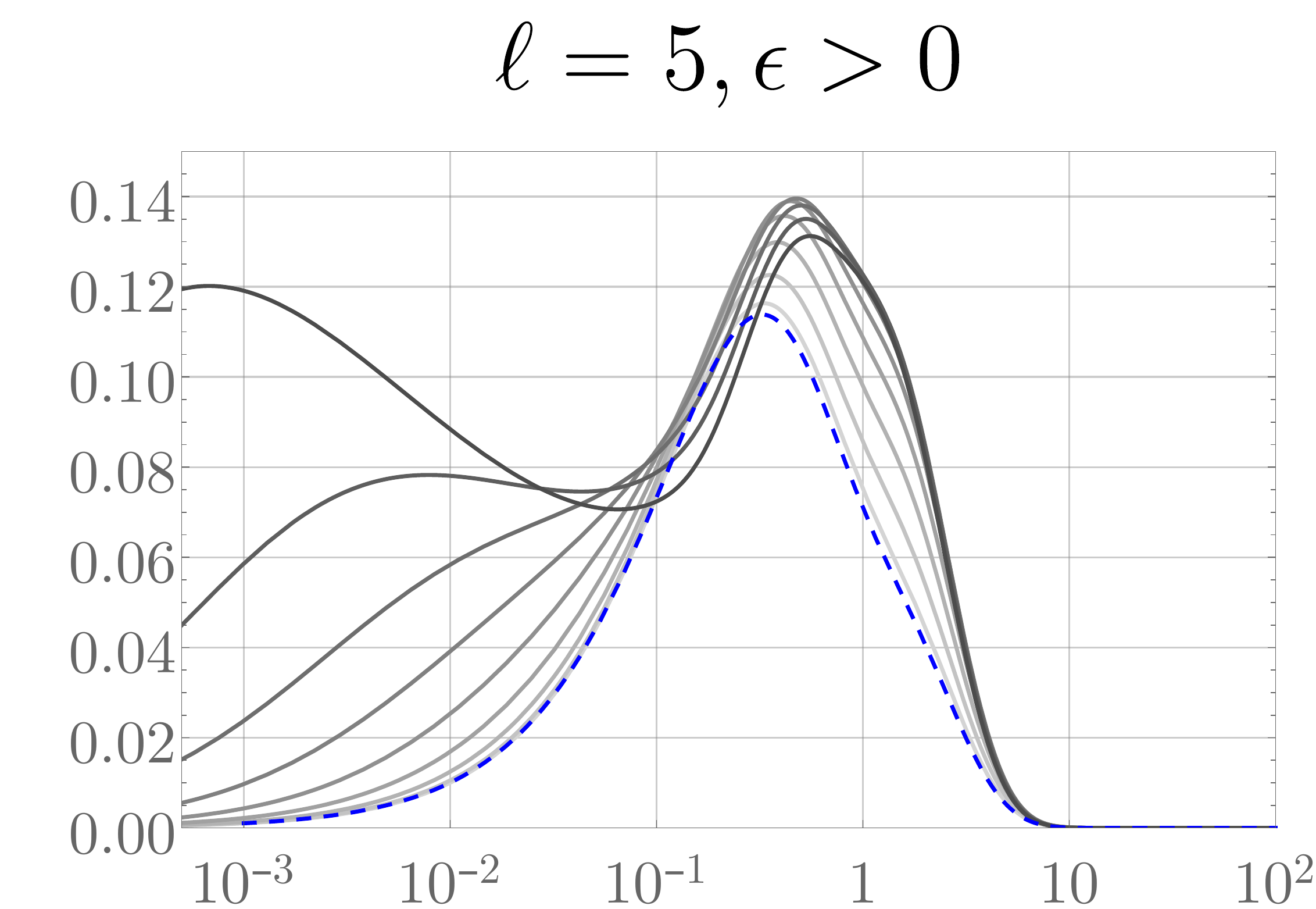}
\includegraphics[width=0.33\textwidth]{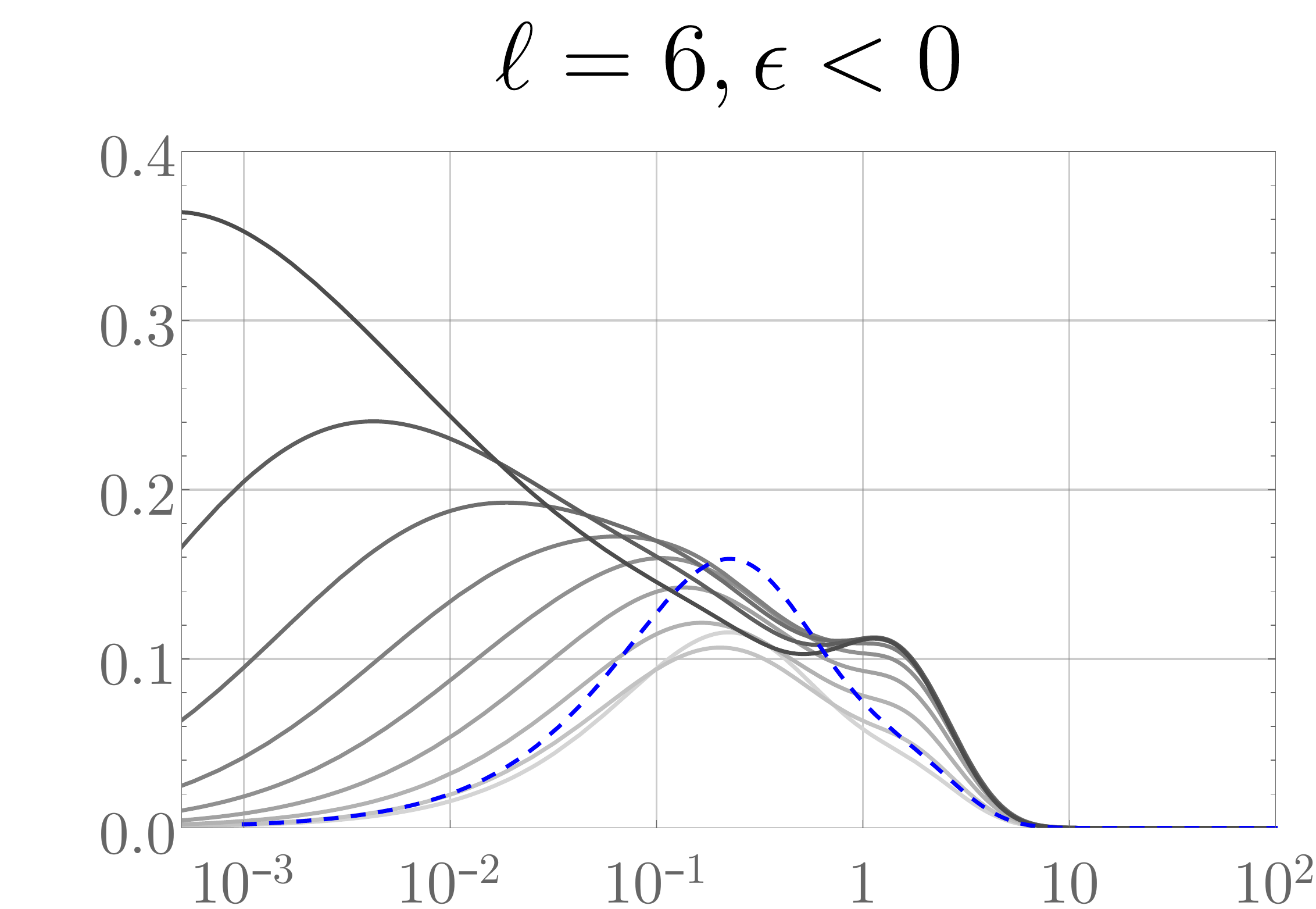}%
\includegraphics[width=0.33\textwidth]{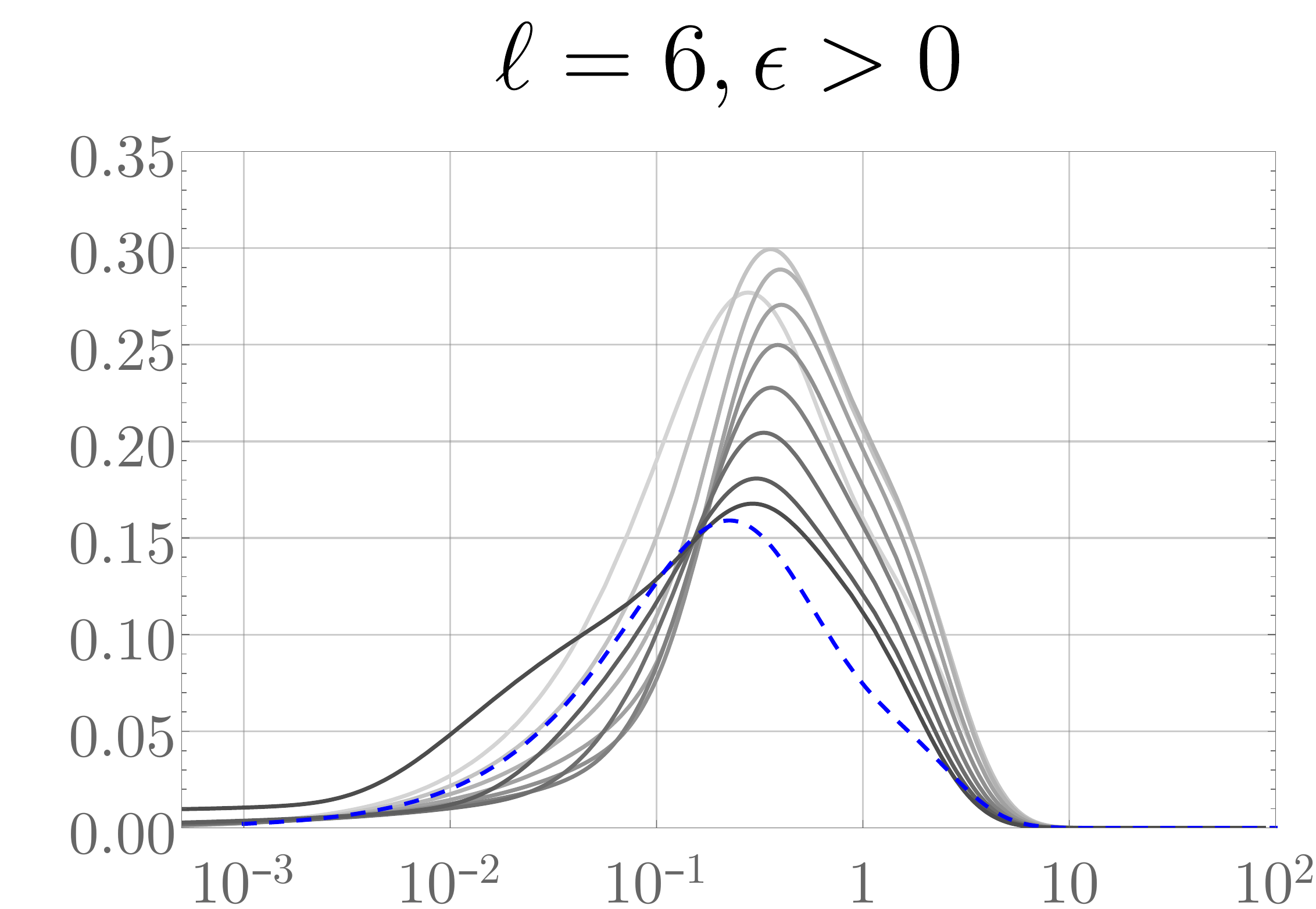}
\caption{The differenced heat kernel~$\Delta K_L(t)$ as a function of~$t$ for the Dirac fermion on the deformed spheres given by~\eqref{eq:Rellembedding}.  Each plot shows the rescaled heat kernel~$-\sigma \Delta K_L/\eps^2$ (except for~$\ell = 1$, which shows~$-\sigma \Delta K_L/\eps^4$), with the dashed blue line corresponding to the perturbative result~\eqref{eq:K2dirac} and the gray lines to the numerical results for the deformations shown in Figure~\ref{fig:embeddings} (hence light to dark gray corresponds to increasing~$|\eps|$, with~$\eps \in [0.9\eps_\mathrm{min}, 0.9\eps_\mathrm{max}]$).}
\label{fig:fermionDeltaK}
\end{figure}

In Figures~\ref{fig:scalarDeltaK} and~\ref{fig:fermionDeltaK}, we show the differenced heat kernels~$\Delta K_L(t)$ for the minimally-coupled scalar and for the Dirac fermion normalized by~$\eps^2$ (or by~$\eps^4$ in the case of~$\ell = 1$) along with the perturbative results derived in Section~\ref{sec:pert}.  Note that we only plot~$\Delta K_L$ down to~$t = 0.0005$; this is because in the small-$t$ regime more and more eigenvalues of~$L$ contribute to~$\Delta K_L$ leading to difficulty in controlling the numerics.  But as discussed above, the small-$t$ regime is controlled by the heat kernel expansion, which guarantees the sign of~$\sigma \Delta K_L$ to be negative there.  We therefore see that~$\sigma \Delta K_L(t)$ is negative for all~$t$ even for large deformations of the sphere.  Interestingly,~$\Delta K_L(t)$ appears to \textit{grow} with~$\eps$ at sufficiently small fixed values of~$t$; this is due to the fact that as the geometry becomes more singular, its Ricci curvature grows, causing the heat kernel coefficient~$\Delta b_4$ defined in~\eqref{eqs:Deltaa4} to grow as well.  This growth is especially pronounced in the deformations with odd~$\ell$ and those with even~$\ell$ and~$\eps < 0$; comparing to Figure~\ref{fig:embeddings}, these deformations all limit towards a connected geometry with a cusp-like defect (the geometries with even~$\ell$ and~$\eps > 0$, on the other hand, pinch off into separate disconnected components as~$\eps \to \eps_\mathrm{max}$).  This growth of~$\Delta K_L$ at small~$t$ should lead to a corresponding growth in the free energy~$\Delta F$; we now investigate this free energy, and then more carefully investigate the divergence structure associated to the limiting singular geometries.

\subsection{Behavior of the Free Energy}
\label{subsec:freeenergynonpert}

At zero mass and temperature, the differenced free energy~$\Delta F_{T = 0} = \Delta E$ may be computed by using~\eqref{eq:Thetazerotemperature} and then integrating the heat kernel with~\eqref{eq:DeltaFheatkernel}; more details on the computation can be found in Appendix~\ref{app:numericalchecks}.  In Figure~\ref{fig:DeltaF} we show~$\Delta E$ for the deformations described above, normalized by the perturbative result~$\Delta E_\mathrm{pert}$.  As expected,~$|\Delta E|$ grows monotonically with increasing~$|\eps|$, though this may not be apparent from Figure~\ref{fig:DeltaF} as the curves are normalized by a factor of~$\eps^2$ (or~$\eps^4$ for~$\ell = 1$) contained in~$\Delta E_\mathrm{pert}$.  Due to the growth of~$\Delta K$ at small~$t$ as~$\eps$ approaches~$\eps_\mathrm{min}$ or~$\eps_\mathrm{max}$, we only show~$\Delta E$ for a range of~$\eps$ within which the error in~$\Delta E$ is no greater than a few percent (this corresponds to~$\eps$ up to~$0.8 \eps_\mathrm{max}$ for~$\ell = 1$ and up to~$0.5\eps_\mathrm{max}$ for~$\ell = 6$).  Nevertheless, the growth in the small-$t$ behavior of the heat kernel makes clear that~$\Delta E$ should continue to grow as the geometry is successively deformed; we will investigate this growth in more detail in the following Section.  For now, let us note the remarkable feature that~$\Delta E/\Delta E_\mathrm{pert}$ looks extremely similar for both the scalar and fermion, despite the fact that the corresponding heat kernels in Figures~\ref{fig:scalarDeltaK} and~\ref{fig:fermionDeltaK} are more substantially different.  It therefore appears that the theory-dependence of~$\Delta E$ is contained almost completely in the perturbative contribution~$\Delta E_\mathrm{pert}$: the ratio~$\Delta E/\Delta E_\mathrm{pert}$ is almost entirely theory-independent (we highlight \textit{almost}: the difference between the curves is larger than numerical error, so they are genuinely different).  This is feature is interestingly reminiscent of the results of~\cite{BobBue17}, which study the free energy of the massless Dirac fermion, the conformally coupled scalar, and holographic CFTs on a squashed Euclidean three-sphere; they found that for small and modest squashings, the free energies of all of these theories agree more closely than should be expected from CFT considerations alone. Indeed, there is a conjecture and good evidence that the subleading term in the perturbative expansion of the free energy in the squashing parameter, determined by the three-point function of the stress tensor, is surprisingly universal for all three-dimensional CFTs \cite{Bueno:2018yzo,Bueno:2020odt}.

\begin{figure}[t]
\centering
\subfloat[][Odd $\ell$]{
\includegraphics[width=0.49\textwidth]{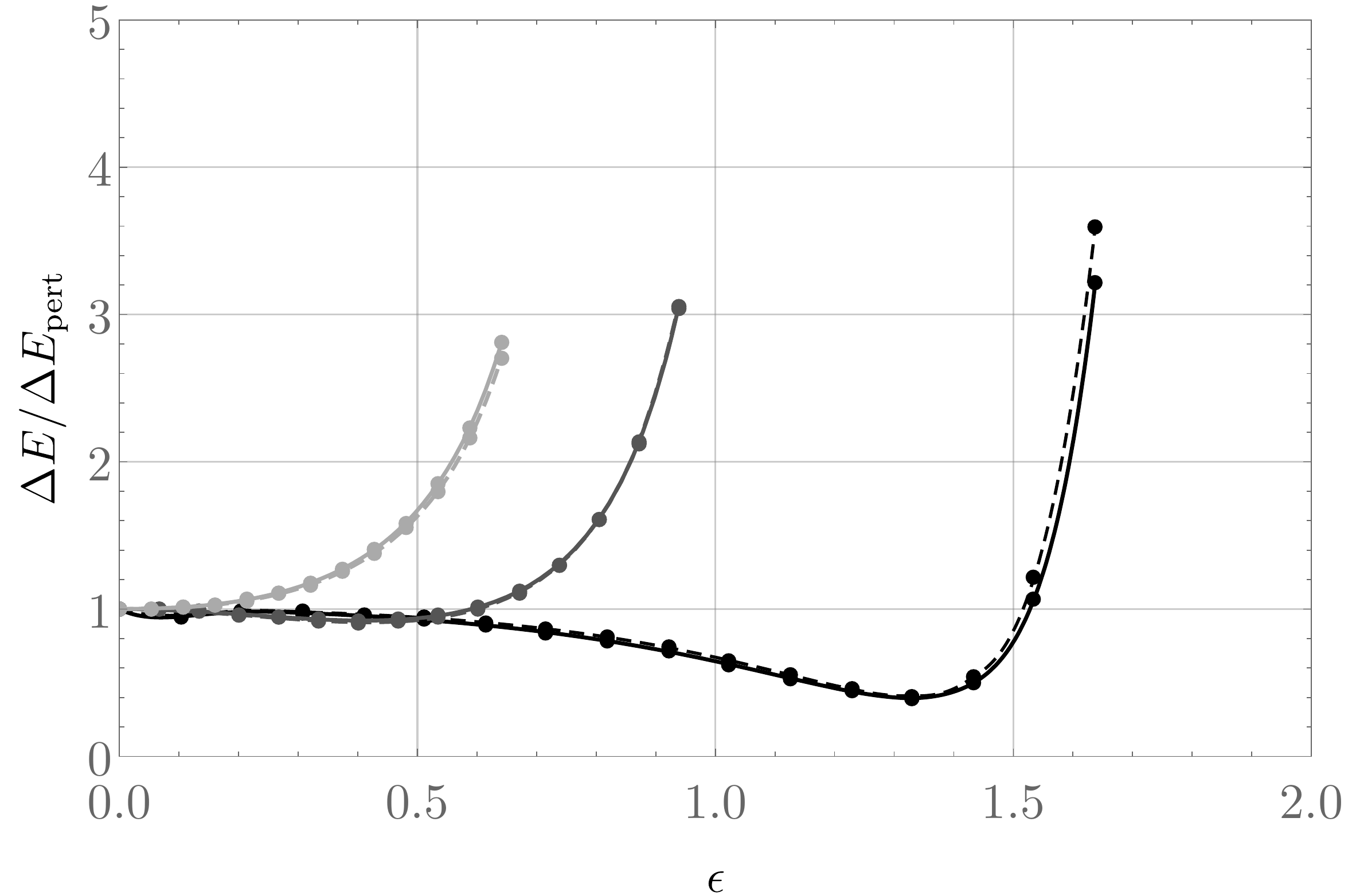}
\label{subfig:oddellDeltaE}
}
\subfloat[][Even $\ell$]{
\includegraphics[width=0.49\textwidth]{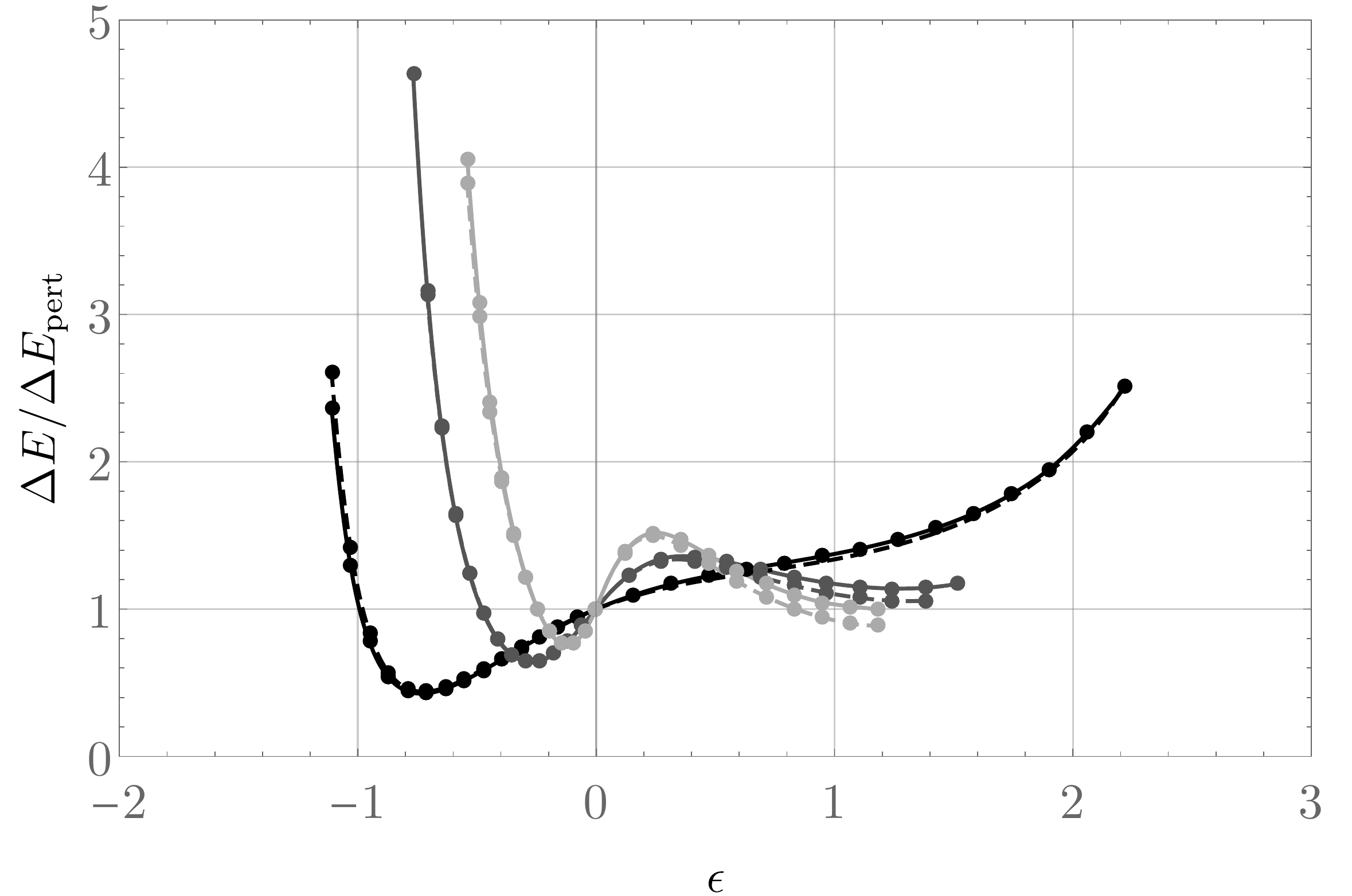}
\label{subfig:evenellDeltaE}
}
\caption{The zero-temperature differenced energy~$\Delta E$ normalized by its perturbative behavior~$\Delta E_\mathrm{pert}$ on the deformed spheres given by~\eqref{eq:NonPerturbativeDeformations} and~\eqref{eq:Rellembedding}.  Solid lines show results for the massless Dirac fermion, while dashed lines are for the massless minimally coupled scalar (points are numerical data; the curves are drawn to guide the eye).  From black to light gray, the curves corresponds to~$\ell = 1, 3, 5$ (left) and~$\ell = 2, 4, 6$ (right).  It is striking that the Dirac fermion and scalar give very similar curves.}
\label{fig:DeltaF}
\end{figure}

In fact, this theory-independence becomes exact in a long-wavelength limit.  Specifically, let~$l$ be the typical curvature scale of the deformed geometry; then the heat kernel coefficient~$\Delta b_{2n}$ scales like~$l^{-2n}$ for~$n \geq 2$.  The heat kernel expansion~\eqref{eq:DeltaKexpansion} then implies that for~$l \gg T^{-1}, M^{-1}$, the free energy~\eqref{eq:DeltaFheatkernel} can be expressed as an expansion in powers of~$1/(Tl)$ or~$1/(M l)$\footnote{In fact, for the scalar such an expansion necessarily requires~$l M \gg 1$, whereas for the fermion it is sufficient for \textit{either}~$M l$ \textit{or}~$T l$ to be large.  This is due to the fact that at large~$T$,~$\Theta_\sigma(T^2 t)$ falls off exponentally for the fermion but approaches a nonzero constant for the scalar.}.  Indeed, for general~$M$,~$T$ we have that~\cite{CheWal18}
\be
\label{eq:DeltaFhighT}
\Delta F  = \sigma \sum_{n = 0}^{\infty} (-1)^n \frac{\Delta b_{2n+4}}{T^{2n+1}} J^{(n)}\left(\frac{M^2}{T^2}\right),
\ee
where~$J^{(n)}$ is the~$n$th derivative of the function given by
\be
J(\zeta) = \frac{1}{2\sqrt{\zeta}} \begin{cases} \coth\left(\frac{\sqrt{\zeta}}{2}\right), & \mbox{scalar}, \\ \tanh\left(\frac{\sqrt{\zeta}}{2}\right), & \mbox{fermion}. \end{cases}
\ee 
For~$M \sim T$,~\eqref{eq:DeltaFhighT} is clearly an expansion in~$1/(T l)^2$.  For~$M \gg T$, it instead becomes
\be
\Delta F = \frac{\sigma}{\sqrt{4\pi} \, M} \sum_{n = 0}^\infty \frac{\Delta b_{2n+4}}{M^{2n}} \, \Gamma\left(n+\frac{1}{2}\right)\left[1 + \Ocal\left(e^{-M/T}\right)\right],
\ee
which is an expansion in~$1/(M l)^2$.  On the other hand, for~$M \ll T$ we have
\be
\Delta F = \sigma \sum_{n=0}^\infty \Delta b_{2n+4} \begin{Bmatrix} \frac{T}{M^2} \frac{n!}{M^{2n}}, & \qquad \mbox{scalar} \\ \frac{(-1)^n J^{(n)}(0)}{T^{2n+1}}, & \qquad \mbox{fermion} \end{Bmatrix} \left[1+ \Ocal\left(M^2/T^2\right)\right],
\ee
which is an expansion in~$1/(M l)^2$ for the scalar and~$1/(T l)^2$ for the fermion.

The point is that as long as~$l \gg T^{-1}, M^{-1}$, the leading-order behavior of the differenced free energy is governed by the lowest heat kernel coefficient~$\Delta b_4$:
\be
\label{eq:DeltaFhighTleading}
\Delta F = \sigma \, \frac{\Delta b_4}{T} \, J\left(\frac{M^2}{T^2}\right) + \cdots,
\ee
where~$\cdots$ denotes subleading terms.  The theory-dependence of~$\Delta F$ can be seen by expanding
\be
\Delta b_4 = \Delta b_4^{(2)} \eps^2 + \Ocal(\eps^3),
\ee
from which we have
\be
\label{eq:DeltaFratiohightemp}
\frac{\Delta F}{\Delta F_\mathrm{pert}} = \frac{\Delta b_4}{\Delta b_4^{(2)} \eps^2} + \cdots.
\ee
But from~\eqref{eqs:Deltaa4}, the ratio~$\Delta b_4/\Delta b^{(2)}_4$ is the same for the fermion and the scalar, so to leading order~$\Delta F/\Delta F_\mathrm{pert}$ is independent of the theory (as well as of the mass and temperature).

\begin{figure}[t]
\centering
\subfloat[][$M = 0$, varying $T$]{
\includegraphics[width=0.49\textwidth]{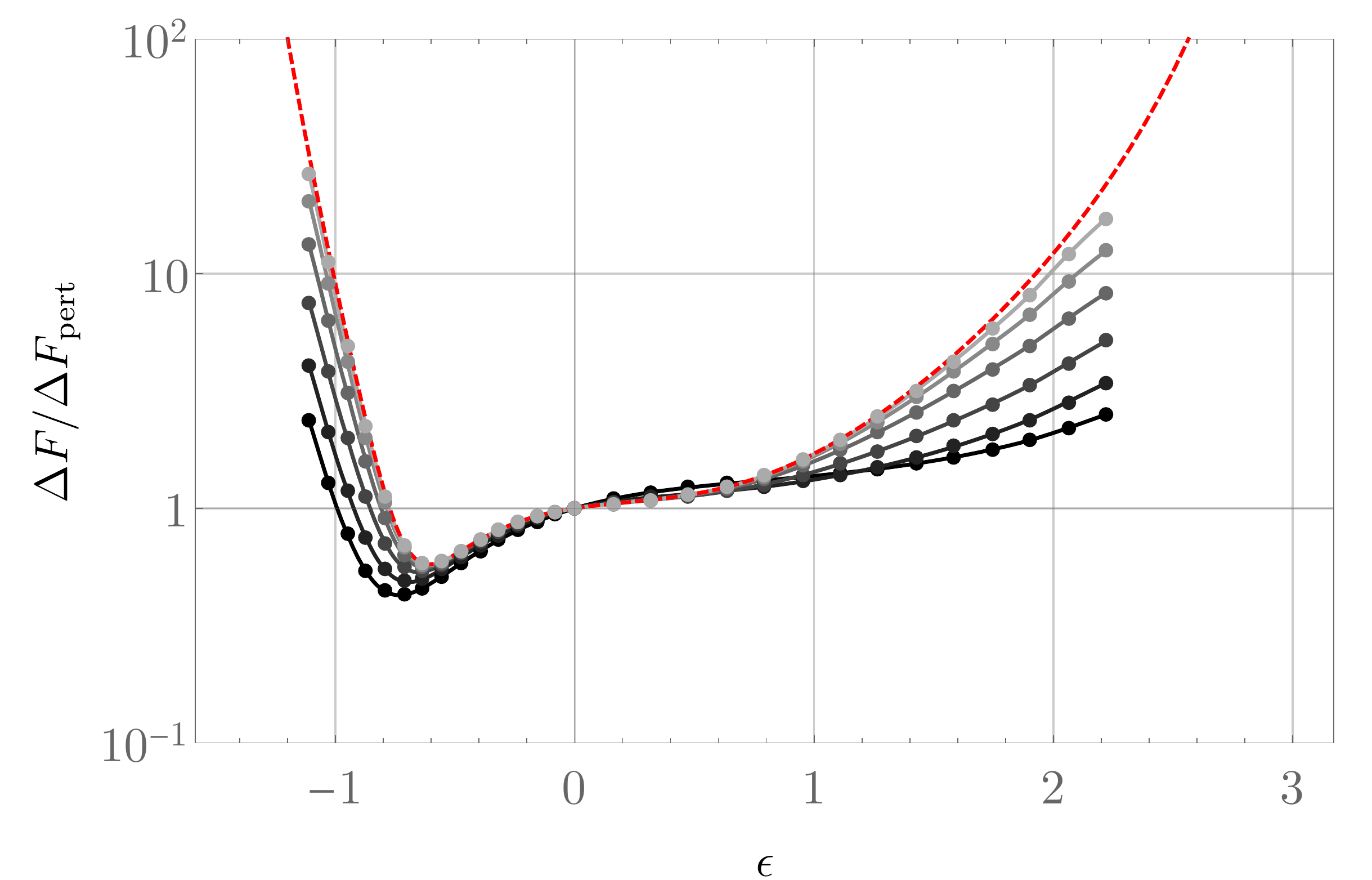}
}
\subfloat[][$T = 0$, varying $M$]{
\includegraphics[width=0.49\textwidth]{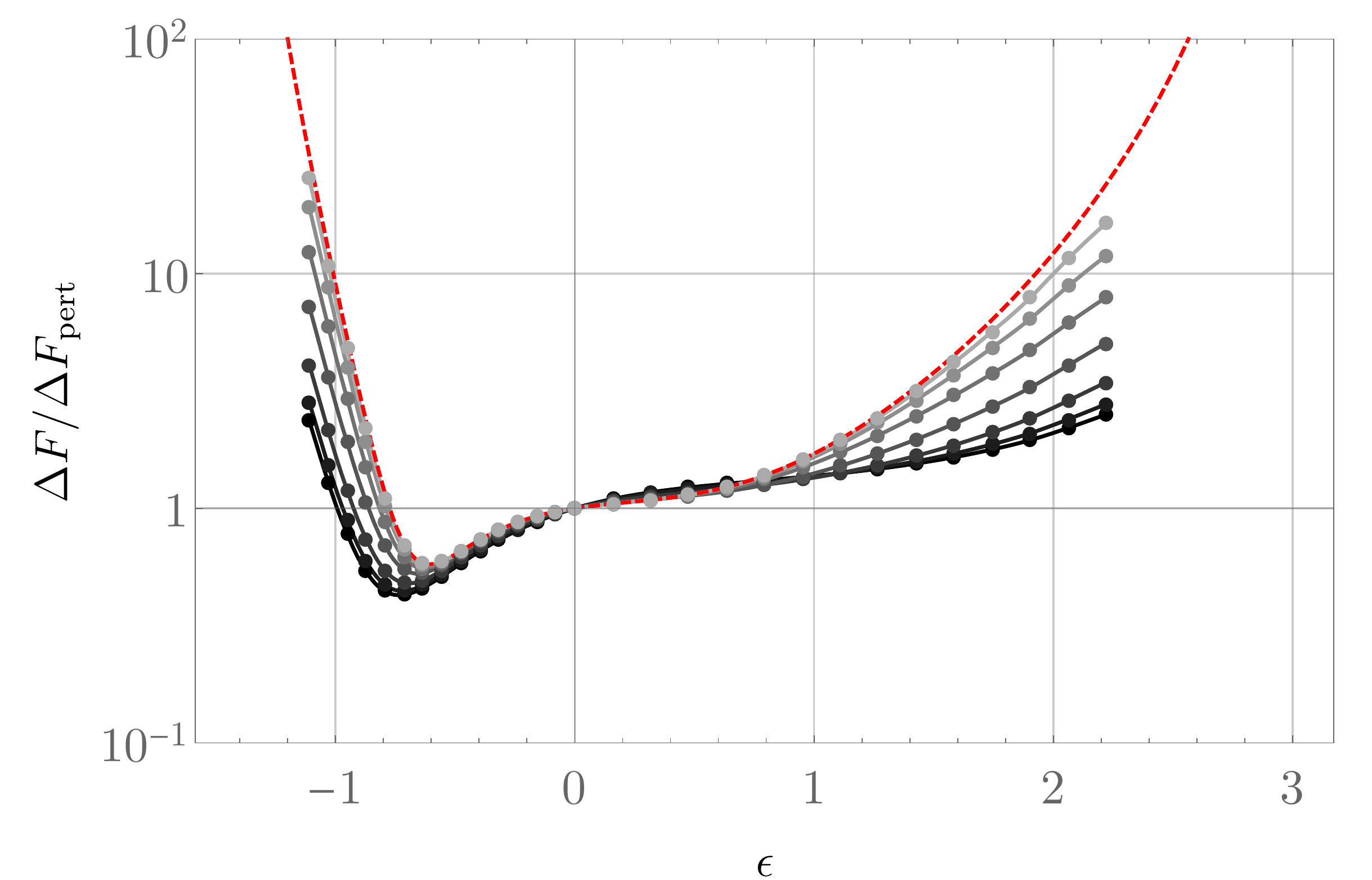}
}
\caption{The ratio~$\Delta F/\Delta F_\mathrm{pert}$ at various temperatures and masses for the Dirac fermion on the deformed spheres given by~\eqref{eq:Rellembedding}, for the representative case~$\ell = 2$.  In both figures, the black curves correspond to the massless, zero-temperature result, while the dashed red line is the long-wavelength behavior given by~\eqref{eq:DeltaFratiohightemp}.  From black to light gray, the left figure shows temperatures~$T = 0$,~$0.5$,~$1$,~$2$,~$4$, and~$8$, while the right figure shows masses~$M = 0$,~$0.5$,~$1$,~$2$,~$4$,~$8$ and~$16$ (points are numerical data; the curves are drawn to guide the eye).  The range of the~$x$-axis goes from~$\eps_\mathrm{min}$ to~$\eps_\mathrm{max}$; the data shown here takes~$|\eps|$ sufficiently small that the numerical error in~$\Delta F$ is no greater than one percent.}
\label{fig:FiniteTemperatureDirac}
\end{figure}

At intermediate masses and temperatures,~$\Delta F$ interpolates between the massless zero-temperature behavior shown in Figure~\ref{fig:DeltaF} and the behavior given by~\eqref{eq:DeltaFhighTleading}.  As a representative example, we show this interpolation in Figure~\ref{fig:FiniteTemperatureDirac} for the case of the fermion and the deformed spheres~\eqref{eq:Rellembedding} with~$\ell = 2$ (results for the scalar and higher~$\ell$ are analogous).  The takeaway is that for any mass and temperature, large deformations of the sphere appear to decrease~$\Delta F$ arbitrarily.  The deformations considered here tend to ``pinch off'' the sphere somewhere, and hence to better understand the behavior of~$\Delta F$ under such extreme deformations, we now examine more closely the behavior of the heat kernel near these transitions.

\section{Towards Singular Geometries}
\label{sec:endstate}

As remarked above, the deformations shown in Figure~\ref{fig:embeddings} fall into roughly two classes: the left two columns (corresponding to odd~$\ell$ and even~$\ell$ with~$\eps < 0$) limit to a connected geometry that ``pinches'' somehwere, while the geometries shown in the right column (corresponding to even~$\ell$ with~$\eps > 0$) tend to disconnect as~$\eps \to \eps_\mathrm{max}$, with the individual connected pieces each potentially having a defect near the transition.  In both classes, we expect the gradient of~$\Delta K_L$ to diverge at~$t = 0$ as~$\eps \to \eps_\mathrm{min,max}$ because the heat kernel coefficient~$\Delta b_4$ diverges as the geometry becomes singular due to the Ricci scalar becoming unbounded near the pinchoff\footnote{This unboundedness of the Ricci scalar renders the asymptotic series~\eqref{eq:DeltaFheatkernel} no longer valid.  The heat kernel may still admit a Frobenius expansion around~$t = 0$ when~$\eps = \eps_\mathrm{min,max}$, but the coefficients in this expansion cannot be given by integrals of successively higher-derivative curvature invariants, because these diverge.}.  However, the behavior of~$\Delta K_L$ at small \textit{nonzero}~$t$ differs between these two classes.  The class with even~$\ell$ and~$\eps > 0$ is perhaps most intuitive: the case~$\ell = 2$ looks like a change in topology from one sphere to two, while for~$\ell \geq 4$ the singular geometry also exhibits conical defects near the transition (in addition to the divergence of the Ricci scalar there).  An isolated conical defect (with no curvature singularity) can be studied analytically, so we begin with a discussion of the associated divergences.

\subsection{Conical Defects}
\label{subsec:cone}

In the vicinity of a conical defect on some manifold~$\Sigma$, the geometry takes the form
\be
\label{eq:ConeMetric}
ds^2 = \left[dr^2 + r^2 d\phi^2\right]\left(1 + \Ocal(r)\right)
\ee
where~$\phi$ has periodicity~$\alpha$ (with~$\alpha = 2\pi$ corresponding to a smooth geometry).  Recall that a conical deficit (corresponding to~$\alpha < 2\pi$) can be embedded in~$\mathbb{R}^3$, while an excess (corresponding to~$\alpha > 2\pi$) cannot.  The differenced free energy, of course, depends only on the intrinsic geometry, so we may still analyze its behavior regardless of the existence of any embedding.  In the presence of such a defect, the corresponding heat kernel expansion exhibits an additional constant term associated to it~\cite{Fursaev1997}:
\be
\label{eq:coneexpansion}
\sigma K_L(t) = \frac{\sigma \mathrm{Vol}[g]}{4\pi t} - \frac{\chi_\Sigma}{12} - \frac{(2\pi-\alpha)^2}{48 \pi \alpha} + \Ocal(t).
\ee
The differenced heat kernel thus satisfies
\be
\label{eq:DeltaKcone}
\sigma \Delta K_{L}(t) = -\frac{(2\pi-\alpha)^2}{48 \pi \alpha} + \Ocal(t),
\ee
which clearly leads to a UV divergence in~$\Delta F$.  Importantly, note that this divergence has fixed sign: it always contributes negatively to~$\sigma \Delta K_L(t)$, and hence to~$\Delta F$.

Interestingly, for a cone (that is, the geometry~\eqref{eq:ConeMetric} with vanishing subleading corrections), the sign of the divergence of the energy depends on whether the defect corresponds to a conical excess or deficit.  For example, in the case of a conformally coupled scalar at zero temperature, the energy density of a cone is~\cite{Dowker1987}
\be
\label{eq:conerho}
\rho \equiv \langle T_{00} \rangle = \frac{G(\alpha)}{r^3},
\ee
where~$G(\alpha) < 0$ for~$\alpha < 2\pi$ and~$G(\alpha) > 0$ for~$\alpha > 2\pi$.  Hence the differenced free energy between a cone and a planar geometry with no conical defect is negatively UV-divergent\footnote{We are ignoring potential IR divergences associated with the fact that a cone is not compact.} when~$\alpha < 2\pi$, and \textit{positively} divergent when~$\alpha > 2\pi$.  One might have na\"ively expected the behavior~\eqref{eq:conerho} to have been universal near conical defects (at least for QFTs with UV fixed points, which are CFTs in the UV), but the heat kernel expansion~\eqref{eq:coneexpansion} shows that the behavior of the stress tensor near such defects must be sensitive to the global properties of~$(\Sigma,g)$ (and in particular, if~$\Sigma$ is compact, it follows from~\eqref{eq:DeltaKcone} that the difference~$\Delta F$ is always negatively UV-divergent, whether the defect is an excess or a deficit).

\begin{figure}[t]
\centering
\includegraphics[width=0.2\textwidth]{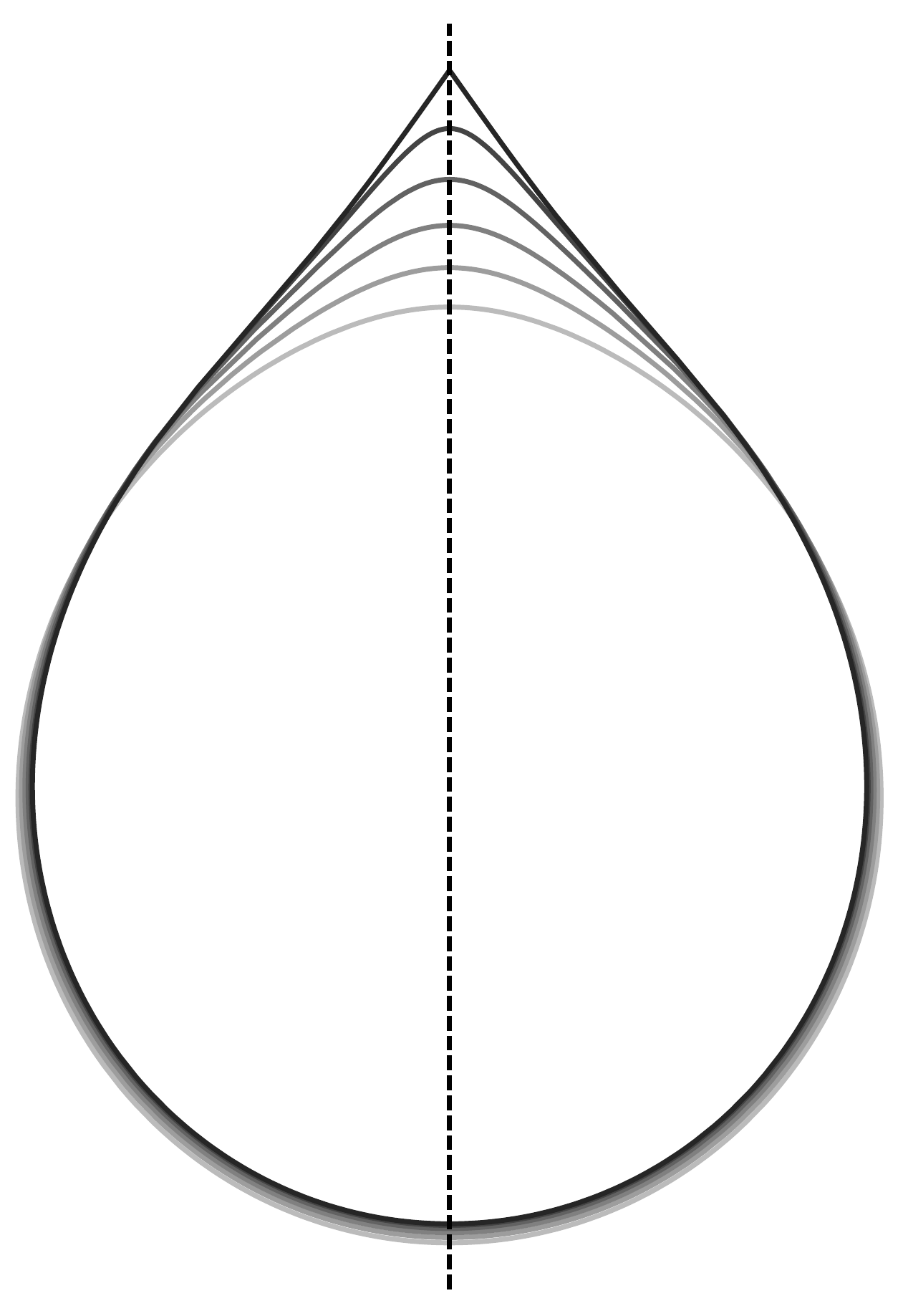}
\caption{Cross-sections of the geometries described by the embedding function~\eqref{eq:Rcone}; these should be rotated around the dotted axis to generate the corresponding deformed spheres.  From gray to black, we show~$\eps = 0$ to~1 in intervals of~$0.2$; the~$\eps = 1$ embedding exhibits a conical defect at the pole.}
\label{fig:conicaldefect}
\end{figure}

\begin{figure}[t]
\centering
\subfloat[][Minimally coupled scalar]{
\includegraphics[width=0.49\textwidth]{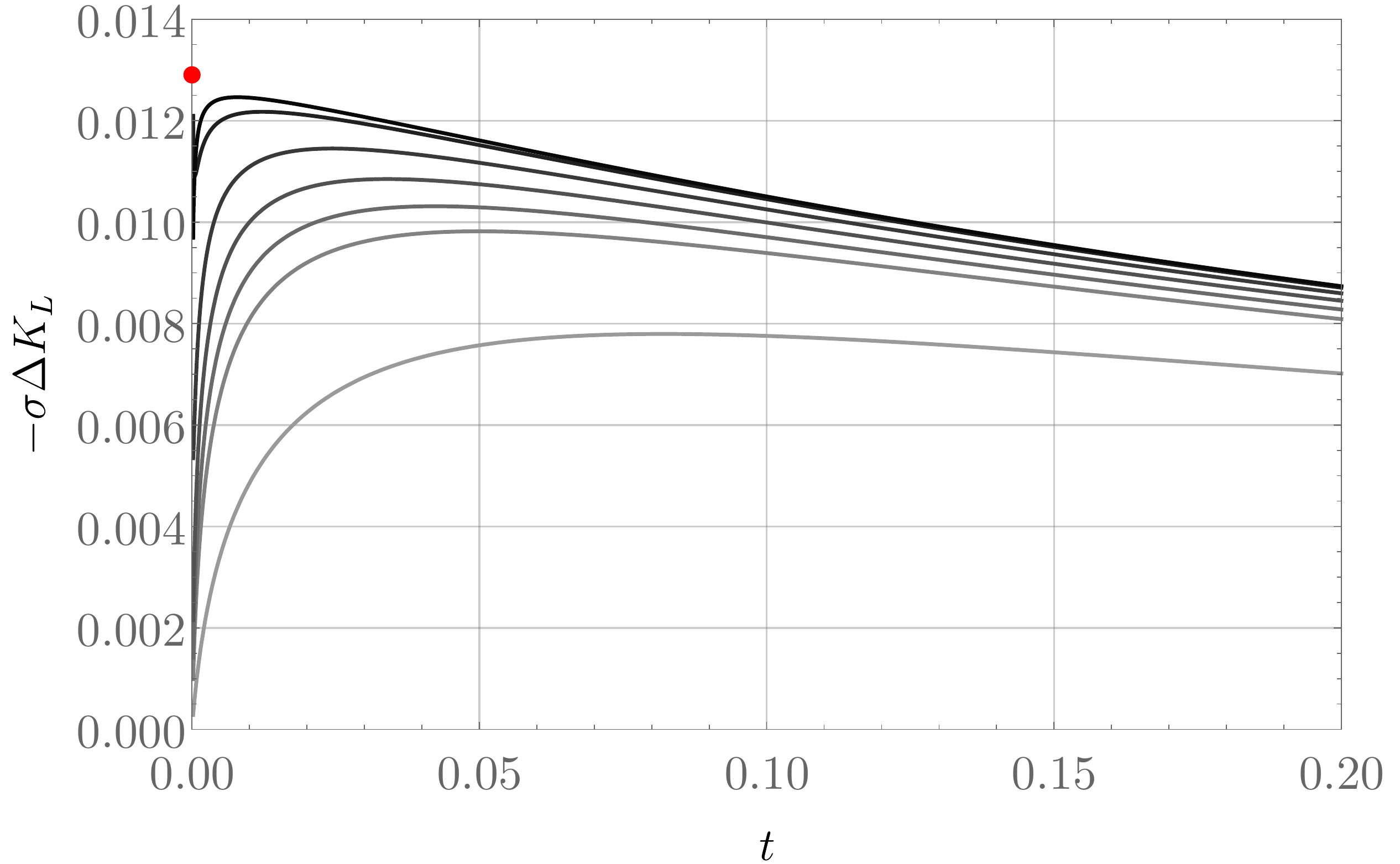}
\label{subfig:ConeKernelsScalar}
}%
\subfloat[][Fermion]{
\includegraphics[width=0.49\textwidth]{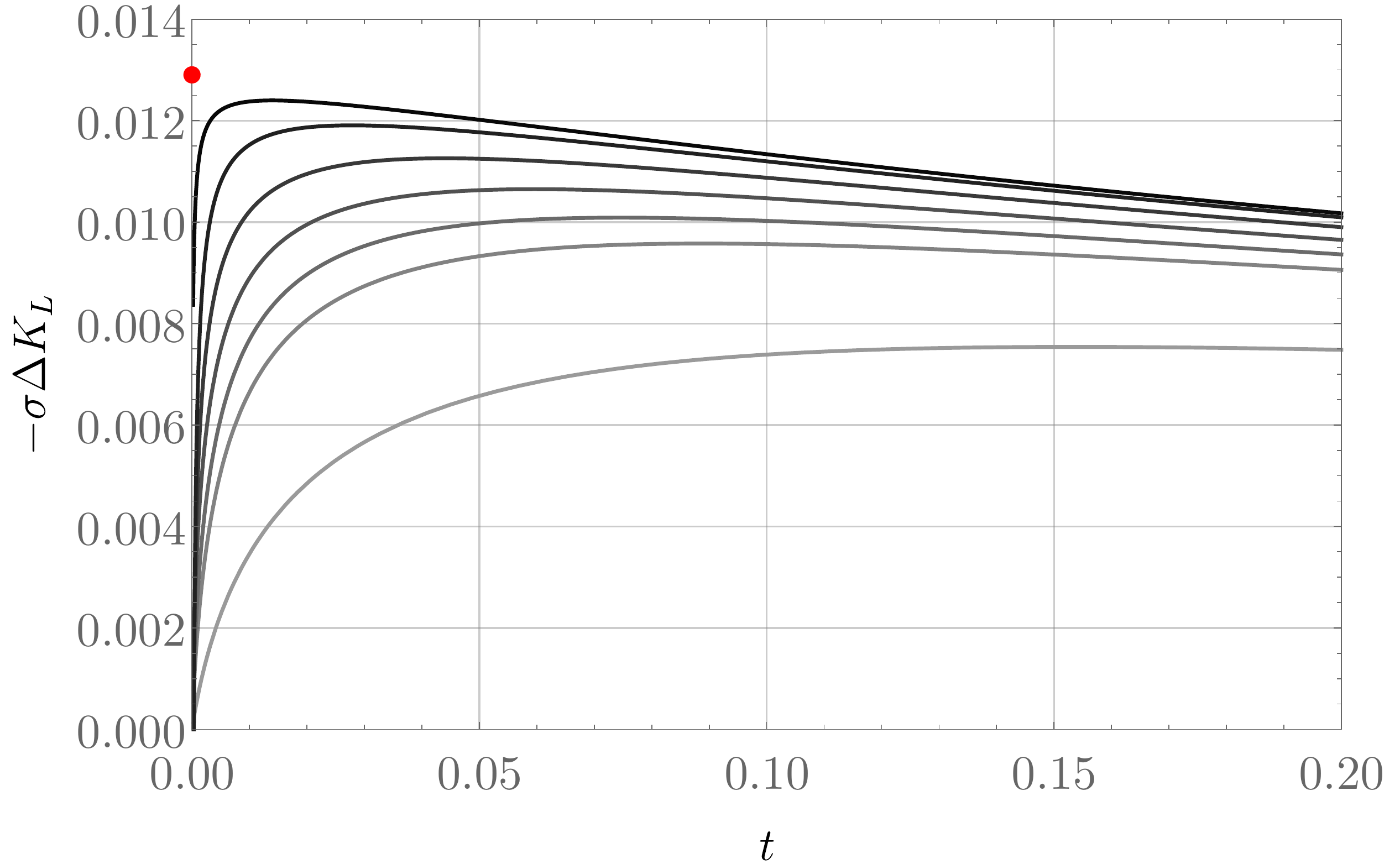}
\label{subfig:ConeKernelsDirac}
}
\caption{The small-$t$ behavior of~$\Delta K_L$ for the minimally coupled scalar (left) and the Dirac fermion (right) on the geometries described by the embedding~\eqref{eq:Rcone}.  From lightest to darkest, the curves correspond to~$\eps = 0.9,0.95,0.96,0.97,0.98,0.99$, and~$0.995$.  The red dot indicates the~$t = 0$ value~\eqref{eq:coneDeltaKlimit} expected on the conical defect geometry corresponding to~$\eps = 1$.}
\label{fig:conedeformations}
\end{figure}

To manifestly illustrate such deformations, as well as to connect to the deformations considered in Section~\ref{sec:nonpert}, consider a one-parameter family of spatial geometries that interpolates from a smooth geometry to one with a conical defect at a pole.  An explicit axisymmetric example of such a family is given by the embedding
\be
\label{eq:Rcone}
R_\eps(\theta) = c_\eps \left(1+ \frac{1}{4} \sqrt{1 + 40 \sin ^2\left(\theta/2\right)}-\frac{1}{4} \sqrt{(1-\eps)^2 + 50\sin ^2\left(\theta/2\right)}\right)^2,
\ee
where~$c_\eps$ is a volume-preserving constant that fixes the volume to~$4\pi$, i.e.~the volume of the round unit sphere.  For any~$\eps < 1$, this geometry is everywhere smooth, and for~$\eps = 1$, it exhibits a conical defect at $\theta = 0$ with  angle~$\alpha = 2\pi/\sqrt{3}$ and a Ricci scalar which is bounded everywhere excluding the defect; see Figure~\ref{fig:conicaldefect}.  For~$\eps < 1$ we may therefore numerically compute the heat kernel as described in the previous section; we show these in Figure~\ref{fig:conedeformations}.  As expected, the differenced heat kernel vanishes linearly at small~$t$ for any~$\eps < 1$, but its gradient there diverges as~$\eps \to 1$.  In the limit~$\eps \to 1$, the heat kernel clearly approaches a function that goes to a \textit{nonzero} value at~$t = 0$ consistent with the expectation from~\eqref{eq:DeltaKcone}:
\be
\label{eq:coneDeltaKlimit}
\lim_{t \to 0^+} \lim_{\eps \to 1^-} \sigma \Delta K_L(t) = -\frac{2\sqrt{3}-3}{36},
\ee
with the right-hand side just the~$\alpha = 2\pi/\sqrt{3}$ case of~\eqref{eq:DeltaKcone}.

\begin{figure}[t]
\centering
\subfloat[][Minimally coupled scalar]{
\includegraphics[width=0.49\textwidth]{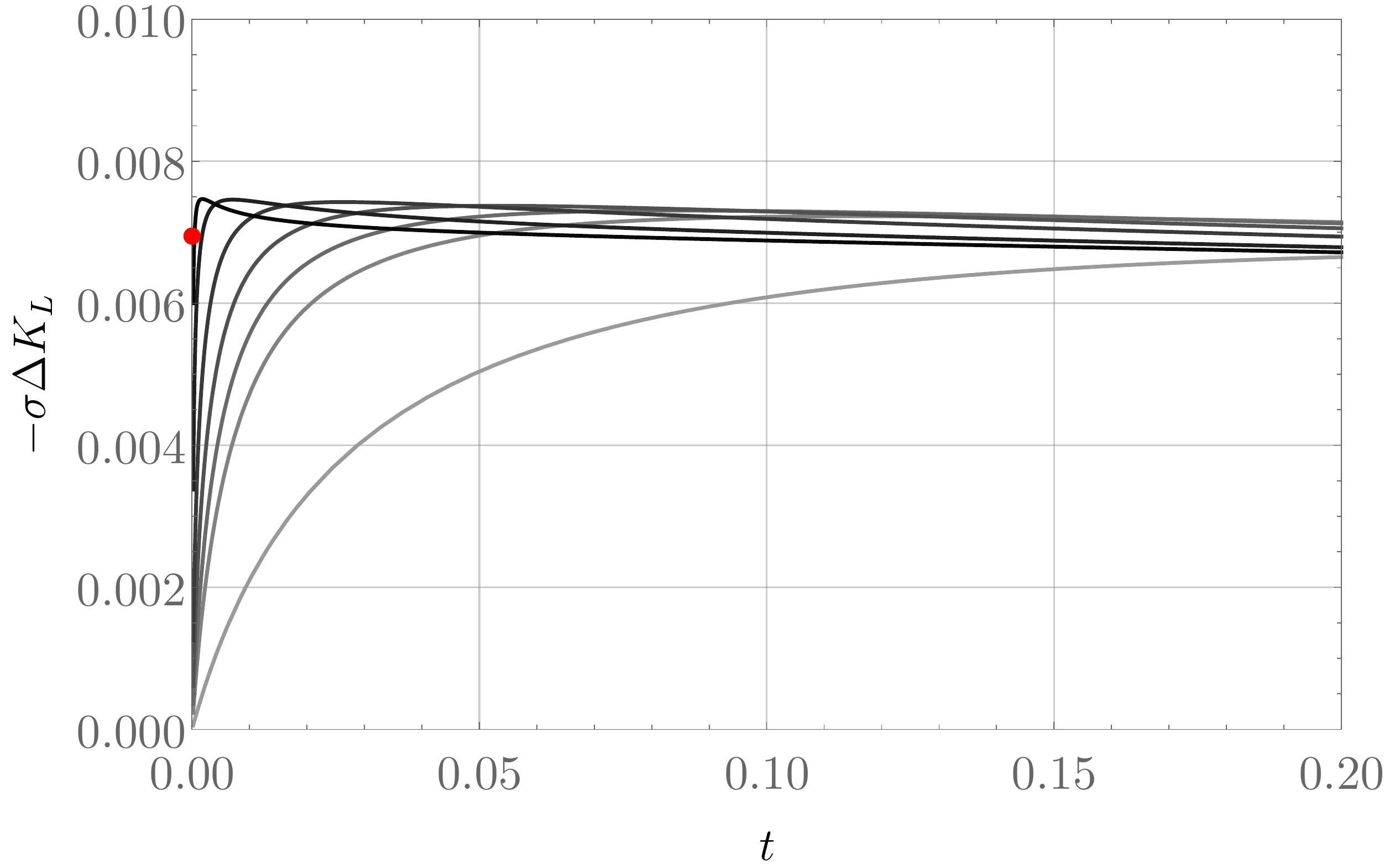}
\label{subfig:ExcessConeKernelsScalar}
}%
\subfloat[][Fermion]{
\includegraphics[width=0.49\textwidth]{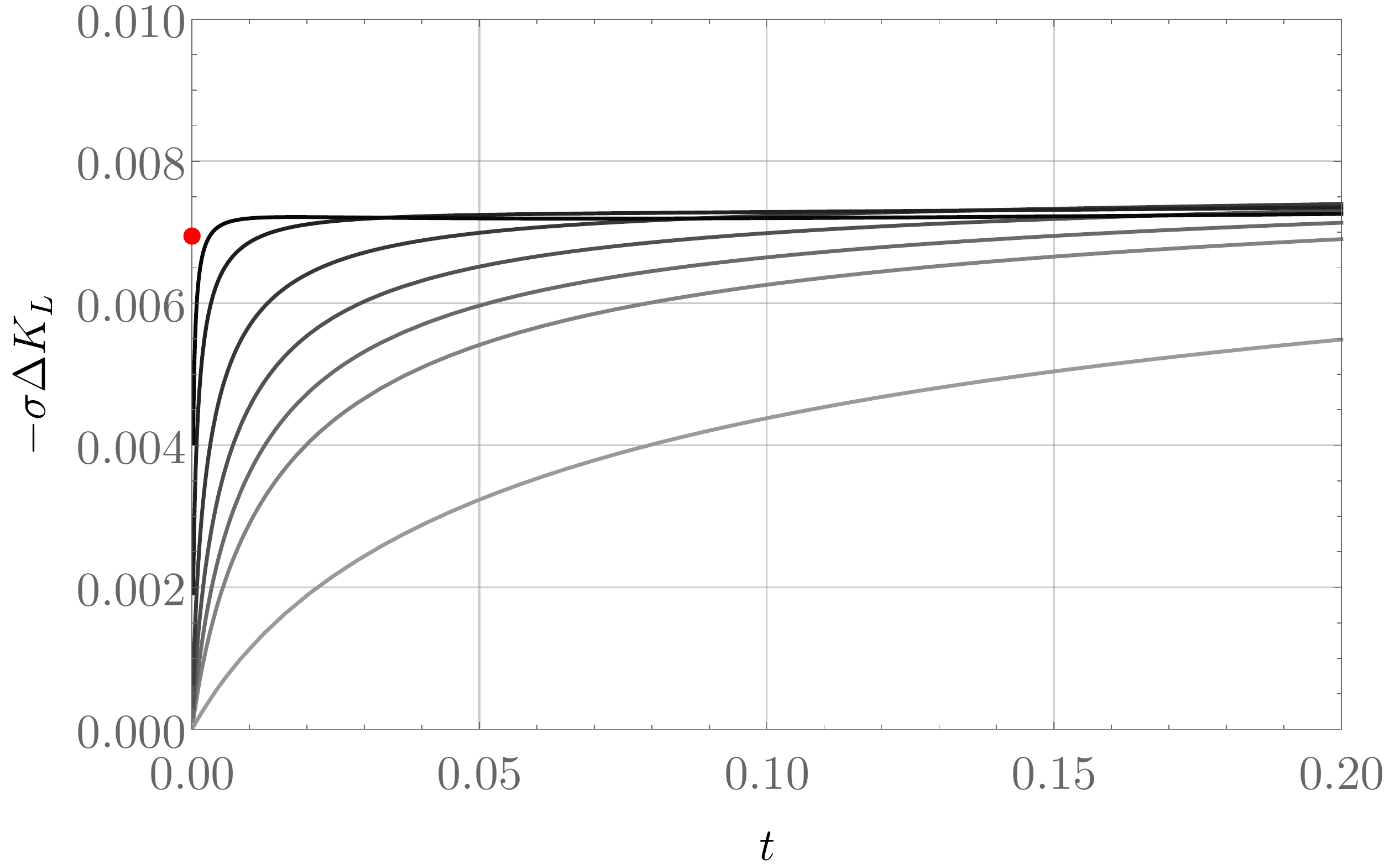}
\label{subfig:ExcessConeKernelsDirac}
}
\caption{The small-$t$ behavior of~$\Delta K_L$ for the minimally coupled scalar (left) and the Dirac fermion (right) on the geometries~\eqref{eq:intrinsiccone} with~$\alpha = 3\pi$.  From lightest to darkest, the curves correspond to~$\eps = 0.9,0.95,0.96,0.97,0.98,0.99$, and~$0.995$.  The red dot indicates the~$t = 0$ value~$-1/144$ expected on the conical defect geometry with~$\eps = 1$.}
\label{fig:excessconedeformations}
\end{figure}

We can investigate a conical excess analogously by specifying the deformed sphere geometry directly rather than considering an embedding.  To that end, consider the family of deformed spheres given by
\be
\label{eq:intrinsiccone}
ds^2 = c_\eps \left( d\theta^2 + e^{2f_\eps} \sin^2\theta\, d\phi^2 \right), \mbox{ with } f_\eps(\theta) = \frac{\ln(\alpha/2\pi)}{\sec^2(\theta/2) + (1-\eps)^2 \csc^2(\theta/2)} ,
\ee
where~$c_\eps$ is again a volume-preserving constant. For~$\eps < 1$, these geometries are smooth, while the~$\eps = 1$ geometry exhibits a conical defect of angle~$\alpha$ at the pole~$\theta = 0$ and a Ricci scalar which is bounded everywhere excluding this pole.  The small-$t$ behavior of the differenced heat kernels for~$\alpha = 3\pi$ is shown in Figure~\ref{fig:excessconedeformations}; note that in the limit~$\eps \to 1$, these too approach the~$t = 0$ value expected from~\eqref{eq:DeltaKcone}.  Morevoer, one again finds~$\sigma \Delta K_L$ appears positive for all~$t$.  In particular, these results confirm that on a topological sphere with a conical defect, both a deficit \textit{and} an excess contribute \textit{negatively} to the free energy, in constract with the expectation from~\eqref{eq:conerho} for planar geometries.
 
\subsection{Even~$\ell$, $\eps > 0$}
\label{subsec:evenellposeps}

\begin{figure}[t]
\centering
\centering
\subfloat[][Minimally coupled scalar]{
\includegraphics[width=0.49\textwidth]{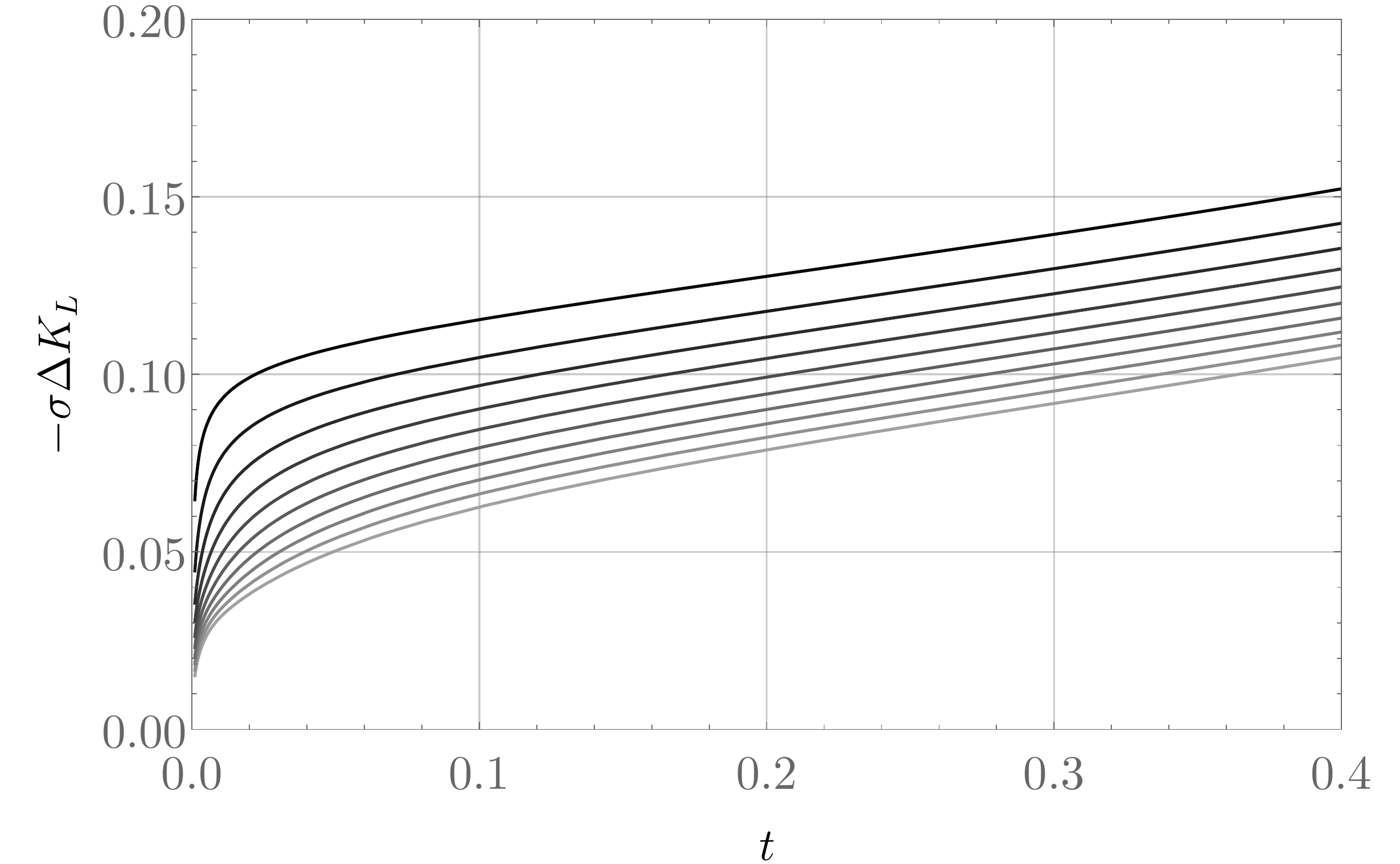}
\label{subfig:evenellsingscalar}
}%
\subfloat[][Fermion]{
\includegraphics[width=0.49\textwidth]{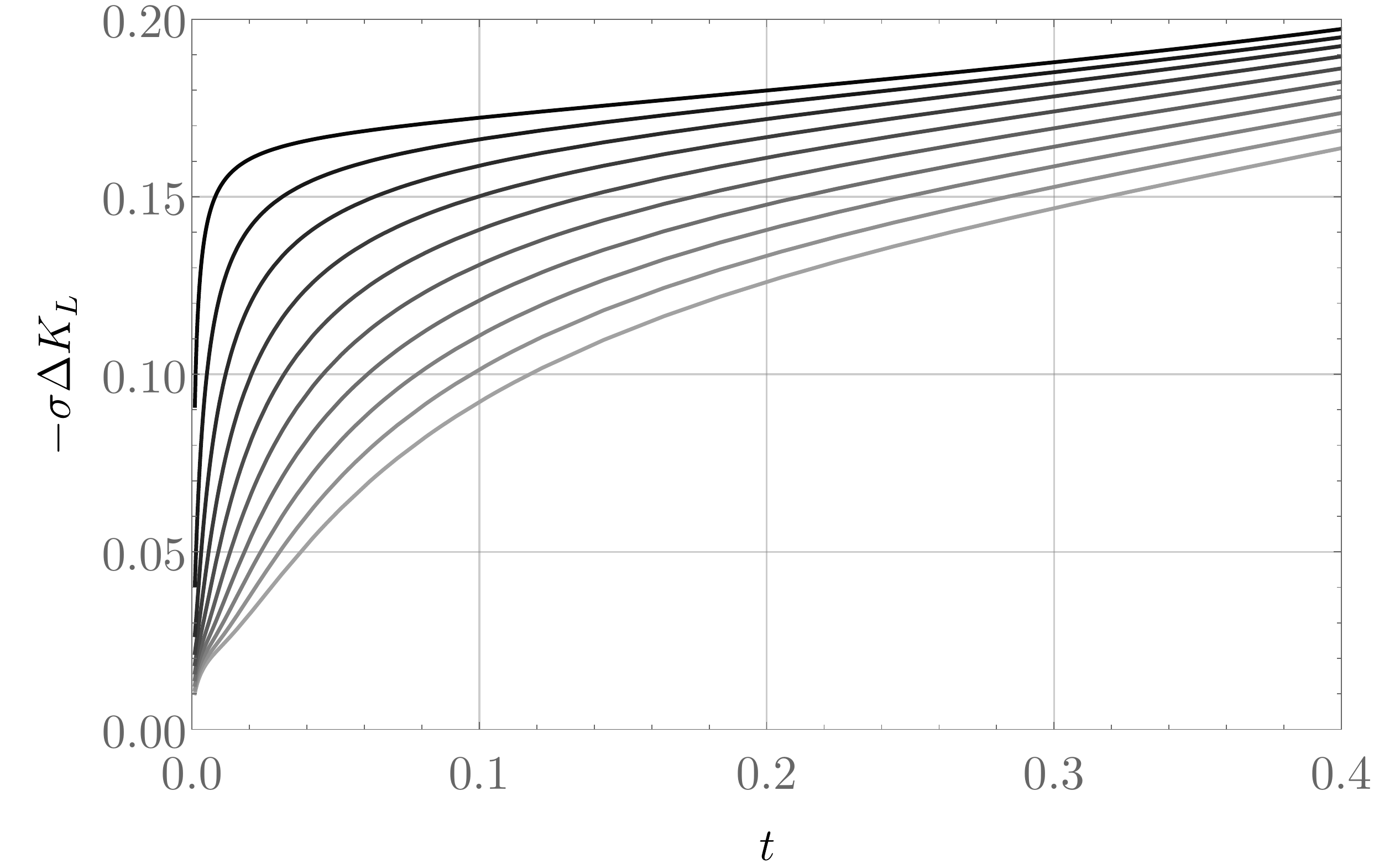}
\label{subfig:evenellsingdirac}
}
\caption{The behavior of~$\Delta K_L$ at small~$t$ for~$\ell = 2$ and~$\eps > 0$ for the minimally coupled scalar (left) and the Dirac fermion (right).  From lightest to darkest, the curves correspond to~$\eps/\eps_\mathrm{max} = 0.9$ to~$0.99$ in intervals of~$0.01$.}
\label{fig:evenellsing}
\end{figure}

Let us now return to the case of the deformed spheres shown in the right-hand column of Figure~\ref{fig:embeddings}.  As a representative example, in Figure~\ref{fig:evenellsing} we show the small-$t$ behavior of~$\Delta K$ for the~$\ell = 2$ deformation~\eqref{eq:Rellembedding} near~$\eps = \eps_\mathrm{max}$.  As expected, the heat kernel always vanishes linearly at~$t = 0$ for any~$\eps < \eps_\mathrm{max}$ but its gradient there diverges as~$\eps \to \eps_\mathrm{max}$.  More interestingly,~$\Delta K$ appears to stabilize to a function that approaches a finite nonzero value at~$t = 0$.  This behavior is quite evident in the case of the fermion, though it is a bit less obvious for the scalar as the successive change in~$\Delta K_L$ appears to grow with each successive step in~$\eps$.

We might hope to understand this behavior by using the heat kernel expansion, but as mentioned above, for~$\eps = \eps_\mathrm{max}$ this expansion breaks down due to the unbounded Ricci scalar near the pinchoff point.  We therefore should not expect the expansion~\eqref{eq:coneexpansion} to capture quantitative details of the small-$t$ behavior near the transition.  However, it is interesting to note that it does capture some qualitative features; for instance, the differenced heat kernel appears to be approaching a function that limits to a nonzero constant at~$t = 0$, similarly to the hear kernels shown in Figures~\ref{fig:conedeformations} and~\ref{fig:excessconedeformations}.  Moreover, note that the~$\ell = 2$,~$\eps = \eps_\mathrm{max}$ geometry does not have a conical defect and can be thought of as a transition from one to two topological spheres.  This transition doubles the Euler characteristic from~$\chi = 2$ to~$\chi = 4$, which from the heat kernel expansion would correspond to a differenced heat kernel of
\be
\sigma \Delta K_L(t) = -\frac{1}{6} + \Ocal(t);
\ee
this limiting value of~$-1/6$ is surprisingly very close to the limiting behavior for the fermion shown in Figure~\ref{subfig:evenellsingdirac}, even though a priori the heat kernel expansion should not be applicable (the scalar heat kernel in Figure~\ref{subfig:evenellsingscalar}, on the other hand, does not appear to approach this limiting value of~$-1/6$, though this is more difficult to verify conclusively because the scalar heat kernel does not appear to be growing linearly in~$\eps$ near~$\eps = \eps_\mathrm{max}$).

\subsection{Odd $\ell$ and even $\ell$, $\eps < 0$}
\label{subsec:oddell}

\begin{figure}[t]
\centering
\centering
\subfloat[][Minimally coupled scalar]{
\includegraphics[width=0.45\textwidth]{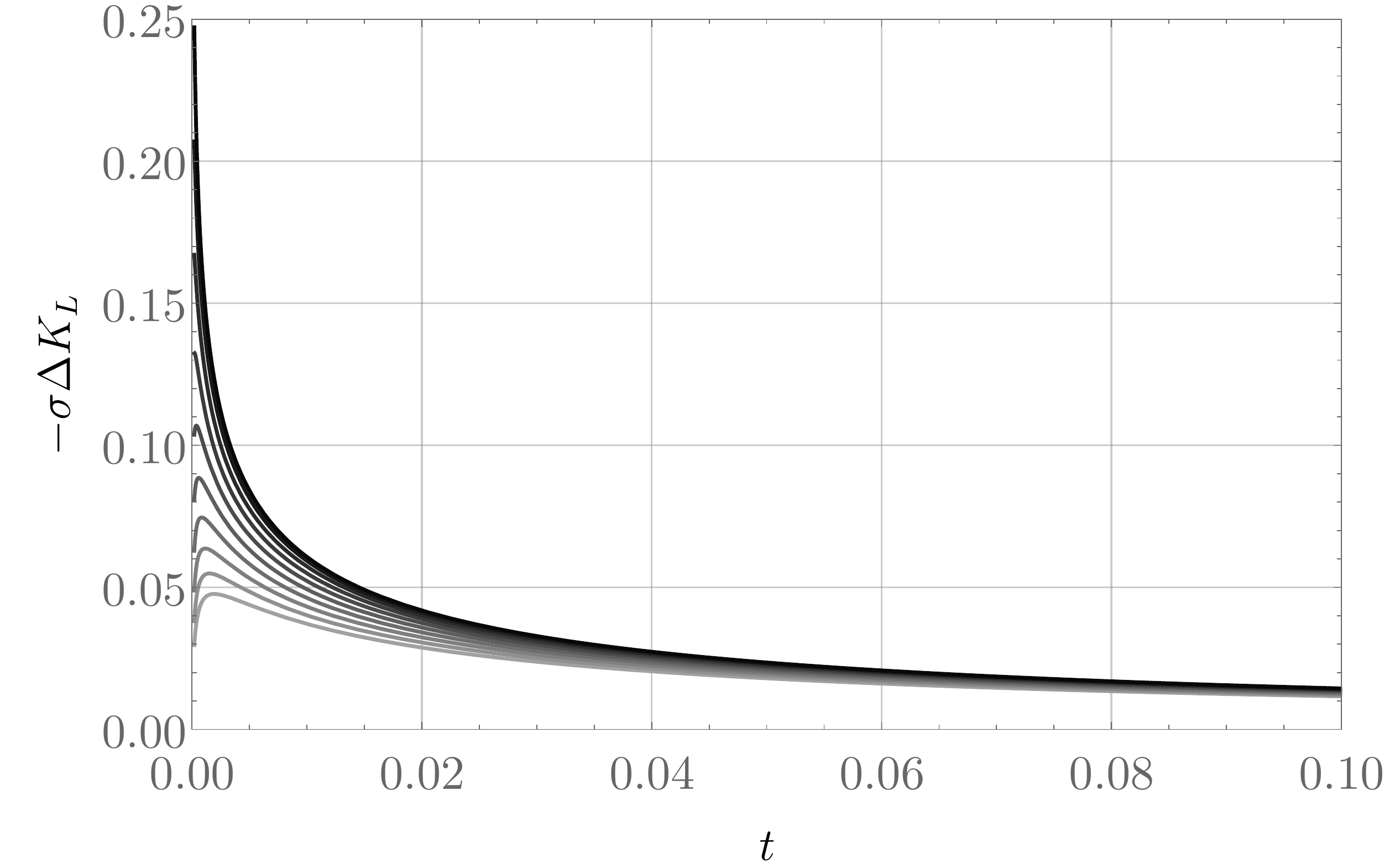}
\label{subfig:oddellDeltaE}
}
\hspace{0.5cm}
\subfloat[][Fermion]{
\includegraphics[width=0.45\textwidth]{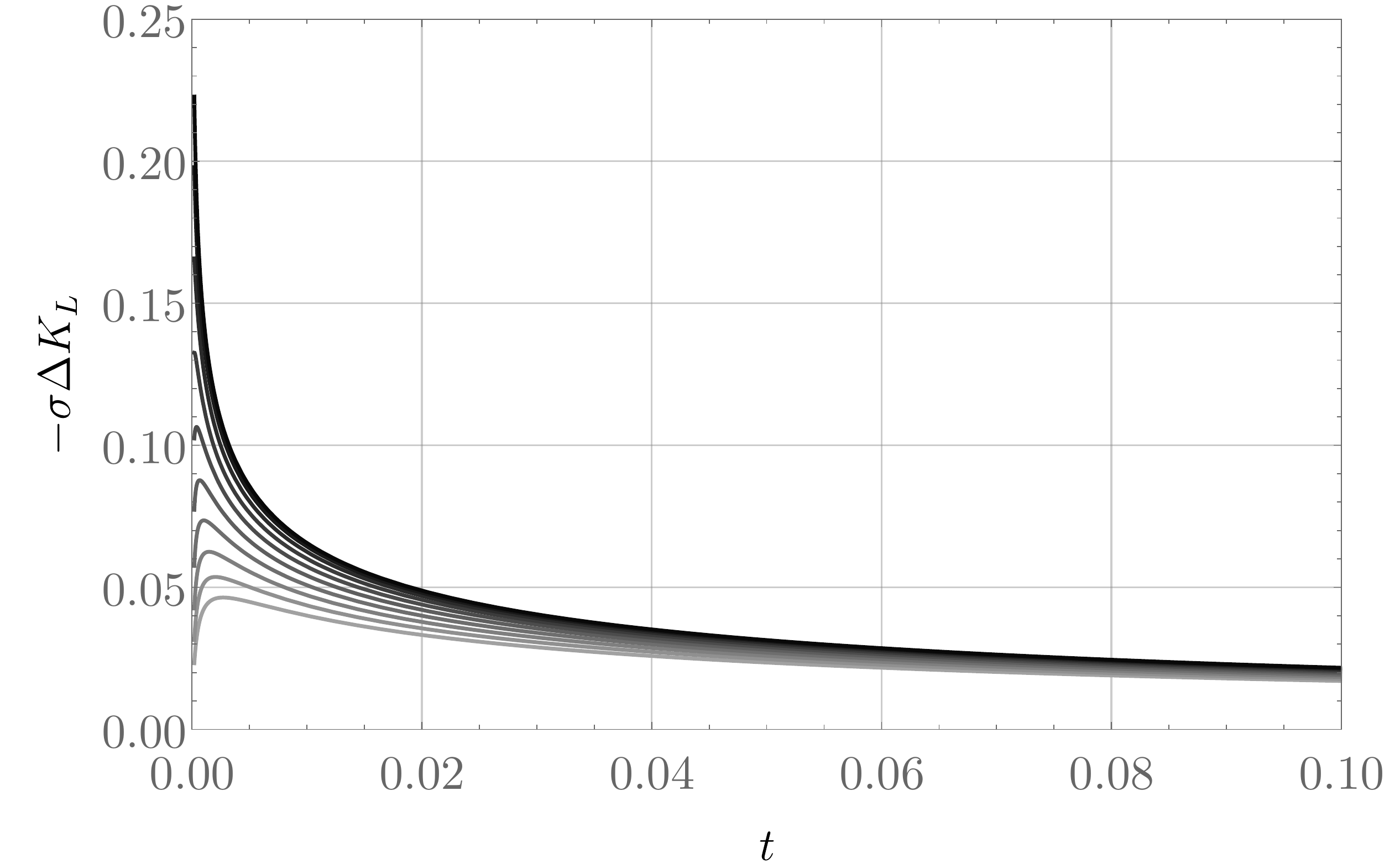}
\label{subfig:evenellDeltaE}
}
\caption{The behavior of~$\Delta K_L$ at small~$t$ for~$\ell = 2$ and~$\eps < 0$ for the minimally coupled scalar (left) and the Dirac fermion (right).  From lightest to darkest, the curves correspond to~$\eps/\eps_\mathrm{min} = 0.9$ to~$0.99$ in intervals of~$0.01$.}
\label{fig:oddellsing}
\end{figure}

\begin{figure}[t]
\centering
\centering
\subfloat[][Minimally coupled scalar]{
\includegraphics[width=0.45\textwidth]{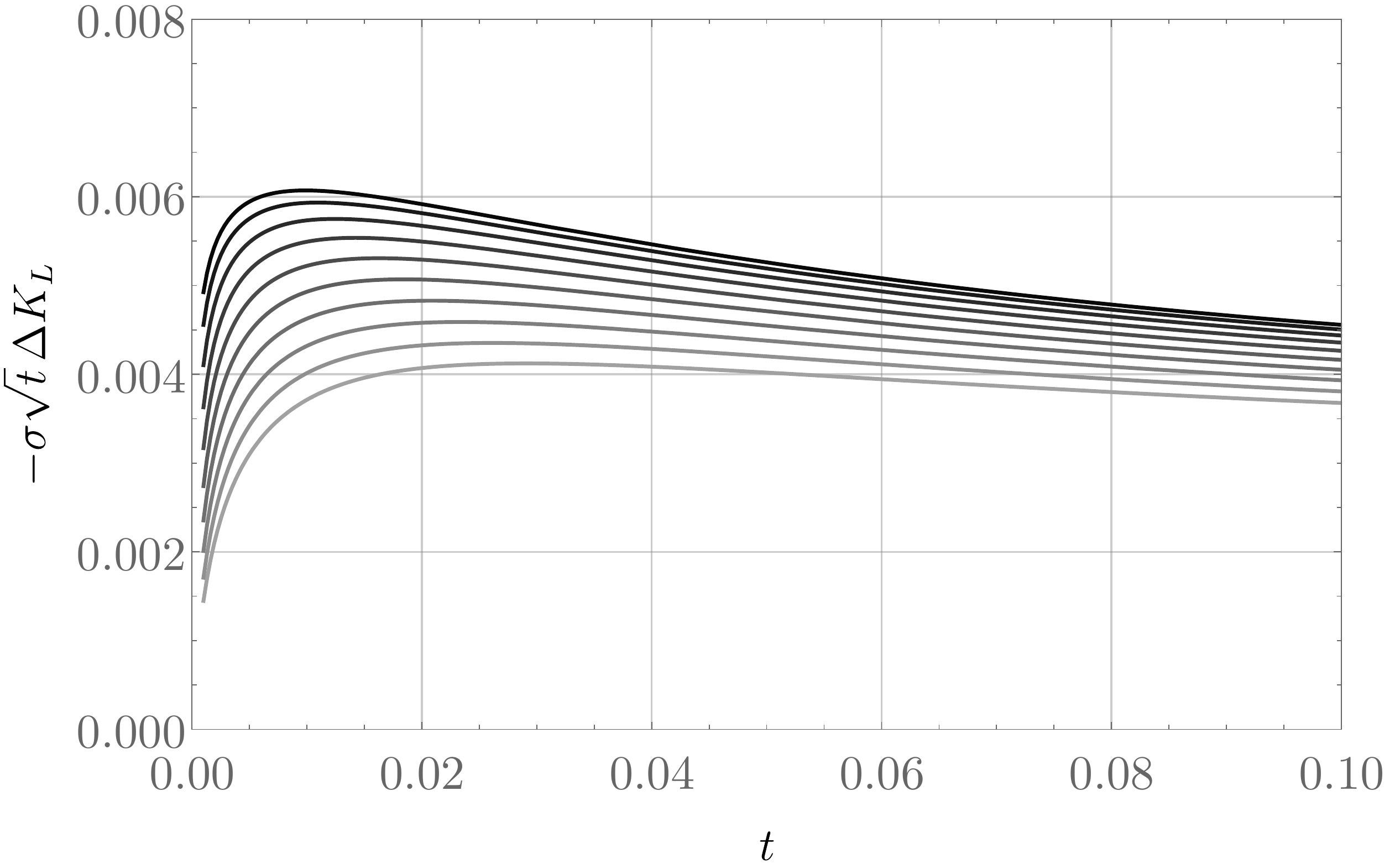}
\label{subfig:oddellDeltaE}
}
\hspace{0.5cm}
\subfloat[][Fermion]{
\includegraphics[width=0.45\textwidth]{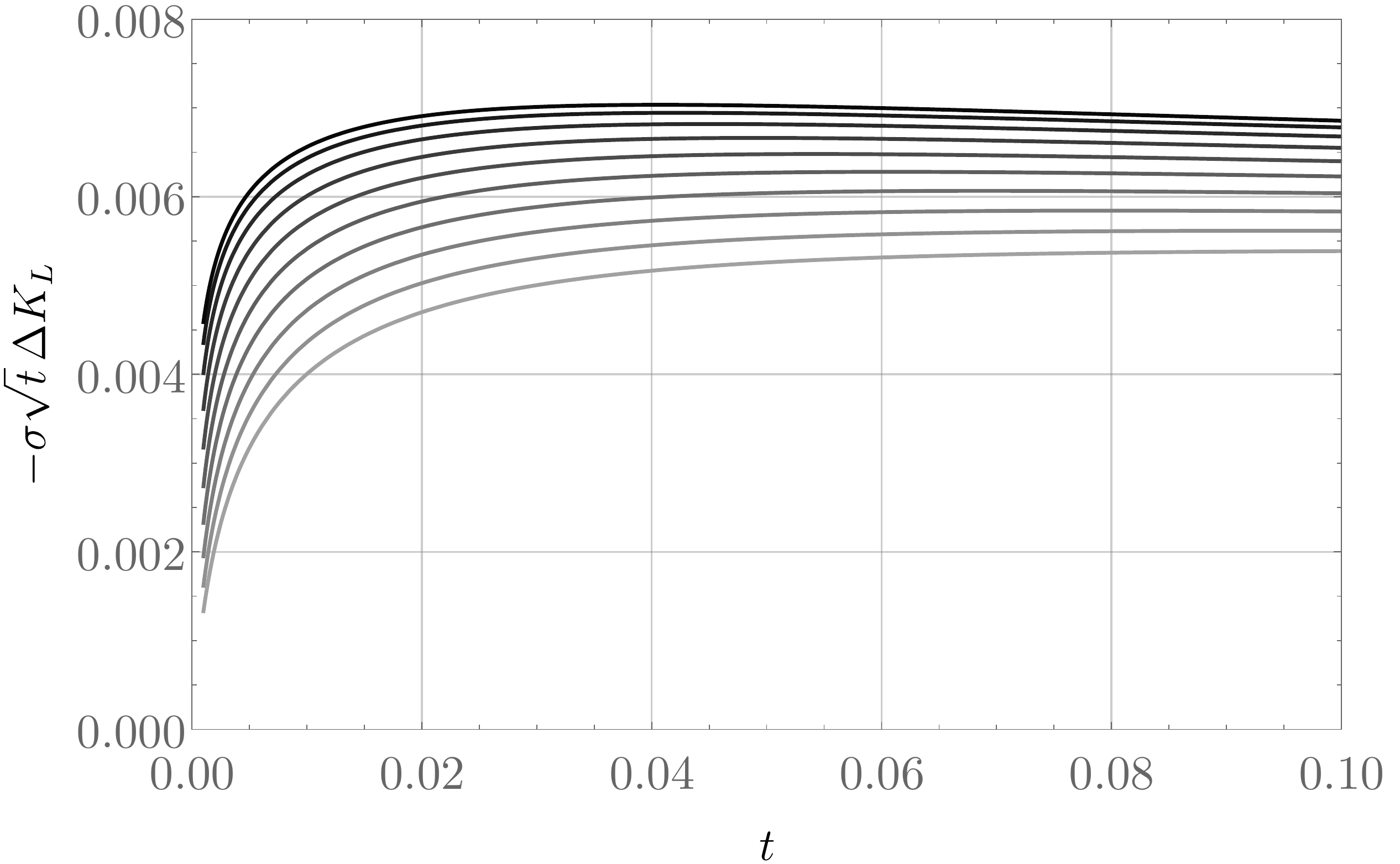}
\label{subfig:evenellDeltaE}
}
\caption{The same as Figure~\ref{fig:oddellsing}, normalized by a factor of~$\sqrt{t}$.  At small-but-not-too-small~$t$,~$\Delta K_L$ appears to go like~$t^{-1/2}$.  It is unclear what happens at much smaller~$t$ due to lack of numerical precision there.}
\label{fig:oddellsingsquareroot}
\end{figure}

For odd~$\ell$ and even~$\ell$ with~$\eps < 0$, the limiting geometry instead has a cusp.  The corresponding behavior of the differenced heat kernels (for~$\ell = 2$) is shown in Figure~\ref{fig:oddellsing}; note that now the heat kernel itself, rather than just its gradient, appears to grow at small~$t$ as the geometry becomes singular.  As shown in Figure~\ref{fig:oddellsingsquareroot}, at intermediate values of~$t$ this growth appears to go roughly like~$t^{-1/2}$, but does not appear to be maintained to arbitrarily small~$t$.  Indeed, the difficulty in inferring the limiting small-$t$ behavior is presumably due to the breakdown of the heat kernel expansion in the singular limit -- that is, it is unclear whether or not~$\sqrt{t} \, \Delta K_L$ vanishes at~$t = 0$ in the singular limit, and therefore whether~$\Delta K_L$ actually approaches a finite nonzero constant at~$t= 0$ like it does for the cone or whether~$\Delta K_L$ genuinely diverges there.  Nevertheless, we note that the scaling as~$t^{-1/2}$ is interesting as such a scaling of the heat kernel is expected on manifolds with boundary~\cite{Vas03}.  This behavior suggests that perhaps the cusp can be interpreted as a sort of boundary.

\subsection{Implications for Graphene-Like Materials}
\label{subsec:grapheneendstate}

The fact that~$\Delta K_L(t)$ approaches a nonzero constant at small~$t$ could have interesting consequences for materials like graphene, which as discussed in Section~\ref{sec:intro} may exhibit a competition between a classical membrane free energy~$\Delta F_c$ and the contribution~$\Delta F_q$ from effective QFT degrees of freedom.  Indeed, note that~\eqref{eq:DeltaFheatkernel} implies that on a geometry with a conical defect,~$\Delta F_q$ has a linear UV divergence:
\be
\label{eq:DeltaFqcone}
\Delta F_q = - B \rho^{-1} + \Ocal(\rho^0),
\ee
where~$\rho$ is a short-distance cutoff that resolves the conical singularity (imposed by restricting to~$t > \rho^2$) and~$B$ is a positive constant.  On the other hand, the Landau free energy~$\Delta F_c$ given in~\eqref{eq:classicalfreeenergy} merely has a logarithmic divergence, which is due to its scale-invariance: near~$\theta = 0$ the mean curvature of the embedding~\eqref{eq:Rcone} with~$\eps = 1$ diverges as~$\theta^{-1}$, and hence
\be
\label{eq:DeltaFccone}
\Delta F_c = \widetilde{B} \ln \rho^{-1} + \Ocal(\rho^0),
\ee
with~$\widetilde{B}$ a positive constant.  Interpreting~$(1-\eps) \sim \rho$ as the resolution parameter of the cone, it therefore follows that~$\Delta F_c$ must grow more slowly with~$\eps$ than~$\Delta F_q$, and hence a deformation with~$\eps$ sufficiently close to~1 will have~$\Delta F_c + \Delta F_q < 0$.  This argument will fail, of course, once~$\eps$ is so close to~1 that UV effects from the ``tip'' of the cone -- which presumably are how the divergences in~\eqref{eq:DeltaFqcone} and~\eqref{eq:DeltaFccone} are resolved -- change the relative growth of~$\Delta F_c$ and~$\Delta F_q$ with~$\eps$.

On the other hand, per the analysis of Section~\ref{subsec:graphene}, for graphene we have that at small~$\eps$,~$\Delta F_c + \Delta F_q > 0$.  So although competition between~$\Delta F_c$ and~$\Delta F_q$ does not render the round sphere locally unstable, it appears that sufficiently large deformations of the round sphere may be preferred to the round sphere itself, even after accounting for the Landau free energy of the membrane.  Whether this is actually the case will depend on the details of when our analysis breaks down.

\section{Conclusion}
\label{sec:conc}

We have provided evidence that for a (minimally or nonminimally coupled) free scalar field and for the Dirac fermion living on~$\mathbb{R} \times \Sigma$, with~$\Sigma$ a two-dimensional manifold with sphere topology and endowed with metric~$g$, the free energy is \textit{maximized} when~$g$ is the metric of the round sphere.  This observation applies to any mass and at any temperature.  We demonstrated this result perturbatively around the round sphere for \textit{any} nontrivial perturbation to the geometry, while for nonperturbative deformations we focused on a class of axisymmetric deformations.  We found, in fact, not just that the free energy difference~$\Delta F$ is negative, but that the (differenced) heat kernel~$\Delta K_L(t)$ itself has fixed sign -- a much stronger result than merely negativity of~$\Delta F$.  We have also shown that the free energy difference~$\Delta F$ between an arbitrary~$g$ and the round sphere metric is unbounded below, diverging as~$g$ develops a conical defect.  This property implies that any dynamics of the membrane driven by this free energy will tend to drive the membrane to a singular geometry (which presumably gets regulated by UV effects).

As an application of our results, we have also briefly investigated their relevance to~$(2+1)$-dimensional crystalline systems like graphene, in which there are several contributions to the free energy.  Specifically, in a Born-Oppenheimer approximation, the free energy we have calculated is that of the low-energy effective field theory of \textit{quantum} excitations propagating on a fixed background determined by the atomic lattice; it simply corresponds to studying QFT on a curved background.  The contribution to the free energy from this background is governed by a \textit{classical} Landau free energy, and in this approximation it is the sum of these two that gives the total free energy of the membrane configuration.  Consequently, understanding whether the negative free energy of QFTs on the membrane is sufficient to render the geometry singular depends on how well this quantum effect can compete with the Landau free energy of the underlying lattice.  In the case of graphene at sufficiently low temperatures, we performed an analysis of this competition and found that with the simplifying assumption of a diffeomorphism invariant Landau free energy, the membrane energy dominates for small perturbations of the round sphere, rendering the round sphere stable.  However, in Section~\ref{subsec:graphene} we provided a more general diagnostic~\eqref{eq:gammaconstraint} for when the QFT free energy can make the round sphere unstable.  This constraint depends on parameters like the bending rigidity, lattice spacing, and details of the effective QFT fields living on the membrane.  Since these parameters are presumably experimentally tunable from one type of crystalline membrane to another, it is plausible that a system can be engineered in which the QFT free energy \textit{does} dominate that of the Landau free energy, and hence one should be able to experimentally observe the preference of such a membrane to deform to a singular effective geometry.  Even without such engineering, the fact that integrating out the QFT fields gives a \textit{non-local} contribution to the free energy (as opposed to the classical contribution from the geometric membrane action, which is local) suggests that perhaps the QFT contribution could be experimentally teased out from the classical piece, even if it never actually dominates the free energy.

More importantly, for \textit{large} deformation of the sphere, we have shown that both~$\Delta F_q$ and~$\Delta F_c$ can be made arbitrarily large as the geometry becomes singular, with~$\Delta F_q$ negative and growing faster than~$\Delta F_c$.  Hence it is conceivable that even if the round sphere is \textit{locally} stable, it is not \textit{globally} stable, and large deformations are preferred.  Verifying whether this is indeed the case requires understanding, for instance, how large~$\eps$ must be in the one-parameter family~\eqref{eq:Rcone} before our analysis breaks down due to UV effects.

Regardless of any competition with a Landau free energy, the results we have presented here prompt several questions, a few of which we would like to highlight.  First, does the differenced heat kernel of \textit{any} free field theory on a deformed sphere always have fixed sign?  This is essentially a purely geometric inquiry, as one can define a heat kernel associated to \textit{any} elliptic differential operator~$L$.  Presumably arguments like the ones we used in Section~\ref{subsec:asymptotics} to show that~$\sigma \Delta K_L$ is negative at sufficiently small and large~$t$ could be used to gain control over the asymptotics of~$\Delta K_L$ for general~$L$, but we do not know how to extend that analysis to intermediate values of~$t$ even in the case of the Dirac fermion and scalar studied here (though we note that renormalized determinants of free field operators are rigorously studied in mathematics~\cite{Okikiolu}).
Nevertheless, we conjecture that (under the condition that the spectrum of~$L$ is positive, to ensure that the free field theory defined by~$L$ is stable) the differenced heat kernel does indeed always have a fixed sign.  Second, is the differenced free energy~$\Delta F$ of \textit{any} unitary, relativistic QFT on a deformed sphere negative?  Note that here we include interacting field theories, for which a heat kernel cannot be defined.  As shown in Table~\ref{tab:results}, this differenced free energy has been shown to be negative for holographic CFTs and (perturbatively) for general CFTs, which is tantalizing evidence that perhaps it is some universal feature of general QFTs.  The generality of this picture suggests that there must be a universal underlying mechanism; it would be extremely interesting to uncover what this mechanism must be.  A third, related, question concerns the quantitative behavior of~$\Delta F$: what underlying mechanism is responsible for the similarity of~$\Delta F$ between the minimally-coupled scalar and the Dirac fermion at small deformation parameter?  As discussed in Section~\ref{subsec:freeenergynonpert}, this mechanism can be understood well in a long-wavelength limit, but we do not yet understand why the zero-temperature curves shown in Figure~\ref{fig:DeltaF} are remarkably similar. Finally, how feasible would it be to engineer materials that actually exhibit this negative~$\Delta F$, either as a genuine instability of the round sphere, or merely as a non-local contribution to an effective description of a membrane's equilibrium dynamics?

\section*{Acknowledgements}

SF acknowledges the support of the Natural Sciences and Engineering Research Council of Canada (NSERC), funding reference number SAPIN/00032-2015; of a grant from the Simons Foundation (385602, AM); and of the hospitality of the KITP, where this work was completed.  LW is supported by the STFC DTP research studentship grant ST/R504816/1.  TW is supported by the STFC grant ST/P000762/1.  This research was supported in part by the National Science Foundation under Grant No.~NSF PHY-1748958.

\appendix

\section{Details on the Perturbative Results}
\label{app:perttheory}

In this Appendix we present the details on the perturbative calculation of the heat kernel for the scalar and the fermion.

\subsection{Spin-Weighted Spherical Harmonics}

In what follows, we will make use of the spin-weighted spherical harmonics~$_s Y_{\ell,m}$.  We refer to the original papers~\cite{NewPen66,GolMac67} for more details and explicit formulae; here we merely list the properties of these functions needed to make this Appendix self-contained.  Essentially, a function~$\eta$ associated to a tensorial structure on the sphere is said to have spin weight~$s$ if under a local rotation of orthonormal frame by angle~$\psi$,~$\eta$ transforms like~$\eta \to e^{is\psi} \eta$.  Scalar fields of course are not tensorial and have spin weight zero, while the components of the Dirac spinor have spin weight~$1/2$.  The spin-weighted spherical harmonics~$_s Y_{\ell,m}$ constitute an orthonormal basis for the space of spin weight-$s$ functions on the sphere (hence the usual spherical harmonics are just the special case of spin weight zero:~$Y_{\ell,m} = \, _0 Y_{\ell,m}$).  The index~$\ell$ takes values~$\ell \in \{s, s+1, s+2,\ldots \}$ with~$s$ a non-negative integer or half-integer, and~$m \in \{-\ell,-\ell+1, \ldots, \ell\}$.  We also note that they obey~$_s Y_{\ell,m}^* = (-1)^{s+m} \, _{-s} Y_{\ell,-m}$ as well as the addition theorem
\be
\label{eq:additiontheorem}
\sum_{m = -\ell}^\ell \, _s Y^*_{\ell,m} \, _s Y_{\ell,m} = \frac{2\ell+1}{4\pi}.
\ee

It will be convenient for later to introduce the spin weight raising and lowering operators~$\edth$ and~$\bar{\edth}$, which act on a function~$\eta$ with spin weight~$s$ as
\begin{subequations}
\label{eqs:edth}
\begin{align}
\edth \eta &= -\sin^s\theta \left(\partial_\theta + i \csc\theta \, \partial_\phi\right) \left(\sin^{-s} \theta \, \eta \right), \\
\bar{\edth} \eta &= -\sin^{-s}\theta \left(\partial_\theta - i \csc\theta \, \partial_\phi\right) \left(\sin^s \theta \, \eta \right);
\end{align}
\end{subequations}
$\edth \eta$ then has spin weight~$s+1$ and~$\bar{\edth} \eta$ has spin weight~$s-1$.  These operators obey the Leibnitz rule (even on products of functions of different spin weights) and hence are bona fide derivative operators, and are also total derivatives in the sense that when~$\edth \eta$  or~$\bar{\edth} \eta$ has spin weight zero, its integral over the round sphere vanishes.  Moreover, they relate spin-weighted spherical harmonics of different spin weight to each other: 
\begin{subequations}
\label{eqs:ladder}
\begin{align}
\edth \, _s Y_{\ell,m} &= \sqrt{(\ell-s)(\ell+s+1)} \, _{s+1}Y_{\ell,m}, \\
\bar{\edth} \, _s Y_{\ell,m} &= -\sqrt{(\ell+s)(\ell-s+1)} \, _{s-1}Y_{\ell,m}.
\end{align}
\end{subequations}
(It then follows that the spin-weighted spherical harmonics with integer~$s$ can be generated from the ordinary spherical harmonics~$Y_{\ell,m}$ by successive applications of~$\edth$.)

The triple overlap of spin-weighted spherical harmonics can be expressed in terms of~$3j$ symbols (see e.g.~\cite{Edmonds,Thompson}\footnote{Technically~\cite{Edmonds,Thompson} only give expressions for the triple integral of Wigner~$D$-matrix elements, but the~$_s Y_{\ell,m}$ are precisely proportional to these matrix elements~\cite{GolMac67}.}):
\begin{multline}
\label{eq:Ytripleoverlap}
\int_{S^2} \, _{s_1} Y_{\ell_1,m_1} \, _{s_2} Y_{\ell_2,m_2} \, _{s_3} Y_{\ell_3,m_3} \\ = \sqrt{\frac{(2\ell_1+1)(2\ell_2+1)(2\ell_3+1)}{4\pi}} \begin{pmatrix} \ell_1 & \ell_2 & \ell_3 \\ -s_1 & -s_2 & -s_3 \end{pmatrix} \begin{pmatrix} \ell_1 & \ell_2 & \ell_3 \\ m_1 & m_2 & m_3 \end{pmatrix},
\end{multline}
where here and in what follows we will leave the volume element $\sin\theta \, d\theta \, d\phi$ in integrals implied.  The~$3j$ symbol~$\begin{pmatrix} \ell_1 & \ell_2 & \ell_3 \\ m_1 & m_2 & m_3 \end{pmatrix}$ vanishes unless the usual rules for angular momentum addition are satisfied: that is,~$m_i \in \{-\ell_i, -\ell_i + 1, \cdots, \ell_i\}$ for each~$i$,~$m_1 + m_2 + m_3 = 0$, the~$\ell_i$ obey the triangle condition~$|\ell_1 - \ell_2| \leq \ell_3 \leq \ell_1 + \ell_2$, and finally~$\ell_1 + \ell_2 + \ell_3$ must be an integer (in fact an even integer if all the~$m_i$ vanish).  Moreover, the~$3j$ symbols have the following properties:
\bea
\sum_{m} (-1)^{\ell-m} \begin{pmatrix} \ell & \ell & \ell' \\ m & -m & 0 \end{pmatrix} &=  \sqrt{2\ell+1} \, \delta_{\ell',0}, \label{subeq:prop1} \\
\sum_{m_1,m_2} \begin{pmatrix} \ell_1 & \ell_2 & \ell_3 \\ m_1 & m_2 & m_3 \end{pmatrix} \begin{pmatrix} \ell_1 & \ell_2 & \ell_3' \\ m_1 & m_2 & m_3' \end{pmatrix} &= \frac{1}{2\ell_3 + 1} \, \delta_{\ell_3,\ell_3'} \delta_{m_3,m_3'} \{\ell_1 \quad \ell_2 \quad \ell_3 \}, \label{subeq:3jmorthog} \\
\sum_{\ell_1,m_1} (2\ell_1 + 1) \begin{pmatrix} \ell_1 & \ell_2 & \ell_3 \\ m_1 & m_2 & m_3 \end{pmatrix} \begin{pmatrix} \ell_1 & \ell_2 & \ell_3 \\ m_1 & m_2' & m_3' \end{pmatrix} &= \delta_{m_2,m_2'} \delta_{m_3,m_3'}, \label{subeq:3jmlorthog} \\
\begin{pmatrix} \ell_2 & \ell_1 & \ell_3 \\ m_2 & m_1 & m_3 \end{pmatrix} = \begin{pmatrix} \ell_1 & \ell_2 & \ell_3 \\ -m_1 & -m_2 & -m_3 \end{pmatrix} &= (-1)^{\ell_1 + \ell_2 + \ell_3} \begin{pmatrix} \ell_1 & \ell_2 & \ell_3 \\ m_1 & m_2 & m_3 \end{pmatrix} \label{subeq:permutation}
\eea
where~$\{\ell_1 \quad \ell_2 \quad \ell_3\} = 1$ if the~$\ell_i$ obey the triangle condition and zero otherwise, every sum over~$m_i$ is understood to run from~$-\ell_i$ to~$\ell_i$, and~\eqref{subeq:permutation} holds for any other odd permutation of the columns.  Finally, integrating~$\edth(\, _{s_1} Y_{\ell_1,m_1} \, _{s_2} Y_{\ell_2,m_2} \, _{s_3} Y_{\ell_3,m_3})$ and using the relations~\eqref{eqs:ladder} and the integration formula~\eqref{eq:Ytripleoverlap} gives the recursion relations
\begin{multline}
\label{eq:3jrecursion}
0 = \sqrt{(\ell_1 \mp s_1) (\ell_1 \pm s_1 + 1)} \begin{pmatrix} \ell_1 & \ell_2 & \ell_3 \\ s_1 \pm 1 & s_2 & s_3 \end{pmatrix}
\\ + \sqrt{(\ell_2 \mp s_2) (\ell_2 \pm s_2 + 1)} \begin{pmatrix} \ell_1 & \ell_2 & \ell_3 \\ s_1 & s_2 \pm 1 & s_3 \end{pmatrix}
\\ + \sqrt{(\ell_3 \mp s_3) (\ell_3 \pm s_3 + 1)} \begin{pmatrix} \ell_1 & \ell_2 & \ell_3 \\ s_1 & s_2 & s_3 \pm 1 \end{pmatrix}.
\end{multline}

\subsection{Scalar}

For the non-minimally coupled scalar, the operator~$L$ was given in~\eqref{eq:Lscalar}, which we repeat here:
\be
L = e^{-2f}\left[-\overline{\grad}^2 + 2\xi\left(1 - \overline{\grad}^2 f\right)\right];
\ee
hence the operators~$L^{(n)}$ introduced in~\eqref{subeq:Ln} are
\bea
L^{(1)} &= 2 f^{(1)} \overline{\grad}^2 - 2\xi\left(2f^{(1)} + \overline{\grad}^2 f^{(1)}\right), \\
L^{(2)} &= 2\left(f^{(2)} - (f^{(1)})^2\right)\overline{\grad}^2 - 2\xi \left[2\left(f^{(2)} - (f^{(1)})^2\right) - 2 f^{(1)} \overline{\grad}^2 f^{(1)} + \overline{\grad}^2 f^{(2)}\right].
\eea
Decomposing~$f^{(1)}$ in spherical harmonics as in~\eqref{eq:fexpansion}, reality of~$f^{(1)}$ requires that~$f^*_{\ell,m} = (-1)^m f_{\ell,-m}$, while the volume preservation condition~\eqref{eq:volpreservation} requires that~$f_{0,0} = 0$.  Then the matrix elements~$\widetilde{L}^{(1)}_{\ell,m,\ell',m'}$ are given by
\bea
\widetilde{L}^{(1)}_{\ell,m,\ell',m'} &= \int_{S^2} Y_{\ell,m}^* L^{(1)} Y_{\ell',m'} \\
			&= -2 \sum_{\ell'',m''} f_{\ell'',m''} \left(\bar{\lambda}_{\ell'} - \xi C_{\ell''} \right) \int_{S^2} Y_{\ell'',m''} Y_{\ell,m}^* Y_{\ell',m'},
\eea
where~$C_\ell = \ell(\ell+1)$ are the eigenvalues of~$-\overline{\grad}^2$ (and the eigenvalues of~$\overline{L}$ are~$\bar{\lambda}_\ell = \ell(\ell+1) + 2\xi$).  Using~\eqref{eq:Ytripleoverlap}, we thus find
\begin{multline}
\label{eq:L1ellm}
\widetilde{L}^{(1)}_{\ell,m,\ell',m'} = -2 (-1)^m \sum_{\ell'',m''} f_{\ell'',m''} \left(\bar{\lambda}_{\ell'} - \xi C_{\ell''} \right) \\ \times \sqrt{\frac{(2\ell+1)(2\ell'+1)(2\ell''+1)}{4\pi}} \begin{pmatrix} \ell & \ell' & \ell'' \\ 0 & 0 & 0 \end{pmatrix} \begin{pmatrix} \ell & \ell' & \ell'' \\ m & -m' & m'' \end{pmatrix}.
\end{multline}
From~\eqref{subeq:prop1} and the fact that~$f_{0,0} = 0$, it then follows that~$\Tr\widetilde{\bm{L}}^{(1)}_{\ell,\ell}$ vanishes, and hence so does the linear correction to the heat kernel:~$\Delta K^{(1)} = 0$.

To compute the second-order correction, we need the traces~$\Tr \widetilde{\bm{L}}^{(2)}_{\ell,\ell}$ and~$\Tr \left(\widetilde{\bm{L}}^{(1)}_{\ell,\ell'}\widetilde{\bm{L}}^{(1)}_{\ell',\ell} \right)$.  To compute the former, we use the addition theorem~\eqref{eq:additiontheorem}, and hence
\bea
\Tr \widetilde{\bm{L}}^{(2)}_{\ell,\ell} &= \frac{2\ell+1}{2\pi} \int_{S^2} \left[-\left(f^{(2)} - (f^{(1)})^2\right) C_\ell - \xi \left(2\left(f^{(2)} - (f^{(1)})^2\right) - 2 f^{(1)} \overline{\grad}^2 f^{(1)} \right)\right], \\
		&= \frac{2\ell+1}{\pi} \int_{S^2} \left(\bar{\lambda}_\ell (f^{(1)})^2 + \xi   f^{(1)} \overline{\grad}^2 f^{(1)} \right), \\
		&= \frac{2\ell+1}{\pi} \sum_{\ell',m'} |f_{\ell',m'}|^2 (\bar{\lambda}_\ell - \xi C_{\ell'}) \label{eq:TrL2scalar},
\eea
where in the first line the Laplacian~$\overline{\grad}^2 f^{(2)}$ vanishes since it's a total divergence, in the second line we used the volume preservation condition~\eqref{eq:volpreservation} to replace the remaining~$f^{(2)}$, and in the final line we used~\eqref{eq:fexpansion}.  To compute~$\Tr \left(\widetilde{\bm{L}}^{(1)}_{\ell,\ell'}\widetilde{\bm{L}}^{(1)}_{\ell',\ell} \right)$, we use~\eqref{eq:L1ellm}:
\bea
\Tr \left(\widetilde{\bm{L}}^{(1)}_{\ell,\ell'}\widetilde{\bm{L}}^{(1)}_{\ell',\ell} \right) &= \sum_{m,m'} \widetilde{L}^{(1)}_{\ell,m,\ell',m'} \widetilde{L}^{(1)}_{\ell',m',\ell,m}, \\
		&= \frac{(2\ell+1)(2\ell'+1)}{\pi} \sum_{\mathclap{\ell_1,m_1,\ell_2,m_2}} \, f_{\ell_1,m_1} f_{\ell_2,m_2} \nonumber 
		\\ & \qquad \times \sqrt{(2\ell_1+1)(2\ell_2+1)} \, (\bar{\lambda}_{\ell'} - \xi C_{\ell_1})(\bar{\lambda}_\ell - \xi C_{\ell_2}) \begin{pmatrix} \ell & \ell' & \ell_1 \\ 0 & 0 & 0 \end{pmatrix} \begin{pmatrix} \ell' & \ell & \ell_2 \\ 0 & 0 & 0 \end{pmatrix}  \nonumber
		\\ & \qquad \qquad \times \sum_{m,m'} (-1)^{m+m'} \begin{pmatrix} \ell & \ell' & \ell_1 \\ m & -m' & m_1 \end{pmatrix} \begin{pmatrix} \ell & \ell' & \ell_2 \\ m & -m' & -m_2 \end{pmatrix}, 
\eea
where we used~\eqref{subeq:permutation}.  Now, because the~$3j$ symbols vanish unless the sum of the~$m$ quantum numbers is zero,~$m$ and~$m'$ must be related by~$m' = m + m_1$.  We may therefore replace the phase~$(-1)^{m+m'}$ with~$(-1)^{m_1}$, which we may combine with~$f_{\ell_1,m_1}$ to give~$f^*_{\ell_1,-m_1}$.  Then using the orthogonality relation~\eqref{subeq:3jmorthog} to evaluate the final sum, we obtain
\begin{multline}
\label{eq:TrL1L1scalar}
\Tr \left(\widetilde{\bm{L}}^{(1)}_{\ell,\ell'}\widetilde{\bm{L}}^{(1)}_{\ell',\ell} \right) = \frac{(2\ell+1)(2\ell'+1)}{\pi} \sum_{\ell_1,m_1} |f_{\ell_1,m_1}|^2 \\ \times  (\bar{\lambda}_{\ell'} - \xi C_{\ell_1})(\bar{\lambda}_\ell - \xi C_{\ell_1}) \begin{pmatrix} \ell & \ell' & \ell_1 \\ 0 & 0 & 0 \end{pmatrix}^2;
\end{multline}
the diagonal part of this result gives
\be
\label{eq:TrL1squaredscalar}
\Tr \left((\widetilde{\bm{L}}^{(1)}_{\ell,\ell})^2\right) = \frac{(2\ell+1)^2}{\pi} \sum_{\ell_1,m_1} |f_{\ell_1,m_1}|^2 (\bar{\lambda}_{\ell} - \xi C_{\ell_1})^2 \begin{pmatrix} \ell & \ell & \ell_1 \\ 0 & 0 & 0 \end{pmatrix}^2.
\ee

We may now insert~\eqref{eq:TrL2scalar},~\eqref{eq:TrL1L1scalar}, and~\eqref{eq:TrL1squaredscalar} into~\eqref{eq:K2perturb} to obtain the second-order correction to the heat kernel; one obtains the expression~\eqref{eq:K2scalar} given in the main text with
\bea
\alpha_{\ell,\ell'} &= -\frac{(2\ell'+1)(\bar{\lambda}_{\ell'} - \xi C_\ell)}{\pi} \left[1+ \sum_{\ell_1,\ell_1 \neq \ell'} \frac{(2\ell_1+1)(\bar{\lambda}_{\ell_1} - \xi C_\ell)}{\bar{\lambda}_{\ell'} - \bar{\lambda}_{\ell_1}} \begin{pmatrix} \ell & \ell' & \ell_1 \\ 0 & 0 & 0 \end{pmatrix}^2 \right], \\
\beta_{\ell,\ell'} &= \frac{(2\ell'+1)^2(\bar{\lambda}_{\ell'} - \xi C_\ell)^2}{2\pi} \begin{pmatrix} \ell & \ell' & \ell' \\ 0 & 0 & 0 \end{pmatrix}^2. \label{subeq:betascalar}
\eea
The sum in the expression for~$\alpha_{\ell,\ell'}$ can be simplified slightly by noting that by~\eqref{subeq:3jmlorthog},
\be
1 = \sum_{\ell_1,m_1} (2\ell_1 + 1) \begin{pmatrix} \ell & \ell' & \ell_1 \\ 0 & 0 & m_1 \end{pmatrix}^2 = \sum_{\ell_1} (2\ell_1 + 1) \begin{pmatrix} \ell & \ell' & \ell_1 \\ 0 & 0 & 0 \end{pmatrix}^2,
\ee
and hence one can write
\begin{multline}
\label{eq:alphascalarrewritten}
\alpha_{\ell,\ell'} = -\frac{(2\ell'+1)(\bar{\lambda}_{\ell'} - \xi C_\ell)}{\pi} \left[(2\ell'+1) \begin{pmatrix} \ell & \ell' & \ell' \\ 0 & 0 & 0 \end{pmatrix}^2 \right. \\ \left. + (\bar{\lambda}_{\ell'} - \xi C_\ell) \sum_{\ell_1,\ell_1 \neq \ell'} \frac{2\ell_1+1}{\bar{\lambda}_{\ell'} - \bar{\lambda}_{\ell_1}} \begin{pmatrix} \ell & \ell' & \ell_1 \\ 0 & 0 & 0 \end{pmatrix}^2 \right].
\end{multline}

The sum over~$\ell_1$ in~$\alpha_{\ell,\ell'}$ is a finite sum due to the triangle condition~$|\ell-\ell'| \leq \ell_1 \leq \ell+\ell'$; by calculating this sum exactly for several values of~$\ell$,~$\ell'$ and using some sequence-finding functions in~\texttt{Mathematica}, we are able to infer the closed-form expression
\be
\label{eq:alphascalarclosed}
\alpha_{\ell,\ell'} = \begin{dcases} \frac{(2\ell'+1)(\bar{\lambda}_{\ell'} - \xi C_\ell)^2}{\pi  C_\ell} \, \frac{\left(\frac{2+\ell}{2}\right)_{\ell'} \left(\frac{\ell}{2}\right)_{-\ell'}}{\left(\frac{3+\ell}{2}\right)_{\ell'} \left(\frac{1+\ell}{2}\right)_{-\ell'}}, & \ell' < \frac{\ell}{2} \\
\left[\frac{(2\ell'+1)(\bar{\lambda}_{\ell'} - \xi C_\ell)^2}{2\pi} \left(H_{\ell'-{\ell\over 2}} - H_{\ell'+{\ell \over 2}} \phantom{\frac{1}{2}}\right.\right. & \\ \left.\left. \hspace{3cm} + H_{\ell'+{\ell+1 \over 2}} - H_{\ell'-{\ell+1 \over 2}} - \frac{2}{2\ell'+1} \right) \right. \\ \left. \hspace{4cm} - \frac{(2\ell'+1)^2(\bar{\lambda}_{\ell'} - \xi C_\ell)}{\pi}\right]\begin{pmatrix} \ell & \ell' & \ell' \\ 0 & 0 & 0 \end{pmatrix}^2 , & \ell' \geq \frac{\ell}{2} \end{dcases}
\ee
where~$(x)_n \equiv \Gamma(x+n)/\Gamma(x)$ are Pochhammer symbols and~$H_n$ are harmonic numbers.  Though we are unable to provide a general derivation, we have verified that this result agrees with~\eqref{eq:alphascalarrewritten} for all values of~$\ell$ and~$\ell'$ from zero to~100.

Finally, note that the~$3j$ symbols in the expressions~\eqref{subeq:betascalar} and~\eqref{eq:alphascalarclosed} are only nonvanishing if~$\ell + 2\ell'$ is an even integer, implying that whenever~$\ell$ is odd,~$\beta_{\ell,\ell'}$ vanishes for all~$\ell'$ and~$\alpha_{\ell,\ell'}$ vanishes for all~$\ell' > \ell/2$.  Thus the case of odd~$\ell$ reproduces the expressions~\eqref{eq:alphabetascalarodd} given in the main text.

\subsection{Dirac Fermion}

For the fermion, we obtain the~$L^{(n)}$ by expanding~$L$ given in~\eqref{eq:LDirac}.  In fact, having introduced the spin weight raising and lowering operators~$\edth$,~$\bar{\edth}$ in~\eqref{eqs:edth} above, it is now natural to re-express~$L$ in terms of them using the fact that~$f$ has spin weight zero and~$L$ acts on the space of functions with spin weight~$1/2$.  We ultimately obtain
\be
L = -e^{-2f} \left[\edth \bar{\edth} + \frac{1}{2}\left(\overline{\grad}^2 f + (\bar{\edth} f) \edth - (\edth f) \bar{\edth}\right) - \frac{1}{4} (\overline{\grad}_a f)^2 \right]
\ee
(we could also write~$\overline{\grad}^2 f = \edth \bar{\edth} f$ and~$(\overline{\grad}_a f)^2 = (\edth f)(\bar{\edth}f )$, but this rewriting will not be needed), and hence the unperturbed operator and the corrections~$L^{(n)}$ are
\bea
\overline{L} &= -\edth \bar{\edth}, \\
L^{(1)} &= 2f^{(1)} \edth \bar{\edth} - \frac{1}{2}\left(\overline{\grad}^2 f^{(1)} + (\bar{\edth} f^{(1)}) \edth - (\edth f^{(1)}) \bar{\edth}\right), \\
L^{(2)} &= 2\left(f^{(2)} + (f^{(1)})^2\right) \edth \bar{\edth} - 2f^{(1)} L^{(1)} + \frac{1}{4} (\overline{\grad}_a f^{(1)})^2,
\eea
where to simplify~$L^{(2)}$ we took~$f^{(2)}$ to be a constant; there is no loss of generality in this simplification, since the purpose of~$f^{(2)}$ is only to ensure that the volume preservation condition~\eqref{eq:volpreservation} can be satisfied for nontrivial~$f^{(1)}$.

From~\eqref{eqs:ladder}, it follows that~$_s Y_{\ell,m}$ is an eigenfunction of~$\edth \bar{\edth}$:
\be
\edth \bar{\edth} \, _s Y_{\ell,m} = -(\ell+s)(\ell-s+1) \, _s Y_{\ell,m}.
\ee
Hence since~$\overline{L}$ acts on the space of functions with spin weight~$1/2$, the~$_{1/2} Y_{\ell,m}$ form a basis of eigenfunctions of~$\overline{L}$ with eigenvalues~$\bar{\lambda}_\ell = (\ell + 1/2)^2$, and we may compute the matrix elements~$\widetilde{L}^{(n)}_{\ell,m,\ell',m'}$ by taking the unperturbed eigenfunctions to be~$\tilde{h}_{\ell,m} = \, _{1/2} Y_{\ell,m}$.  Proceeding in this manner, first we obtain (again by expanding~$f^{(1)}$ in spherical harmonics as in~\eqref{eq:fexpansion})\footnote{Technically for the fermion we should be careful to denote the limits of summation, since the index~$\ell$ can either range over all non-negative integers, as it does for the decomposition of~$f^{(1)}$ in terms of spin weight-zero spherical harmonics, or over positive half-odd integers~$1/2, 3/2, \ldots$, as for the eigenfunctions~$\tilde{h}_{\ell,m}$.  We assume it is clear from context what the appropriate limits of each summation should be.}
\begin{multline}
\label{eq:L1tildefermionraw}
\widetilde{L}^{(1)}_{\ell,m,\ell',m'} = \sum_{\ell'', m''} f_{\ell'',m''} \left[\left(-2\bar{\lambda}_{\ell'} + \frac{1}{2} C_{\ell''}\right) \int_{S^2} \, _{1/2} Y^*_{\ell,m} Y_{\ell'',m''} \, _{1/2} Y_{\ell',m'} \right. \\ \left. -\frac{1}{2} \int_{S^2}\, _{1/2} Y^*_{\ell,m} \left( (\bar{\edth} Y_{\ell'',m''})(\edth \, _{1/2} Y_{\ell',m'}) - (\edth Y_{\ell'',m''})(\bar{\edth} \, _{1/2} Y_{\ell',m'}) \right) \right].
\end{multline}
We may then use the rules~\eqref{eqs:ladder} to replace the derivatives of the spherical harmonics with spherical harmonics of different spin weights, and finally (using~$_s Y^*_{\ell,m} = (-1)^{s+m} \, _{-s} Y_{\ell,-m}$) we may use~\eqref{eq:Ytripleoverlap} to perform the remaining integrals, thereby obtaining
\begin{multline}
\label{subeq:L1tildefermion}
\widetilde{L}^{(1)}_{\ell,m,\ell',m'} = \frac{1}{2} \sum_{\ell'',m''} f_{\ell'',m''} (-1)^{m+1/2} \sqrt{\frac{(2\ell+1)(2\ell'+1)(2\ell''+1)}{4\pi}} \\ \times \begin{pmatrix} \ell & \ell' & \ell'' \\ -m & m' & m'' \end{pmatrix} A_{\ell,\ell',\ell''},
\end{multline}
where to slightly compactify notation we have defined
\bea
A_{\ell,\ell',\ell''} &\equiv \sqrt{(\bar{\lambda}_{\ell'}-1) C_{\ell''}} \begin{pmatrix} \ell & \ell' & \ell'' \\ 1/2 & -3/2 & 1 \end{pmatrix} - \sqrt{\bar{\lambda}_{\ell'} C_{\ell''}} \begin{pmatrix} \ell & \ell' & \ell'' \\ 1/2 & 1/2 & -1 \end{pmatrix} \nonumber \\ 
& \hspace{6cm} + \left(-4\bar{\lambda}_{\ell'} + C_{\ell''} \right) \begin{pmatrix} \ell & \ell' & \ell'' \\ 1/2 & -1/2 & 0 \end{pmatrix}, \label{subeq:Araw} \\
			&= \left(\bar{\lambda}_\ell - 3\bar{\lambda}_{\ell'} + 2(-1)^{\ell+\ell'+\ell''} \sqrt{\bar{\lambda}_\ell \bar{\lambda}_{\ell'}} \right) \begin{pmatrix} \ell & \ell' & \ell'' \\ 1/2 & -1/2 & 0 \end{pmatrix}, \label{subeq:Acompact}
\eea
with the second expression obtained from the first by using~\eqref{eq:3jrecursion} and~\eqref{subeq:permutation}.

As for the scalar, it follows immediately from~\eqref{subeq:prop1} and the fact that~$f_{0,0} = 0$ that~$\Tr\widetilde{\bm{L}}^{(1)}_{\ell,\ell} = 0$, and hence~$\Delta K^{(1)} = 0$ as well.  To get the second order term, we first compute
\bea
\Tr \widetilde{\bm{L}}^{(2)}_{\ell,\ell} &= \frac{2\ell+1}{16\pi} \int_{S^2} (\overline{\grad}_a f^{(1)})^2 - 2\sum_m \int_{S^2} \, _{1/2} Y^*_{\ell,m} f^{(1)} L^{(1)} \, _{1/2} Y_{\ell,m}, \\
		&= \frac{2\ell+1}{16\pi} \sum_{\ell'',m''} C_{\ell''} |f_{\ell'',m''}|^2 - 2\sum_{m,\ell',m'} \widetilde{L}^{(1)}_{\ell',m',\ell,m} \int_{S^2} \, _{1/2} Y^*_{\ell,m} f^{(1)} \, _{1/2} Y_{\ell',m'},
\eea
where to get the first line we used the addition formula~\eqref{eq:additiontheorem} and the volume preservation condition~\eqref{eq:volpreservation}, and to get to the second line we integrated by parts in the first integral and inserted a resolution of the indentity in terms of the~$_{1/2} Y_{\ell',m'}$ in the second.  The remaining integral can be performed using~\eqref{eq:Ytripleoverlap}, followed by using the expression~\eqref{subeq:L1tildefermion} and the orthogonality relations~\eqref{subeq:3jmorthog} and~\eqref{subeq:3jmlorthog} to collapse most of the sums (note that the orthogonality relation~\eqref{subeq:3jmlorthog} is implemented most directly by writing~$A_{\ell,\ell',\ell''}$ in the longer form~\eqref{subeq:Araw}); the final result is
\be
\Tr \widetilde{\bm{L}}^{(2)}_{\ell,\ell} = \frac{2\ell+1}{16\pi} \sum_{\ell'',m''} |f_{\ell'',m''}|^2 \left(16 \bar{\lambda}_\ell - 3 C_{\ell''}\right).
\ee
The same manipulations, again using~\eqref{subeq:L1tildefermion}, also yield
\be
\Tr \left(\widetilde{\bm{L}}^{(1)}_{\ell,\ell'}\widetilde{\bm{L}}^{(1)}_{\ell',\ell} \right) = \frac{(2\ell+1)(2\ell'+1)}{16\pi} \sum_{\ell'',m''} |f_{\ell'',m''}|^2 A_{\ell,\ell',\ell''} A_{\ell',\ell,\ell''},
\ee
and hence also
\be
\Tr \left((\widetilde{\bm{L}}^{(1)}_{\ell,\ell})^2\right) = \frac{(2\ell+1)^2}{16\pi} \sum_{\ell'',m''} |f_{\ell'',m''}|^2 A^2_{\ell,\ell,\ell''}.
\ee
Inserting these into the expression for the second-order correction to the heat kernel and using~\eqref{subeq:Acompact}, we thus obtain~\eqref{eq:K2dirac} with
\begin{subequations}
\label{eqs:alphabetafermionraw}
\begin{align}
\alpha_{\ell,\ell'} &= -\frac{2\ell'+1}{16\pi} \left[16\bar{\lambda}_{\ell'} - 3C_\ell \phantom{\frac{1}{2}} \right. \nonumber \\ & 
\left. \qquad + \sum_{\ell'',\ell'' \neq \ell'} \frac{2\ell''+1}{\bar{\lambda}_{\ell'} - \bar{\lambda}_{\ell''}} 
\left(\bar{\lambda}_{\ell'} - 3 \bar{\lambda}_{\ell''} + 2(-1)^{\ell+\ell'+\ell''}\sqrt{\bar{\lambda}_{\ell'} \bar{\lambda}_{\ell''}} \right) \right. \nonumber \\ &
\left. \qquad \qquad \qquad \times \left(\bar{\lambda}_{\ell''} - 3 \bar{\lambda}_{\ell'} + 2(-1)^{\ell+\ell'+\ell''}\sqrt{\bar{\lambda}_{\ell'} \bar{\lambda}_{\ell''}} \right) \begin{pmatrix} \ell & \ell' & \ell'' \\ 0 & 1/2 & -1/2 \end{pmatrix}^2 \right], \label{subeq:Diracalpharaw} \\
\beta_{\ell,\ell'} &= \frac{(2\ell'+1)^2\bar{\lambda}_{\ell'}^2}{4\pi} \, \left(1+(-1)^\ell \right) \begin{pmatrix} \ell & \ell' & \ell' \\ 0 & 1/2 & -1/2 \end{pmatrix}^2. \label{subeq:Diracbetaraw}
\end{align}
\end{subequations}
Note that as for the scalar,~$\beta_{\ell,\ell'}$ vanishes whenever~$\ell$ is odd.  The sum in~$\alpha_{\ell,\ell'}$ is again a finite sum since the~$3j$ symbol vanishes unless~$|\ell-\ell'| \leq \ell'' \leq \ell+\ell'$; by evaluating the sum exactly for various values of~$\ell$,~$\ell'$ we are able to infer the expression
\be
\label{eq:alphafermionclosed}
\alpha_{\ell,\ell'} = \begin{dcases} - \frac{(2\ell'+1)^3}{16\pi} \, \frac{\left(\frac{2+\ell}{2}\right)_{\ell'+1/2} \left(\frac{2+\ell}{2}\right)_{-(\ell'+1/2)} }{\left(\frac{1+\ell}{2}\right)_{\ell'+1/2} \left(\frac{1+\ell}{2}\right)_{-(\ell'+1/2)}}, & \ell' < \frac{\ell}{2} \\
-\frac{(1+2\ell') \, \Gamma\left(\frac{1+\ell}{2}\right)^2(1+(-1)^\ell)}{32\pi^2 \, \Gamma\left(\frac{2+\ell}{2}\right)^2}\left[16\ell'(\ell'+1) - \ell(\ell+1)    \phantom{\frac{\left(\frac{1}{2}\right)}{\left(\frac{1}{2}\right)}} \right. & \\ \left. \qquad + 4 + \frac{16}{\pi} \sum_{\ell'' = 0}^{\frac{\ell-2}{2}} \frac{\left(\ell''+\frac{1}{2}\right)^4 \left(\frac{\ell}{2}-\ell''\right)_{1/2} \left(\frac{2+\ell}{2} + \ell''\right)_{1/2}}{\left( \left(\ell'+\frac{1}{2}\right)^2 - \left(\ell''+\frac{1}{2}\right)^2\right)^2}\right], & \ell' \geq \frac{\ell}{2} \end{dcases}
\ee
which for odd~$\ell$ reproduces the result~\eqref{eq:alphabetafermionodd} given in the main text.  Though we are not able to give a derivation of this result, we have checked its agreement with~\eqref{subeq:Diracalpharaw} for all values of~$\ell$ up to~100 and all values of~$\ell'$ up to~$201/2$.

\section{Flat Space Scaling Limit}
\label{app:flatspace}

Here we provide some more details on the flat-space scaling limit performed in Section~\ref{subsec:flat}.  First, to obtain the limiting behavior~\eqref{eq:aflatspace}, note that for odd~$r_0 k$ we have from~\eqref{eq:K2scalinglimit}
\be
\label{eq:ar0k}
a_{r_0 k}(t) = \frac{t}{r_0^2} \sum_{k'}^{k/2} e^{-(k')^2 t} \alpha_{r_0 k, r_0 k'} = r_0^2 \, t \int_0^{k/2} dk' \, e^{-(k')^2 t} H(k,k') + \Ocal(r_0),
\ee
where we have defined~$H(k,k') = \lim_{r_0 \to \infty} \alpha_{r_0 k, r_0 k'}/r_0^3$ and re-expressed the sum as an integral in the~$r_0 \to \infty$ limit.  The function~$H(k,k')$ can be obtained from the closed-form expressions~\eqref{eq:alphabetascalarodd} and~\eqref{eq:alphabetafermionodd} by using the fact that for~$x > 0$,
\be
\frac{\Gamma(x+1/2)}{\Gamma(x)} = \sqrt{x} + \Ocal(x^{-1/2}),
\ee
from which we find
\be
H(k,k') = \begin{dcases} \frac{2k'((k')^2 -\xi k^2)^2}{\pi k \sqrt{k^2 - 4(k')^2}}, & \mbox{scalar} \\
-\frac{(k')^3 \sqrt{k^2 - 4(k')^2}}{2\pi k}, & \mbox{fermion} \end{dcases}
\ee
Using these expressions to perform the integral in~\eqref{eq:ar0k}, we recover~\eqref{eq:aflatspace} as promised.  We now remove the restriction that~$r_0 k$ be odd.

Next, to obtain~\eqref{eq:fflatspace}, we must keep track of how the mode decomposition of~$f^{(1)}$ in spherical harmonics behaves in the~$r_0 \to \infty$ scaling limit.  To this end, we need the appropriate scaling behavior of the spherical harmonics; to obtain it, first define as above~$k = \ell/r_0$ and in addition~$k_y = m/r_0$.  Expressing the spherical harmonics in terms of Legendre polynomials~$P_\ell^m(x)$, we immediately have
\be
Y_{k r_0, k_y r_0}\left(\frac{\pi}{2} + \frac{x}{r_0}, \frac{y}{r_0}\right) \to N_{k r_0, k_y r_0} P_{r_0 k}^{r_0 k_y}\left(\frac{x}{r_0}\right) e^{i k_y y},
\ee
where~$N_{k r_0, k_y r_0}$ is a normalization constant.  The functional form of~$P_{r_0 k}^{r_0 k_y}$ as~$r_0 \to \infty$ can be inferred from the scaling limit of Legendre's differential equation (as well as from the symmetry properties of~$P_\ell^m(x)$ about~$x = 0$), from which we then find that
\be
Y_{k r_0, k_y r_0}\left(\frac{\pi}{2} + \frac{x}{r_0}, \frac{y}{r_0}\right) \to \widetilde{N}_{k r_0, k_y r_0} \cos\left(\sqrt{k^2 - k_y^2} \, x \right) e^{i k_y y}
\ee
for even~$(k+k_y)r_0$, and the same expression with a sine (instead of a cosine) for odd~$(k+k_y) r_0$.  Here~$\widetilde{N}_{k r_0, k_y r_0}$ is some new constant, which can be obtained up to an overall phase by the normalization condition
\bea
1 &= \sum_{\ell',m'} \int Y^*_{\ell',m'}(\theta,\phi) Y_{\ell,m}(\phi,\theta) \sin\theta \, d\theta \, d\phi \\
	&\to \int_0^\infty dk' \int_{-k'}^{k'} dk_y' \, \widetilde{N}^*_{k' r_0, k_y' r_0} \widetilde{N}_{k r_0, k_y r_0} \\ &\hspace{3cm} \times \int_{-\infty}^\infty dx \, dy \, \cos\left(\sqrt{(k')^2 - (k_y')^2} \, x \right) \cos\left(\sqrt{k^2 - k_y^2} \, x \right) e^{i(k_y - k_y') y} \nonumber
\eea
for any~$\ell = k r_0$ and~$m = k_y r_0$, where the domain of integration in~$k'$ and~$k_y'$ comes from the restriction that~$\ell' > 0$ and~$-\ell' \leq m' \leq \ell'$.  Performing the integrals finally gives that up to an overall phase,
\be
Y_{r_0 k, r_0 k_y}\left(\frac{\pi}{2} + \frac{x}{r_0}, \frac{y}{r_0}\right) \to \frac{1}{\pi} \frac{\sqrt{k}}{(k^2 - k_y^2)^{1/4}} \cos\left(\sqrt{k^2 - k_y^2} \, x \right) e^{i k_y y}
\ee
for even~$(k+k_y)r_0$ and the same expression with a sine for odd~$(k+k_y) r_0$.

Hence in the~$r_0 \to \infty$ scaling limit, the coefficients~$f_{\ell,m}$ give the desired expression~\eqref{eq:fflatspace}:
\bea
f_{r_0 k, r_0 k_y} &= \int Y^*_{r_0 k, r_0 k_y}(\theta,\phi) f(\theta,\phi) \sin\theta \, d\theta \, d\phi, \\
		&= \frac{1}{r_0^2} \int Y^*_{r_0 k, r_0 k_y}\left(\frac{\pi}{2} + \frac{x}{r_0}, \frac{y}{r_0}\right) f(x,y) \sin\left(\frac{\pi}{2} + \frac{x}{r_0}\right) \, dx \, dy, \\
		&\to \frac{1}{2\pi r_0^2} \frac{\sqrt{k}}{(k^2 - k_y^2)^{1/4}} \int dx \, dy \left(e^{i\sqrt{k^2-k_y^2} \, x} \pm e^{-i\sqrt{k^2-k_y^2} \, x}\right) e^{-ik_y y} f(x,y), \\
		&= \frac{2\pi}{r_0^2} \frac{\sqrt{k}}{(k^2 - k_y^2)^{1/4}} \left( \hat{f}(\sqrt{k^2 - k_y^2}, k_y) \pm \hat{f}(-\sqrt{k^2 - k_y^2}, k_y)\right),
\eea
with the upper (lower) sign for even (odd)~$(k+k_y)r_0$ (and we are neglecting an overall phase that will cancel out).  We remind the reader that~$\hat{f}(k_x,k_y)$ is the Fourier transform of~$f^{(1)}(x,y)$, and the assumption that~$f^{(1)}(x,y)$ vanishes at large~$(x,y)$ (i.e.~$f^{(1)}(\theta,\phi)$ vanishes away from~$(\theta = \pi/2, \phi = 0)$) is what allows us to take the scaling limit of the integrand before evaluating the integral.

Finally, decomposing~$\Delta K^{(2)}$ into contributions based on the parity of~$(k+k_y) r_0$ as
\be
\Delta K^{(2)} \to \Delta K^{(2)}_\mathrm{even} + K^{(2)}_\mathrm{odd},
\ee
we have
\bea
\Delta K^{(2)}_\mathrm{even,odd} &= \frac{t}{r_0^2} \, \sum_{\mathclap{\substack{k,k_y \\ \mathrm{even,odd} \, (k+k_y) r_0}}} \, \frac{k^5 I(k^2 t)}{\sqrt{k^2 - k_y^2}} \left| \hat{f}(\sqrt{k^2 - k_y^2}, k_y) \pm \hat{f}(-\sqrt{k^2 - k_y^2}, k_y) \right|^2, \\
		&\to \frac{t}{2} \int_0^\infty dk \int_{-k}^k dk_y \, \frac{k^5 I(k^2 t)}{\sqrt{k^2 - k_y^2}} \left| \hat{f}(\sqrt{k^2 - k_y^2}, k_y) \pm \hat{f}(-\sqrt{k^2 - k_y^2}, k_y) \right|^2
\eea
with the upper (lower) sign for~$\Delta K^{(2)}_\mathrm{even}$ ($\Delta K^{(2)}_\mathrm{odd}$).  Note that the factor of~$1/2$ comes from the fact that for fixed~$k r_0$, we are only summing over~$k_y r_0$ with a given parity, and thus the spacing in~$k_y$ is~$2/r_0$.  We now first switch the integrals around by taking the range of the~$k_y$ integral to be~$(-\infty, \infty)$ and the range of the~$k$ integral to be~$(|k_y|, \infty)$, after which we change to a new variable~$k_x = \sqrt{k^2 - k_y^2}$ which has range~$(0, \infty)$.  Since~$dk_x = (k/k_x) dk$, we thus have
\bea
\Delta K^{(2)}_\mathrm{even,odd} &\to \frac{t}{2} \int_{-\infty}^\infty dk_y \int_0^\infty dk_x \, k^4 I(k^2 t) \left| \hat{f}(k_x, k_y) \pm \hat{f}(-k_x, k_y) \right|^2, \\
		&= \frac{t}{2} \int d^2 k \, k^4 I(k^2 t) \left( \left| \hat{f}(\vec{k}) \right|^2 \pm \hat{f}^*(-k_x, k_y) \hat{f}(k_x,k_y)\right),
\eea
with the latter expression obtained by expanding out the square and redefining~$k_x \to -k_x$ as appropriate.  Hence adding~$\Delta K^{(2)}_\mathrm{even}$ and~$\Delta K^{(2)}_\mathrm{odd}$ we obtain the flat-space expression~\eqref{eq:K2flatlimit}.

\section{Numerical Method}
\label{app:numericalchecks}

In this Appendix, we describe the numerical methods used to compute the heat kernels and free energy in Section~\ref{sec:nonpert} and Section~\ref{sec:endstate}.  We will always restrict to axisymmetric deformations of the sphere; almost all will be metrics of the form~\eqref{eq:NonPerturbativeDeformations} obtained from an embedding~$r = R(\theta)$, though we will also consider metrics of the form~\eqref{eq:intrinsiccone} to allow us to consider conical excesses (which cannot be embedded in~$\mathbb{R}^3$).  In what follows, it will be convenient to work with the function~$f = \ln R$, so that the deformed metric~\eqref{eq:NonPerturbativeDeformations} can be written as
\be
\label{eq:metricf}
ds^2 = e^{2f} \left[\left(1+f'(\theta)^2\right)d\theta^2 + \sin^2\theta \, d\phi^2 \right].
\ee
We will always use spherical coordinates~$\{\theta,\phi\}$ with ranges~$\theta \in [0,\pi]$ and~$\phi \in [0,2\pi)$.

To compute the differenced heat kernel we use the form given in~\eqref{eq:trace},
\be
\Delta K_L(t) = \sum_I \left(e^{-t \lambda_I} - e^{-t \bar{\lambda}_I}\right).
\ee
Due to the axisymmetry, the eigenfunctions of~$L$ are separable and can thus be written as~$h(\theta,\phi) = w(\theta) e^{im\phi}$, where~$m$ takes integer values for the scalar and half-integer values for the fermion.  We then have that~$L (w e^{-im\phi}) = e^{-im\phi} D_m w$, with~$D_m$ a second-order ordinary differential operator given explicitly on the geometry~\eqref{eq:metricf} by
\begin{multline}
\label{eq:Dmdef}
D_m w = e^{-2f}\left[-\frac{w''}{1 + (f')^2} + \frac{f'' f' - \cot\theta \left( 1 + (f')^2\right)}{\left(1 + (f')^2\right)^2} \, w' \right. \\ \left. \phantom{-\frac{h''}{1 + (f')^2}} + \left( A(\theta) + m B(\theta)  + m^2 \csc^2\theta \right) w \right],
\end{multline}
where for the scalar
\bea
A(\theta) &= -2 \xi \, \frac{\left(f'' - (f')^2 - 1\right) \left(1 - \cot\theta f'\right)}{\left(1 + (f')^2\right)^2}, \\
B(\theta) &= 0,
\eea
while for the fermion
\bea
A(\theta) &= - \frac{1}{2} \, \frac{\left(f'' - (f')^2 - 1\right) \left(1 - \cot\theta f'\right)}{\left(1 + (f')^2\right)^2} + \frac{1}{4} \frac{\left( \cot\theta + f' \right)^2}{1 + (f')^2}, \\
B(\theta) &= - \frac{\csc\theta \left( \cot\theta + f' \right)}{\sqrt{ 1 + (f')^2 }}.
\eea
The operator~$D_m$ on the conical geometry~\eqref{eq:intrinsiccone} can be obtained analogously, so we do not explicitly write it here.

We index the eigenvalues~$\lambda_I$ as follows.  Since~$\lambda_I$ is an eigenvalue of~$L$ if and only if it is also an eigenvalue of~$D_m$ for some allowed~$m$, it is natural to take~$m$ to index the corresponding subspaces of eigenvalues.  Within each subspace (that is, for each fixed~$m$), we then introduce an integer~$l \geq 0$ to index the eigenvalues of the operator~$D_m$ in ascending order.  We therefore label the eigenvalues as~$\lambda_{m,l}$: for any given~$m$, the~$\lambda_{m,l}$ for~$l = 0, 1, \ldots$ are all the eigenvalues of~$D_m$.  With this notation, the eigenvalues of~$\overline{L}$ (on the sphere with unit radius) are given by
\be
\bar{\lambda}_{m,l} = \begin{cases} (|m| + l)(|m| + l + 1) + 2\xi, & \mbox{scalar}, \\ \left(|m| + l + \frac{1}{2}\right)^2, & \mbox{fermion}; \end{cases}
\ee
in other words, the usual quantum number~$\ell$ is replaced by~$|m| + l$, enforcing that for a fixed~$m$,~$\ell \geq |m|$.

The eigenvalues~$\lambda_{m,l}$ are determined numerically by discretizing the operators~$D_m$ over the interval~$\theta \in [0, \pi]$ using standard pseudospectral differencing with a Chebyshev grid of~$N+2$ lattice points including the two boundary points~$\theta = 0,\pi$ (see for example~\cite{Trefethen2000}).  One subtlety is the appropriate treatment of the poles~$\theta = 0$ and $\pi$.  Regularity of the metric~\eqref{eq:metricf} requires that~$f'(0) = 0 = f'(\pi)$, and hence an expansion of~$f$ around these points has a vanishing linear term.  Using~\eqref{eq:Dmdef} for the scalar field, it then follows from a Frobenius expansion that near~$\theta = 0$ any solution to~$D_m w = \lambda_{m,l} w$ admits a regular behavior that goes like
\be
w = \theta^{|m|} \left(w_0 + \theta^2 w_2 + \cdots \right)
\ee
with~$w_0 \neq 0$, in addition to a singular behavior that goes like~$\theta^{-|m|}$.  At the other pole, we have an analogous behavior:
\be
w = \left( \pi - \theta \right)^{|m|} \left( \tilde{w}_0 +  \left( \pi - \theta \right)^2 \tilde{w}_2 + \cdots \right).
\ee
Hence when we difference the operators~$D_m$ with~$m \neq 0$, we impose Dirichlet boundary conditions at the poles, while when we difference the operator~$D_0$ we impose Neumann boundary conditions.  On the other hand, for the fermion we instead have the allowed behaviors
\bea
w &= \theta^{\left| m - \frac{1}{2} \right|} \left( w_0 + \theta^2 w_2 + \cdots \right), \\
w &= \left( \pi - \theta \right)^{\left| m + \frac{1}{2} \right|} \left( \tilde{w}_0 +  \left( \pi - \theta \right)^2 \tilde{w}_2 + \cdots \right).
\eea
Thus for~$| m | \ne 1/2$ we discretize with Dirichlet boundary conditions at both poles, while for~$m=1/2$ we take a Neumann boundary condition at $\theta =0$ and Dirichlet at $\theta = \pi$, and likewise for $m = -1/2$ a Neumann condition at $\theta = \pi$ and Dirichlet at $\theta = 0$.

Thus for each~$m$ we obtain an~$N \times N$ matrix representing the discretization of~$D_m$; for large~$N$, the~$N$ eigenvalues of this matrix should approximate the eigenvalues~$\lambda_{m,l}$ for sufficiently low~$l < N$.  Of course, a finite~$N$ will not be able to keep track of eigenvalues with~$m$ too large, so some cutoff on~$m$ must be imposed.  A natural one is suggested by the spherical harmonics: on the round sphere, we might wish to keep all eigenvalues up to a fixed~$\ell = |m| + l$; since~$l < N$, the strongest constraint is obtained by considering the lowest allowed~$|m|$, which fixes a cutoff~$\ell < N$.  Implementing this same cutoff procedure on the deformed sphere leads us to keeping all eigenvalues satisfying~$|m| + l < N$: for each allowed (i.e.~integer or half-integer)~$m$ with~$|m| < N$, we compute the eigenvalues of the discretized operator~$D_m$ and keep only the ones with~$l < N - |m|$.  The actual computation of the eigenvalues of the discretized~$D_m$ is conveniently done with the Arnoldi algorithm which is implemented in the {\tt Mathematica} matrix eigenvalue finder.  We also note that for the minimally coupled scalar (i.e.~$\xi = 0$), we explicitly drop the lowest eigenvalue~$m=0$,~$l = 0$ because as discussed in Section~\ref{subsec:asymptotics} it is the same for both~$L$ and $\overline{L}$ and thus cancels exactly in the differenced heat kernel.

For a given~$N$, a truncated differenced heat kernel can then be defined:
\be
\Delta K_L^{(N)}(t) = \sum_{|m| < N} \, \sum_{l < N-|m|} \left(e^{-t\lambda_{m,l}^{(N)}} - e^{-t \bar{\lambda}_{m,l}}\right),
\ee
where~$\lambda^{(N)}_{m,l}$ are the eigenvalues of the discretized operators~$D_m$, as described above.  Increasing~$N$ should yield a better approximation to the exact heat kernel.  We expect this approximation to be best at large~$t$, in which the sum is dominated by the smallest eigenvalues, while the approximation should fail for sufficiently small~$t$, when many eigenvalues make nontrivial contributions to the sum.  Since the heat kernel time~$t$ can be thought of as an inverse square of a length scale, we expect that for a fixed~$N$ the agreement should fail for~$t$ smaller than order $\sim \ell_\mathrm{max}^{-2} \sim N^{-2}$.

However, the differenced free energy \textit{is} sensitive to the small-$t$ behavior of the differenced heat kernel.  To accurately compute the free energy, we therefore implement a cutoff time~$t_\mathrm{cut}$ above which we integrate~\eqref{eq:DeltaFheatkernel} with the truncated heat kernel~$\Delta K_L^{(N)}$, and below which we integrate~\eqref{eq:DeltaFheatkernel} using the leading-order behavior~$\Delta b_4 t$ from the heat kernel expansion~\eqref{eq:DeltaKexpansion}.  For each~$N$ and choice of cutoff~$t_\mathrm{cut}$, this gives an approximation to the free energy:
\begin{multline}
\label{eq:cutoffDeltaF}
\Delta F^{(N,t_\mathrm{cut})} = \sigma T \left( \Delta b_4 \int_0^{t_\mathrm{cut}} dt \, e^{-M^2 t} \Theta_\sigma(T^2 t) \right. \\ \left. + \int_{t_\mathrm{cut}}^\infty \frac{dt}{t} \, e^{-M^2 t} \Theta_\sigma(T^2 t) \Delta K_L^{(N)}(t) \right).
\end{multline}
The accuracy of this approximation relies on~$\Delta K^{(N)}_L$ being well-approximated by the linear behavior~$\Delta b_4 t$ around~$t = t_\mathrm{cut}$, so that~$\Delta F^{(N,t_\mathrm{cut})}$ is in fact independent of~$t_\mathrm{cut}$.  With our choice of~$N = 600$, the truncated heat kernel~$\Delta K^{(N)}_L$ gives a good approximation down to~$t \sim 2 \times 10^{-4}$.  For moderate deformations of the sphere (up to around~$\eps = 0.5$-$0.7$ for the deformations~\eqref{eq:Rellembedding}, depending on~$\ell$),~$\Delta K^{(N)}_L$ agrees well with the leading-order behavior~$\Delta b_4 t$ around this lowest value of~$t$, and we may therefore compute the free energy as described.  In this case, typically we take~$t_\mathrm{cut} = 2.5 \times 10^{-4}$, and then varying~$t_\mathrm{cut}$ gives an estimate of the systematic error in~$\Delta F^{(N,t_\mathrm{cut})}$ (for all plots in the main text, this error is no greater than a few percent).  For larger deformations, however,~$\Delta K^{(N)}_L$ is \textit{not} well-approximated by the leading-order behavior of the heat kernel expansion around~$t \sim 2\times 10^{-4}$, and therefore we are unable to accurately compute the differenced free energy for such deformations.

We now discuss in more detail the convergence of~$\Delta K^{(N)}_L$ with~$N$, agreement with the heat kernel expansion at small~$t$, and agreement with the perturbative results for small deformations of the sphere.

\subsection{Convergence}

Since we are using pseudospectral differencing, we expect the error in a given eigenvalue to fall exponentially with~$N$ until a limit from machine precision is reached.  Since the truncated differenced heat kernel is constructed directly from the eigenvalues, it too should converge to~$\Delta K_L$ exponentially until hitting machine precision.  This convergence is fast at large $t$, since there~$\Delta K_L$ is sensitive to only the smallest eigenvalues, whereas at small~$t$ convergence (which is still exponential) requires larger~$N$ to achieve the same accuracy. 

\begin{figure}[t]
\centering
\subfloat[][Minimally coupled scalar]{
\includegraphics[width=0.49\textwidth]{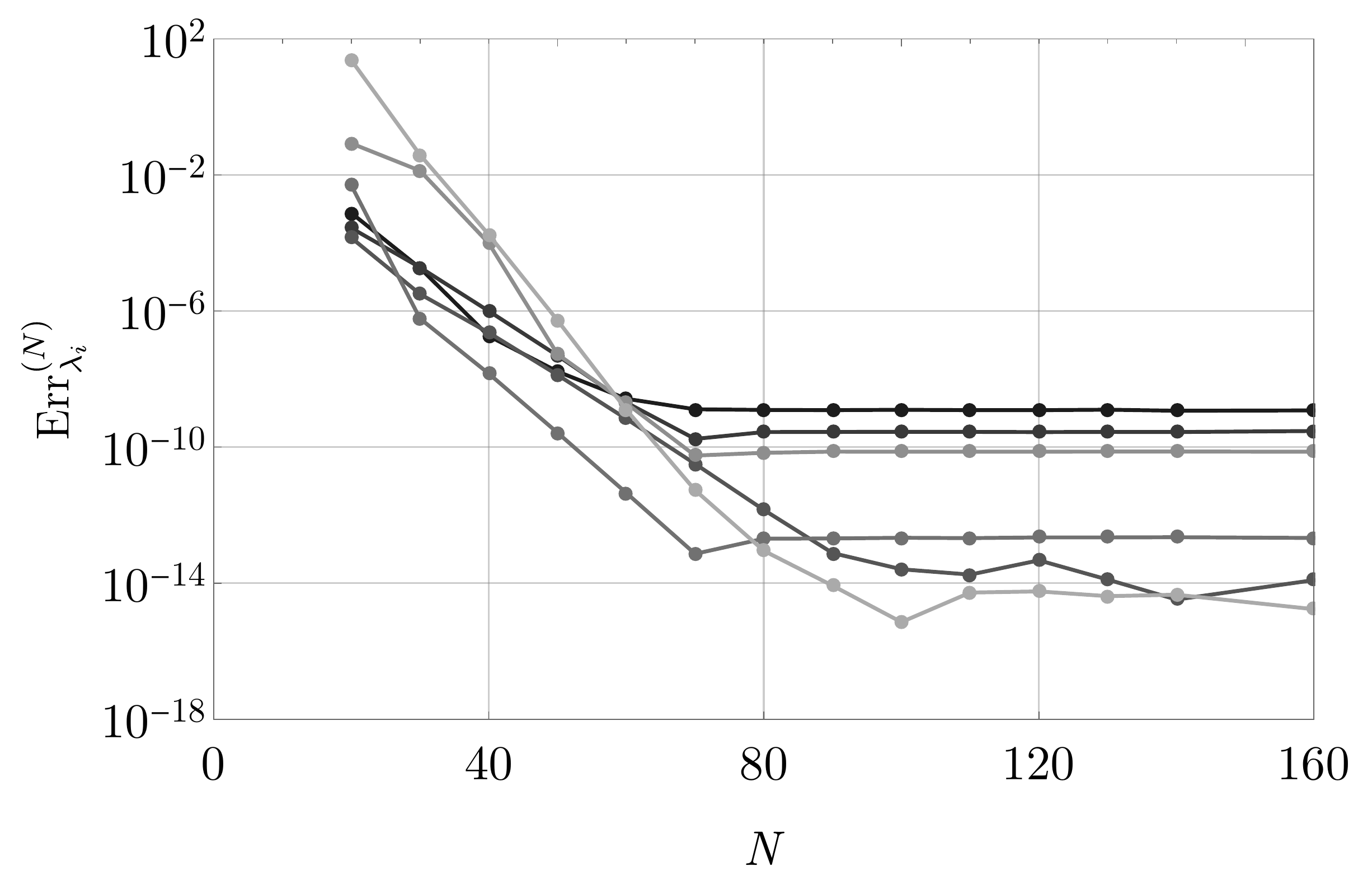}
}%
\subfloat[][Fermion]{
\includegraphics[width=0.49\textwidth]{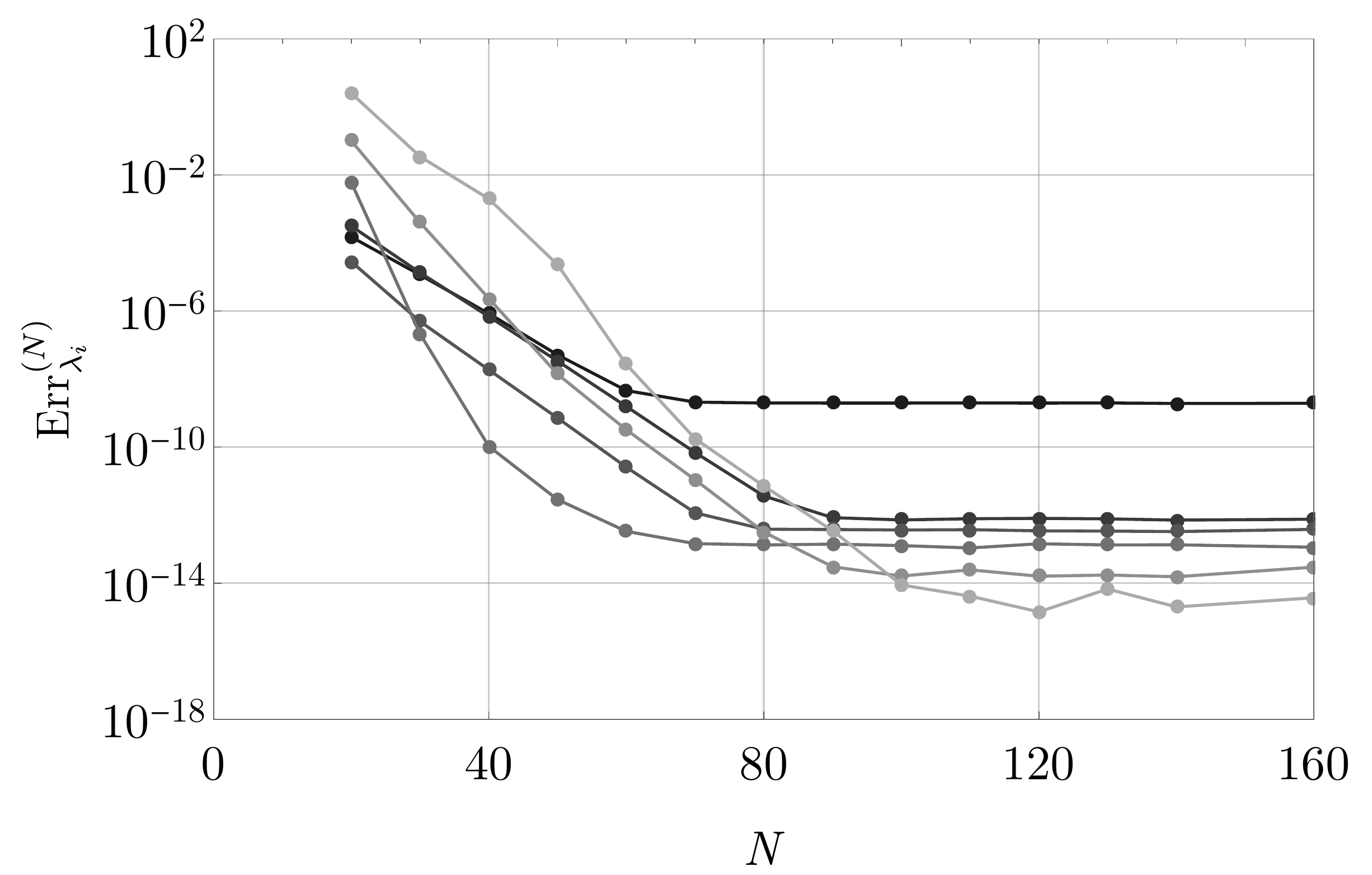}
}
\caption{Plots of estimated fractional error~\eqref{eq:fractionalerrorlambda} for the eigenvalues of~$L$ on the geometry given by the embedding~\eqref{eq:Rellembedding} with~$\ell = 3$ and~$\eps/\eps_\mathrm{max} = 0.5$.  From dark to light gray, each curve corresponds to the~$i^\mathrm{th}$ eigenvalue of~$L$ with~$i = 2, 10, 25, 100, 200, 400$.}
\label{fig:EvalConvergence}
\end{figure}

To exhibit this convergence, let us order the eigenvalues of~$L$ in ascending order; for a given resolution~$N$, we then define the fractional error in the~$i^\mathrm{th}$ eigenvalue as
\be
\label{eq:fractionalerrorlambda}
\mathrm{Err}^{(N)}_{\lambda_i} = \left| \frac{ \lambda^{(N)}_i - \lambda^{(N_\mathrm{max})}_i }{ \lambda^{(N_\mathrm{max})}_i } \right|,
\ee
where the maximum resolution we use is~$N_\mathrm{max} = 600$.  We plot this fractional error in Figure~\ref{fig:EvalConvergence} for several eigenvalues in the geometry corresponding to the~$\ell = 3$,~$\eps = 0.5\eps_\mathrm{max}$ embedding~\eqref{eq:Rellembedding}.  This corresponds to a non-linearly deformed sphere, although one that still is not very close to being singular.  We see that all the eigenvalues converge exponentially with~$N$ until reaching machine precision around~$N \sim 100$.  As we would expect, it is the lower eigenvalues that suffer most from machine precision limitations in terms fractional error since they have a smaller absolute value (roughly the magnitude of the eigenvalues goes as~$\lambda_i \sim i$).  The data indicate that at the resolution~$N_\mathrm{max} = 600$ used in this paper, the eigenvalues have a fractional error less than~$\sim 10^{-8}$ compared to their exact values.

\begin{figure}[t]
\centering
\subfloat[][Minimally coupled scalar]{
\includegraphics[width=0.49\textwidth]{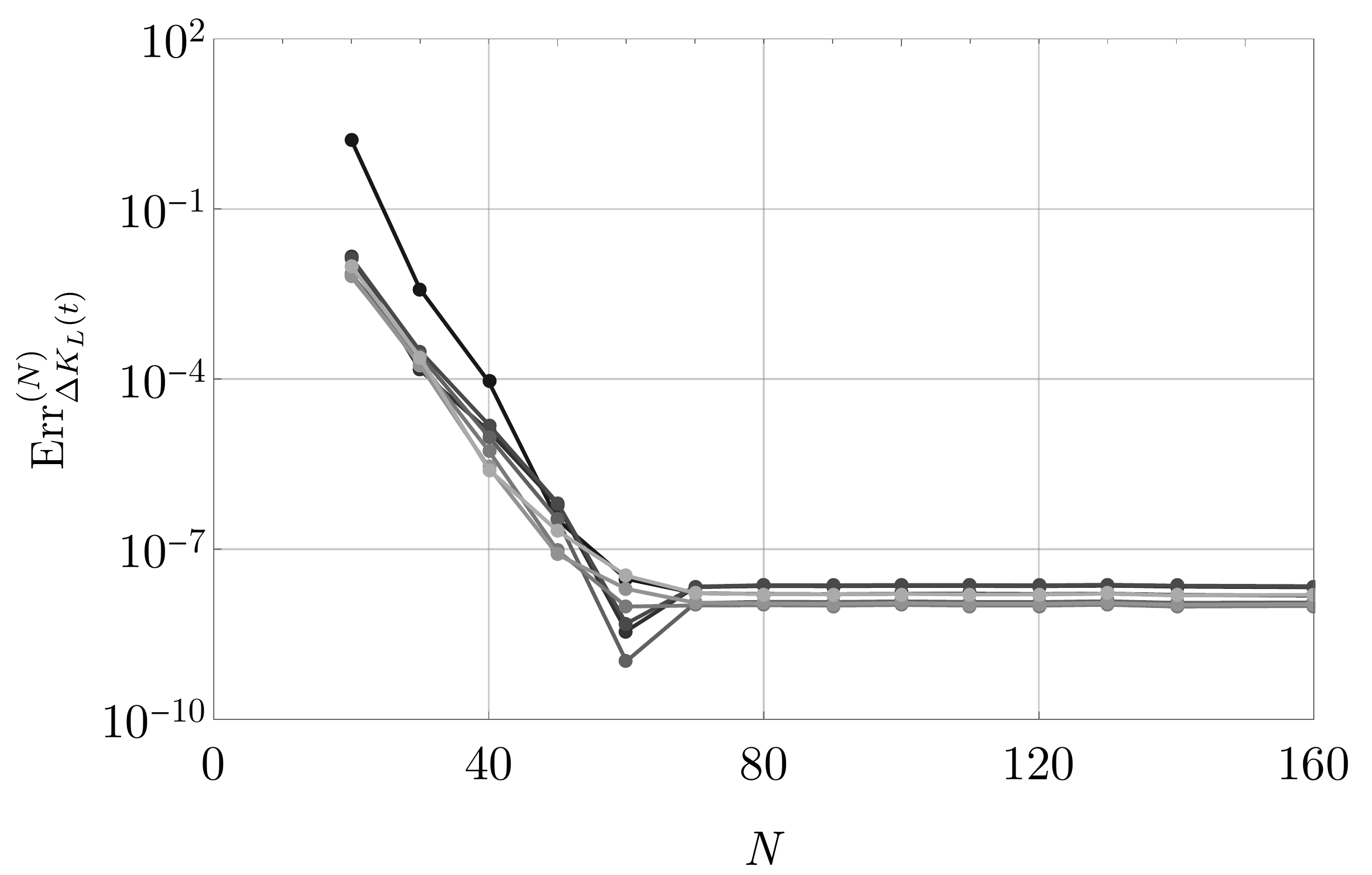}
}%
\subfloat[][Fermion]{
\includegraphics[width=0.49\textwidth]{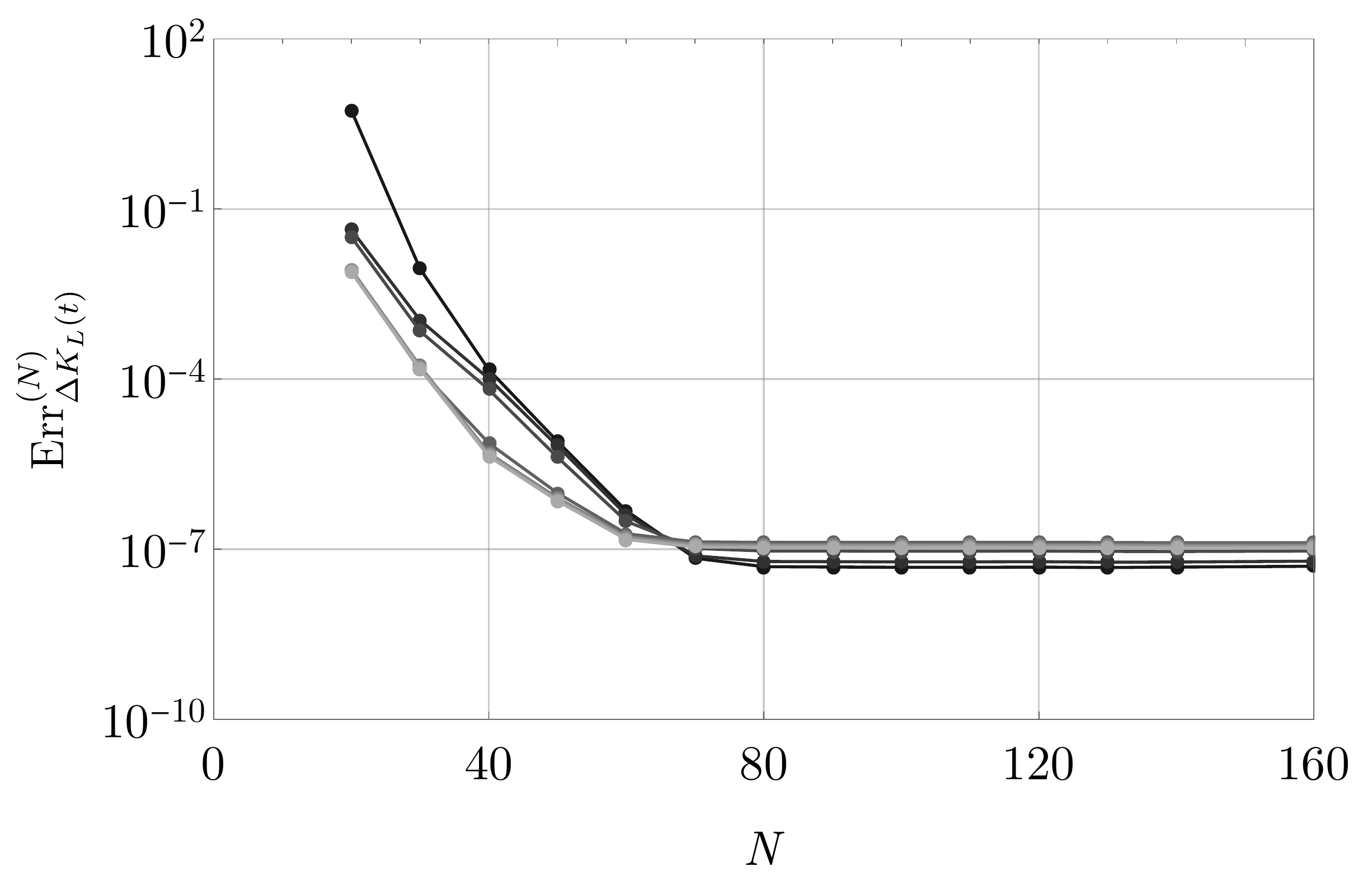}
}
\caption{The fractional error~\eqref{eq:fractionalerrorDeltaK} in the differenced heat kernel on the geometry given by the embedding~\eqref{eq:Rellembedding} with~$\ell = 3$ and~$\eps/\eps_\mathrm{max} = 0.5$; from dark to light gray, the curves correspond to~$t = 0.05, 0.1, 0.2, 1, 2, 4, 8$.}
\label{fig:HKConvergence}
\end{figure}

We may likewise define the fractional error in the differenced heat kernel as
\be
\label{eq:fractionalerrorDeltaK}
\mathrm{Err}^{(N)}_{\Delta K_L(t)} = \left| \frac{ \Delta K^{(N)}_L(t) - \Delta K^{(N_\mathrm{max})}_L(t) }{ \Delta K^{(N_\mathrm{max})}_L(t) } \right|.
\ee
This fractional error is shown in Figure~\ref{fig:HKConvergence} (in the same~$\ell = 3$,~$\eps = 0.5\eps_\mathrm{max}$ geometry~\eqref{eq:Rellembedding}) for several different values of~$t$.  Again, we observe initial exponential convergence before we become machine precision limited by~$N \sim 80$.  Note that smaller~$t$ requires a larger $N$ to reach the same accuracy, but the rate of convergence is roughly independent of~$t$.  We can estimate that for $t > 0.05$ the fractional error in the differenced heat kernel at~$N_\mathrm{max} = 600$ is better than $\sim 10^{-7}$, which is commensurate with the error in the individual eigenvalues.

\subsection{Comparison to Heat Kernel Expansion and to Perturbative Results}

\begin{figure}[t]
\centering
\subfloat[][Minimally coupled scalar]{
\includegraphics[width=0.47\textwidth]{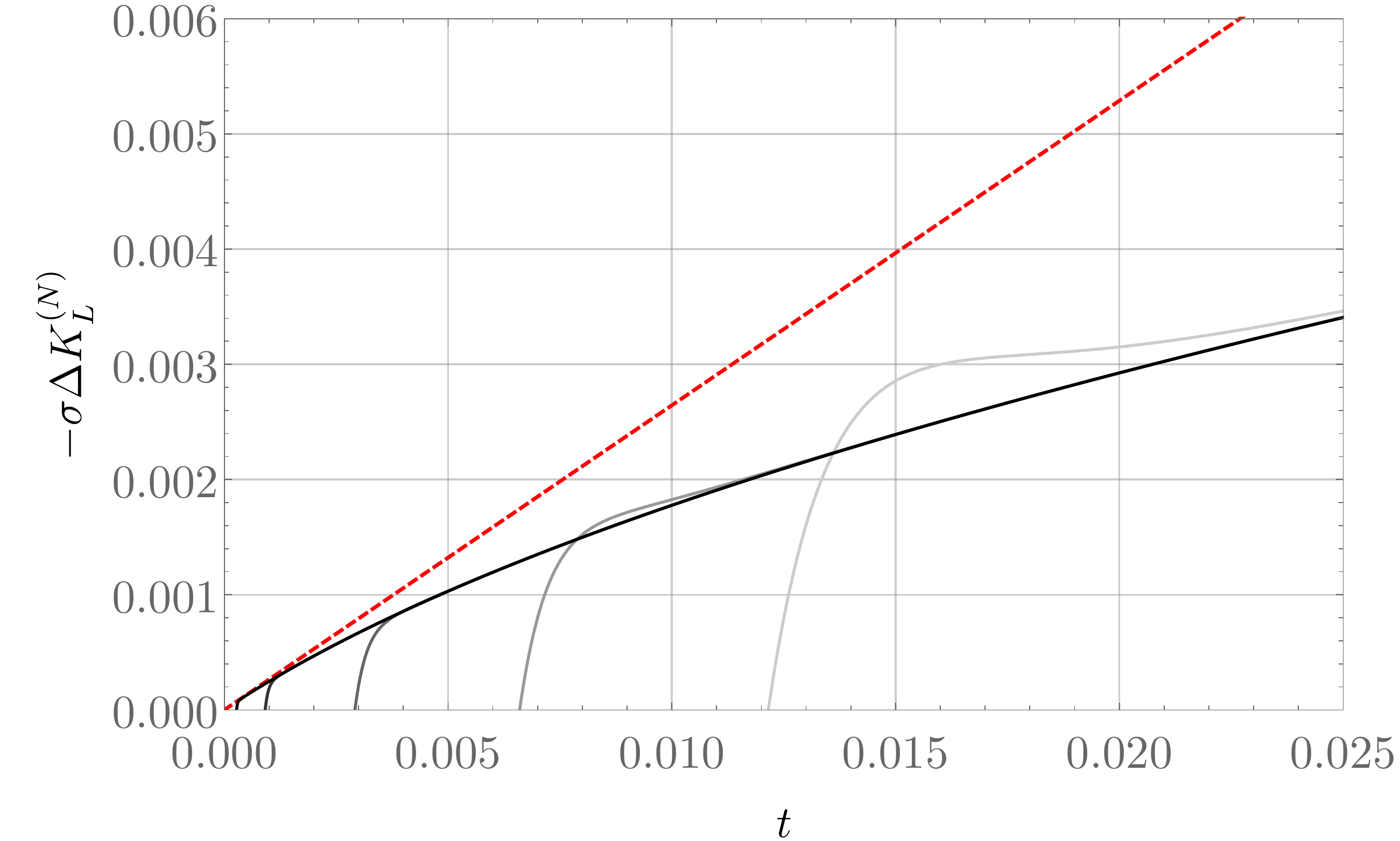}
}%
\subfloat[][Fermion]{
\includegraphics[width=0.47\textwidth]{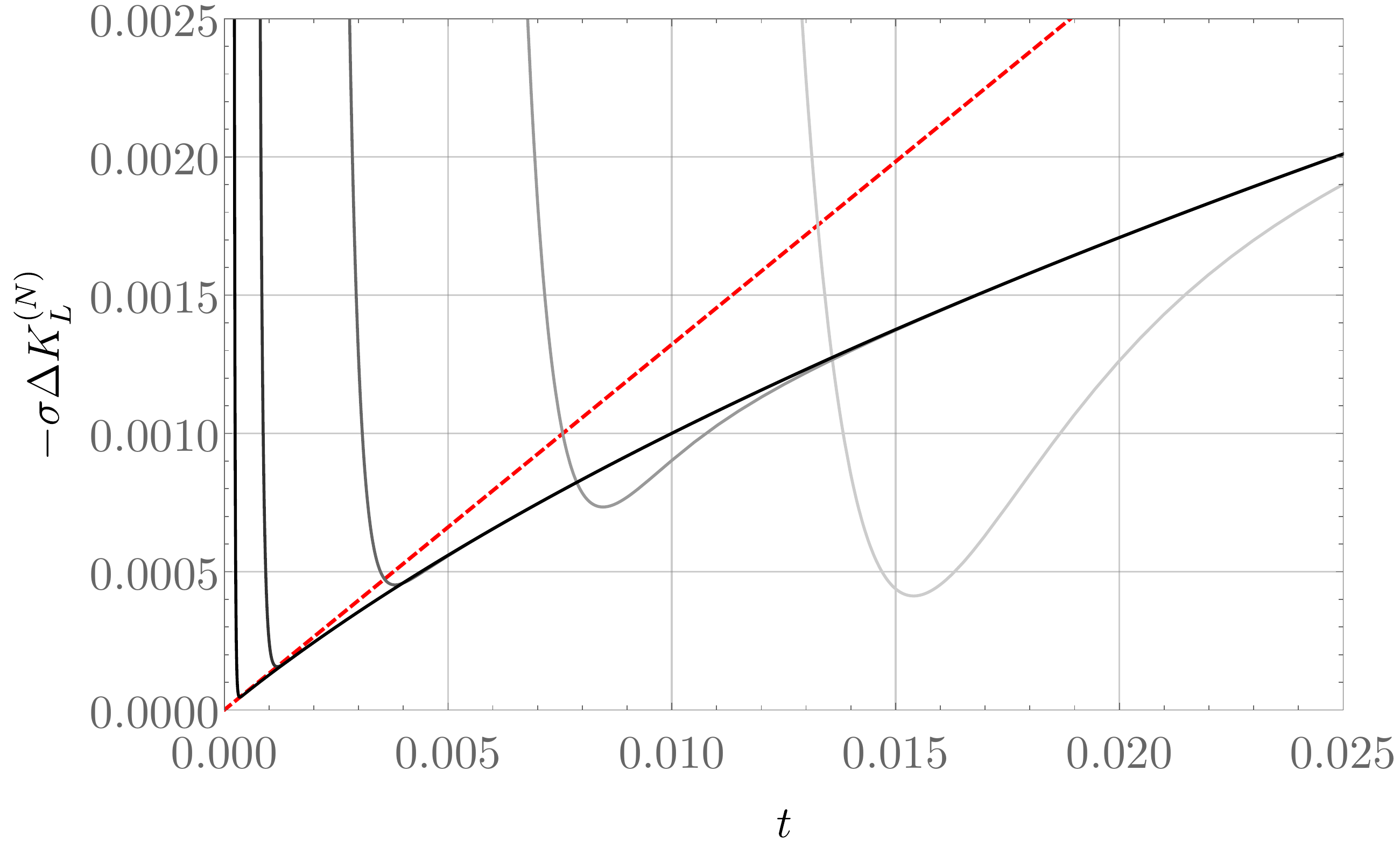}
}
\caption{Small-$t$ behavior of the truncated differenced heat kernel~$\Delta K^{(N)}_L$ for various~$N$; as in Figures~\ref{fig:EvalConvergence} and~\ref{fig:HKConvergence}, here we show the result for the geometry given by the embedding~\eqref{eq:Rellembedding} with~$\ell = 3$ and~$\eps/\eps_\mathrm{max} = 0.5$.  From light to dark gray, the curves correspond to~$N = 40, 60, 100, 200, 400$, while the dashed red line shows the linear behavior~$\Delta b_4 t$ expected from the heat kernel expansion.  Note that the linear behaviour only approximates the differenced heat kernel for quite small $t$, and we need to take~$N \gtrsim 400$ to reach this linear regime.}
\label{fig:HKExp}
\end{figure}

In addition to allowing us to compute the differenced free energy via~\eqref{eq:cutoffDeltaF} as described above, verifying that the heat kernel approaches the behavior predicted from the heat kernel expansion at small~$t$ also provides a check of our numerical methods.  To that end, in Figure~\ref{fig:HKExp} we compare the small-$t$ behavior of~$\Delta K^{(N)}_L$ for various~$N$ to the linear behavior~$\Delta b_4 t$ expected from the heat kernel expansion; again we are taking the~$\ell = 3$,~$\eps = 0.5\eps_\mathrm{max}$ embedding~\eqref{eq:Rellembedding} as a typical example.  There are two features to highlight.  First, even the lowest value~$N = 40$ recovers the heat kernel well above~$t \sim 0.1$, but computing the heat kernel accurately at very small~$t$ clearly requires using larger values of~$N$.  In particular, with the choice of~$N_\mathrm{max} = 600$ used in this paper, we can reliably compute the heat kernel down to~$t \sim 2 \times 10^{-4}$ (with some variation depending on the deformation).  Second, while the linear approximation~$\Delta b_4 t$ does agree with the truncated heat kernel for sufficiently large~$N$, for even moderate deformations of the sphere this agreement is only valid for very small~$t$ (for the case shown here, the fractional error between the linear behavior and the heat kernel is less than about two percent for~$t < 5 \times 10^{-4}$, but grows much larger for larger~$t$).  For larger deformations this agreement moves to smaller and smaller~$t$, eventually leaving the domain in which we can reliably approximate the exact heat kernel.  It is for this reason that~\eqref{eq:cutoffDeltaF} cannot be used to approximate the differenced free energy for very large deformations.

\begin{figure}[t]
\centering
\subfloat[][Minimally coupled scalar]{
\includegraphics[width=0.49\textwidth]{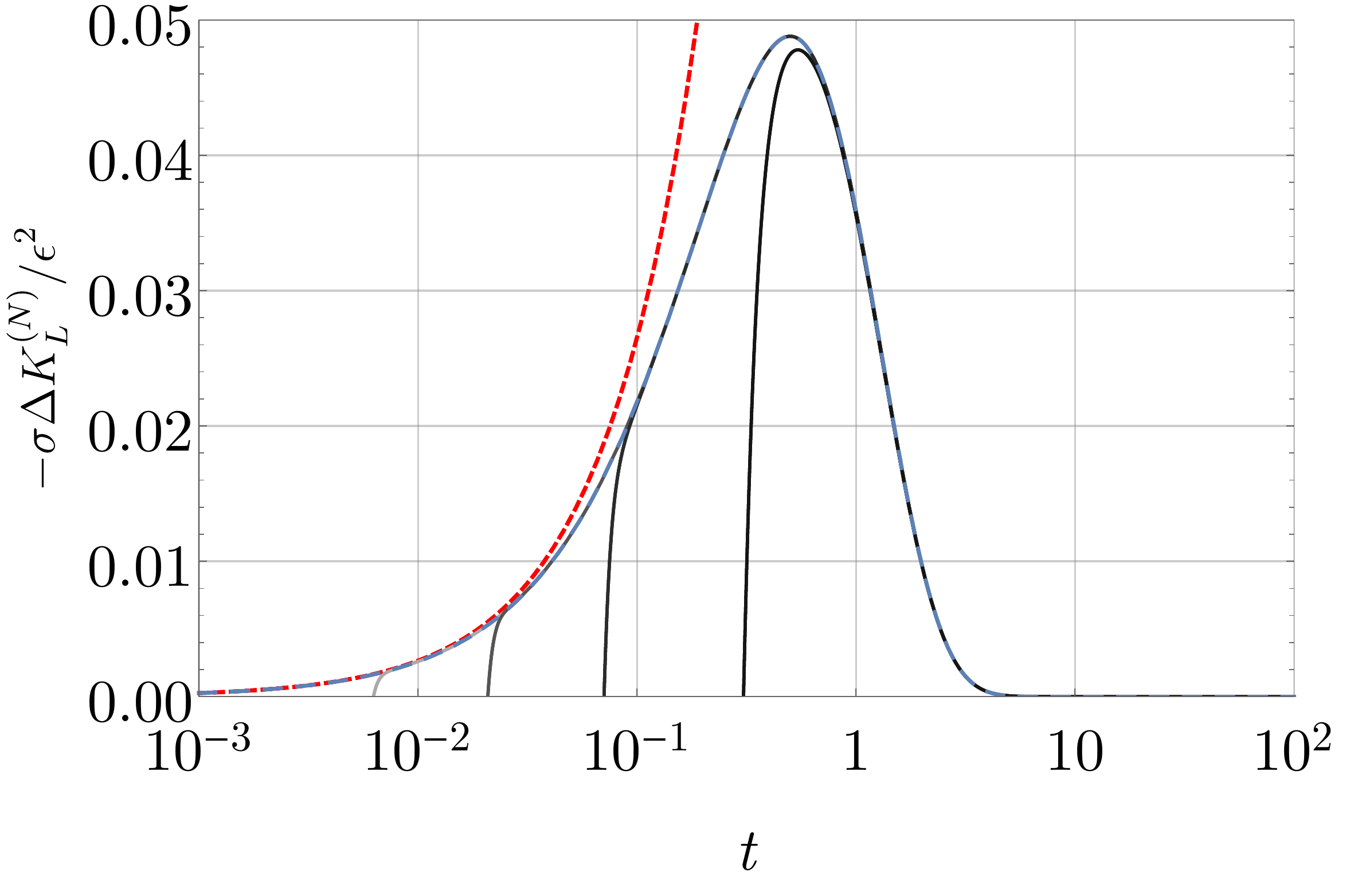}
}%
\subfloat[][Fermion]{
\includegraphics[width=0.49\textwidth]{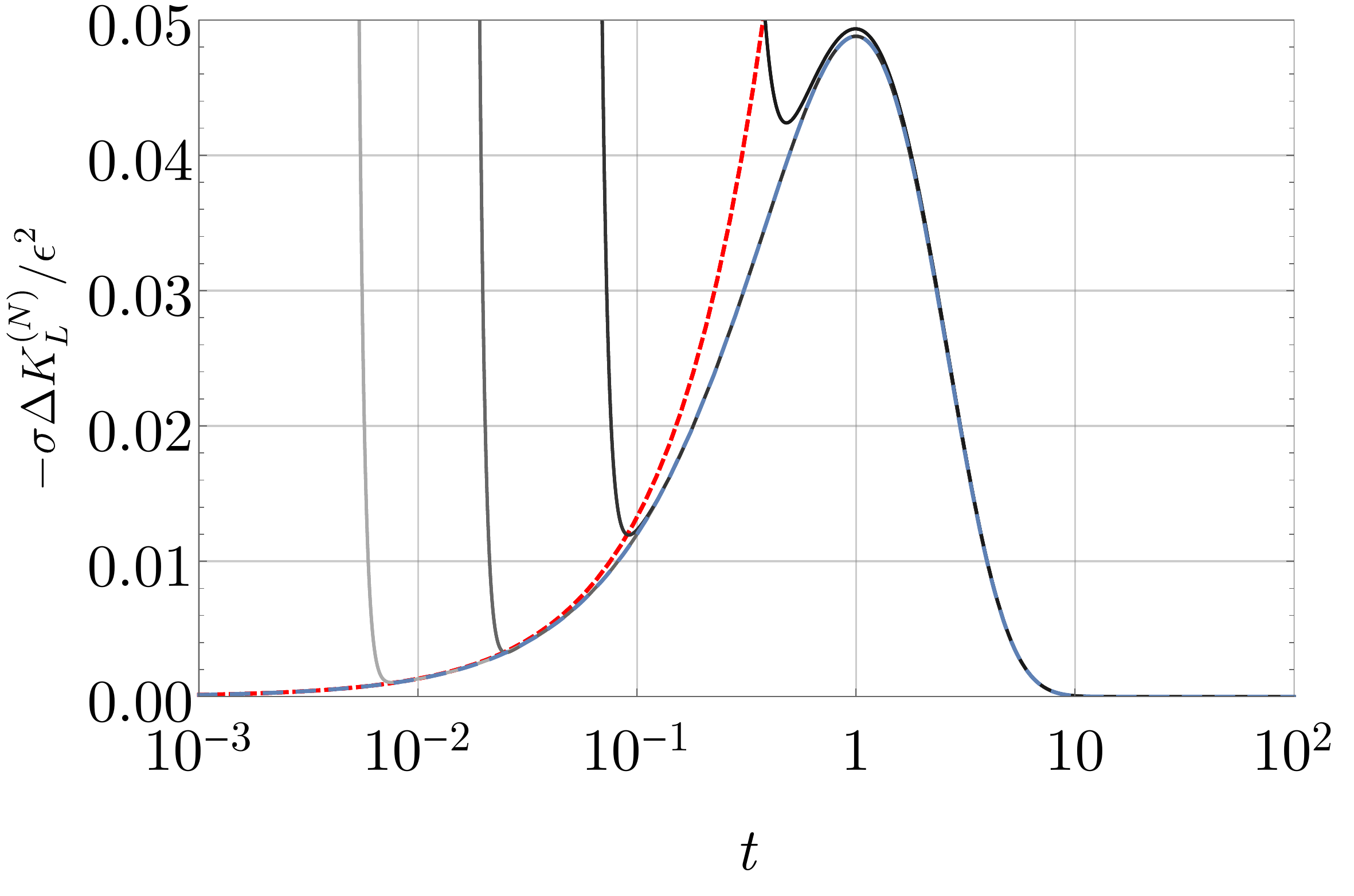}
}
\caption{The convergence of the numerically-computed truncated heat kernel~$\Delta K^{(N)}_L$ to the perturbative results, for the~$\ell = 3$ deformation~\eqref{eq:Rellembedding} with~$\eps = 0.01$.  The blue dashed curves show the perturbative heat kernels~\eqref{eq:K2scalar} and~\eqref{eq:K2dirac}, while the dark to light gray curves show the truncated differenced heat kernel for~$N = 10, 20, 40, 80$.  As in Figure~\ref{fig:HKExp} we see that accuracy at small $t$ requires larger $N$.  We also show the leading linear behavior~$\Delta b_4 t$ from the heat kernel expansion (dashed, red).}
\label{fig:HKPert}
\end{figure}

As an additional check of our numerical method, we may compare the truncated heat kernel for very small deformations of the sphere to the perturbative heat kernels~\eqref{eq:K2scalar} and~\eqref{eq:K2dirac}.  We show this agreement in Figure~\ref{fig:HKPert}, again for the~$\ell = 3$ embedding~\eqref{eq:Rellembedding} but now only with a weak deformation of~$\eps = 0.01$.  Even for modest~$N$,~$\Delta K_L^{(N)}$ is very close to the perturbative result for reasonably large~$t$.  Increasing~$N$ gives agreement with the perturbative results to smaller~$t$, as expected.  We also show a comparison with the leading-order heat kernel expansion; unlike the moderate deformation~$\eps/\eps_\mathrm{max} = 0.5$ shown in Figure~\ref{fig:HKExp}, here we see good agreement with the expected linear behavior up to almost~$t \sim 0.1$.

\bibliographystyle{jhep}
\bibliography{biblio}

\end{document}